\pdfoutput=1

\documentclass[onecolumn, 10pt]{IEEEtran}
\usepackage{amsfonts}
\usepackage{amsmath}
\usepackage{algorithmic}
\usepackage{algorithm}
\usepackage{overpic}
\usepackage{slashbox}
\usepackage{epsfig}
\usepackage{latexsym}
\usepackage{amssymb}
\usepackage{amsmath}
\usepackage{graphicx}
\usepackage{subfigure}
\usepackage{url}
\usepackage{dsfont}
\usepackage{caption}
\usepackage{mathrsfs}
\usepackage{multirow}
\usepackage{cite}
\usepackage{epstopdf}
\usepackage[usenames]{color}
\usepackage{array}
\usepackage{bbm}
\usepackage{diagbox}
\usepackage[thinc]{esdiff}
\usepackage{accents}

\renewcommand{\baselinestretch}{1.2} 

\ifodd 1

\newcommand{\com}[1]{\textbf{\color{red} (#1)}}

\else

\newcommand{\com}[1]{}
\newcommand{\response}[1]{}
\fi

\newtheorem{que}{Key Question}

\newtheorem{gam}{Game}
\newtheorem{prop}{Proposition}

\newtheorem{thm}{Theorem}

\newtheorem{defin}{Definition}

\DeclareMathOperator*{\argmax}{\arg\max}

\begin{document}
\bibliographystyle{IEEEtran}

\title{Mechanism Design for Blockchain Order Books against Selfish Miners }
\author{Yunshu Liu and Lingjie Duan, \emph{Senior Member, IEEE}\footnote{Y. Liu is with the School of Cybersecurity, Northwestern Polytechnical University, Xi'an 710129, China (email: liuyunshu93@gmail.com). 
		
		L. Duan is with the Pillar of Engineering Systems and Design, Singapore University of Technology and Design, Singapore 487372 (email: lingjie\_duan@sutd.edu.sg). 
		
		This work is supported in part by the Ministry of Education, Singapore, under its Academic Research Fund Tier 2 Grant with Award no. MOE-T2EP20121-0001; in part by SUTD Kickstarter Initiative (SKI) Grant with no. SKI 2021\_04\_07; and in part by the Joint SMU-SUTD Grant with no. 22-LKCSB-SMU-053.
}}

\date{\today}

\maketitle


\begin{abstract}
	In blockchain-based order book systems, buyers and sellers trade assets, while it is miners to match them and include their transactions in the blockchain. It is found that many miners behave selfishly and myopically, prioritizing transactions with high fees and ignoring many desirable matches that could enhance social welfare. Existing blockchain mechanisms fail to address this issue by overlooking miners' selfish behaviors. To our best knowledge, this work presents the first analytical study to quantify and understand buyer and seller transaction fee choices and selfish miners' transaction matching strategies, proving an infinitely large price of anarchy (PoA) for social welfare loss. To mitigate this, we propose an adjustable block size mechanism that is easy to implement without altering the existing decentralized protocols and still allows buyers and sellers to freely decide transaction fees and miners to selfishly match. The analysis is challenging, as pure strategy Nash equilibria do not always exist, requiring the analysis of many buyers' or sellers' interactive mixed-strategy distributions. Moreover, the system designer may even lack information about each buyer's or seller's bid/ask prices and trading quantities. Nevertheless, our mechanism achieves a well-bounded PoA, and under the homogeneous-quantity trading for non-fungible tokens (NFT), it attains a PoA of 1 with no social welfare loss. We implement our mechanism on a local instance of Ethereum to demonstrate the feasibility of our approach. Experiments based on the realistic dataset demonstrate that our mechanism achieves social optimum for homogeneous-quantity trading like NFT. It can enhance social welfare up to 3.7 times compared to the existing order book benchmarks for heterogeneous-quantity trading of Bitcoin tokens. It exhibits robustness against random variations in buyers and sellers. 
\end{abstract}


\maketitle

\section{Introduction } \label{sec:intro}

\subsection{Motivations}
\label{sec:mot}

Blockchain-based order books (BBOB) have garnered significant attention due to their ability to facilitate secure, decentralized transactions in various markets, including decentralized exchanges (DEXs) and non-fungible tokens (NFTs) \cite{SIG3}. The market value of BBOBs continues to rise, with a current market cap of \$864 million in 2024 \cite{Market_value}. Within these systems, buyers aim to purchase assets at the lowest possible price, while sellers seek to sell at the highest price. Both parties set transaction fees to incentivize miners, who then decide which transactions to match and confirm on the blockchain. 

Despite fast development, current BBOBs face significant challenges. Many existing protocols suffer from poor performance, where the executed trade price deviates from the expected price, leading to inefficient trades \cite{Order_book_issue}\cite{SIG2}. One of the key reasons for the current BBOB design's poor performance is miners' self-interested or myopic behavior. Miners may prioritize those transactions with higher fees, even if these result in lower social welfare, thereby misaligning individual incentives with the collective benefits of the system \cite{gebraselase2021effect}.

Addressing this issue requires research efforts in two areas. First, buyers and sellers continuously experience benefit loss and complain about the system's poor performance caused by miners \cite{Miner_issue1}\cite{Miner_issue2}, yet our assessment of the potential loss is limited. It is crucial to assess the extent to which miners' self-interested actions degrade system performance, which motivates the following key question:
\begin{que}
	How much do miners' self-interested actions worsen the BBOB system's trading performance? 
\end{que}
Answering Key Question 1 requires characterizing and understanding the strategic interactions between self-interested buyers, sellers, and miners. Our rigorous game-theoretical analysis reveals that miners' self-interested or myopic actions can lead to arbitrarily large social welfare losses. With the understanding of this huge damage and its condition to happen, there is a strong need to design a mechanism mitigating the adverse effects of self-interested miner behavior. The mechanism should be practical and operate within existing blockchain protocols, e.g., Bitcoin Cash \cite{Simple_mechanism}. This leads us to ask the next key question:
\begin{que}
	How to design an easy-to-implement mechanism to maximize the social welfare of the BBOB system?
\end{que}
The challenge for addressing Key Question 2 lies in not forcing buyers, sellers, and miners to adopt a completely new protocol or even mechanism. Therefore, we will turn to key BBOB parameters that can be controlled easily and adjusted flexibly by the system, focusing on block size design. By optimizing this parameter, we aim to create a practical and effective mechanism in enhancing social welfare.

Our work addresses the incentive issues in BBOB systems. Several existing studies focus on various user-related aspects within this area (e.g., \cite{daian2020flash,schnaubelt2022deep,hautsch2024building,heimbach2022sok,victor2021detecting,raheman2021architecture}). Daian \emph{et al.} \cite{daian2020flash} analyzed the frontrunning issue among buyers and sellers in BBOB systems. Schnaubelt \emph{et al.} \cite{schnaubelt2022deep} proposed a reinforcement learning method to optimize users' order placements in BBOB systems. Hautsch \emph{et al.} \cite{hautsch2024building} examined arbitrage opportunities for users. Heimbach \emph{et al.} \cite{heimbach2022sok} designed a mechanism to mitigate the frontrunning issue. Victor \emph{et al.} \cite{victor2021detecting} detected and quantified wash trading behavior in the system. Raheman \emph{et al.} \cite{raheman2021architecture} designed an automated agent to collect revenue from the BBOB system. However, the existing literature on mechanism design in BBOB systems has not adequately considered miners' incentives. Our work fills in this gap by analyzing the impact of miners' incentives and proposing an incentive mechanism to mitigate these effects.

\vspace{-5pt}
\subsection{Contributions}
\vspace{-2pt}
In this paper, we consider the strategic interaction among buyers, sellers, and miners within the system designer's mechanism design framework. Specifically, the system designer first determines the block size to maximize social welfare. Subsequently, buyers and sellers decide their transaction fees. Finally, miners select transactions to match and include in the blockchain based on these fees.


As such, we need to tackle several intricate technical challenges in the BBOB system. We model the complex and coupled dynamic game-theoretical interactions among self-interested buyers, sellers, and miners. The analysis is further complicated by the non-existence of pure strategy Nash equilibria under certain conditions, requiring the analysis of many buyers' or sellers' interactive mixed-strategy distributions. What is worse, the mechanism design may lack buyers' and sellers' detailed trading information, making the problem quite challenging. Our mechanism must be practical, easy to implement, and compatible with existing blockchain protocols. These challenges highlight our research's novelty and significant technical contributions to the field.

Our key results and contributions are summarized as follows:
\begin{itemize}
	
	\item \emph{First analytical work to curb selfish miners in BBOB system}: Prior literature's mechanisms (e.g., \cite{daian2020flash,schnaubelt2022deep,hautsch2024building,heimbach2022sok,victor2021detecting,raheman2021architecture}) do not adequately address this selfish miners' created problem. To the best of our knowledge, this work presents the first analytical study to quantify the selfish behavior of miners and proactively design the simplest possible mechanism to mitigate the issue.
	
	\item \textit{Infinitely large efficiency damage by selfish miners}: Our dynamic game-theoretical analysis addresses the complex and coupled interactions among buyers, sellers, and miners. Our findings reveal that miners' self-interested actions can cause the price of anarchy (PoA) to equal infinity, meaning arbitrarily low social welfare. This is because miners make poor transaction matches to maximize their fee collection.
	
	\item \emph{Simple and effective mechanism via limiting block size}: We propose an adjustable block size (ABS) mechanism that balances the need to reduce miners' tendency to ignore desirable matches while maintaining high throughput. The mechanism design is challenging due to the non-existence of pure strategy Nash equilibrium, which results in buyers' and sellers' strategies being interactive mixed-strategy distributions. Moreover, the system designer may even lack information about each buyer's or seller's bid/ask prices and trading quantities. Nevertheless, our mechanism achieves asymptotically social optimum under homogeneous-quantity matching for NFT and provides a bounded approximation ratio under heterogeneous-quantity matching.
	
	\item \emph{Performance evaluation and implementation}: 
	We implement the ABS mechanism on an Ethereum blockchain testbed, validating its practical applicability. Based on the testbed, we conduct extensive experiments based on the actual BBOB dataset. For homogeneous-quantity trading of NFT, our mechanism achieves social optimum. For heterogeneous-quantity trading of Bitcoin tokens, our mechanism enhances social welfare up to 3.7 times compared to the existing benchmark and achieves at least 60\% of the social optimum. Moreover, we find that miners' non-selfish behavior may help to enhance social welfare. Our mechanism also demonstrates robustness in scenarios with random variations in the numbers of buyers and sellers.
	
\end{itemize}


The rest of the paper is organized as follows. 
Section \ref{sec:system} introduces the system model. Sections \ref{sec:analysis} and \ref{sec:general} analytically answer Key Questions 1 and 2, respectively. We evaluate the system performance in Section \ref{sec:per} and conclude this paper in Section~\ref{sec:conclusion}.

\section{System Model and Problem Formulation }\label{sec:system}
In this section, we describe the system model of the BBOB system to match buyers and sellers by miners. We first introduce the buyers, sellers, and miners' decision models in Section \ref{sub:energy_blockchain}. We then formulate their dynamic game interaction in Section \ref{sub:payoff}.
\subsection{Decision models of Buyers, Sellers, and Miners in BBOB}\label{sub:energy_blockchain}
\begin{figure}[!h]
	\vspace{-2mm}
	\centerline{\includegraphics[width=5.9cm]{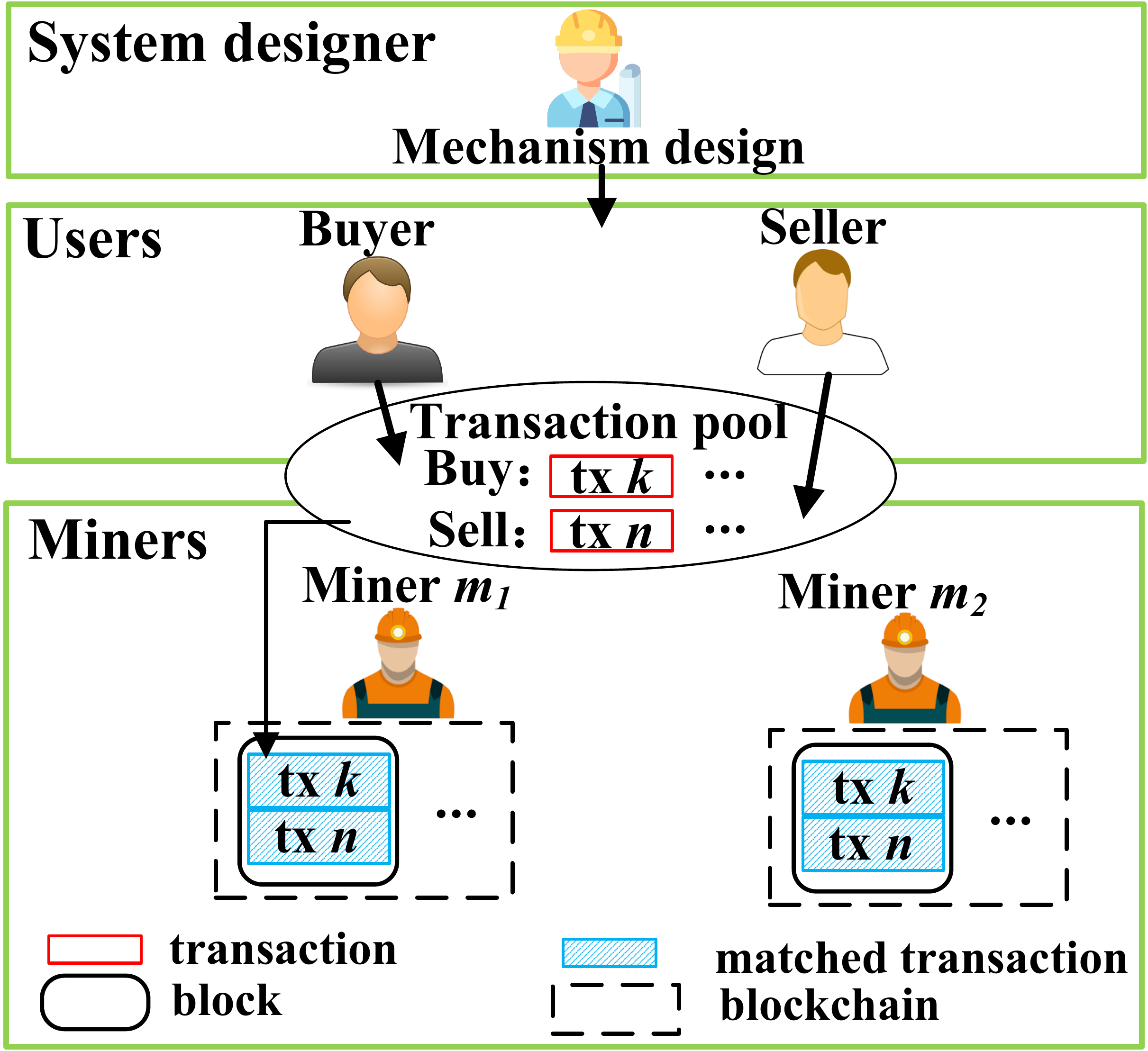}}
	\vspace{-1mm}
	\caption{The blockchain-based order book to match buyer and seller by miners under the system designer's initially designed mechanism.}
	\vspace{-1mm}
	\label{BC_process}
\end{figure}
We consider the BBOB system as illustrated in Fig. \ref{BC_process}. Specifically, the system designer first determines the mechanism under the blockchain protocols for participants. Then, buyers and sellers decide on their transaction fees. Finally, miners select transactions to match and include in the blockchain. 

{\color{blue}Next, we will present the detailed model. Table \ref{table_Notation} summarizes the key notations of this paper.}

\begin{table}[h]
	\caption{\color{blue}Key notations.}
	\label{table_Notation}
	\normalsize 
	{\color{blue}
		\begin{tabular}{|l|l|}
			\hline
			Notations &  Physical meaning \\
			\hline
			$f_k^{\rm buy}$ & Buyer $k$'s transaction fee decision\\
			\hline
			$f_n^{\rm sell}$ & Seller $n$'s transaction fee decision\\
			\hline
			$\mathcal{X}_m^t$ & Miner $m$'s transaction selection in round $t$ \\
			\hline
			\multirow{2}{*}{$x_{m,kn}^t$} & if miner $m$ matches buyer $k$ with seller $n$\\
			& when mining block $t$ \\
			\hline
			$A$ & Block size\\
			\hline
			$\mathcal{K}(K)$ & Set (number) of buyers\\
			\hline
			$b_k$ & Buyer $k$'s asset purchase proposal\\
			\hline
			\multirow{2}{*}{$\overline{b} (\underline{b})$} & Upper (lower) bound of asset purchase \\
			& and selling\\
			\hline
			$R_k$ & Buyer $k$'s utility of unit asset purchase\\
			\hline
			$\mathcal{N}(N)$ & Set (number) of sellers\\
			\hline
			$q_n$ & Seller $n$'s asset selling proposal\\
			\hline
			$C_n$ & Seller $n$'s cost of unit asset selling\\
			\hline
			$\mathcal{M}(M)$ & Set (number) of miners\\
			\hline
			$\alpha_m$ & Mining power of miner $m$\\
			\hline
			$\mathcal{T}(T)$ & Set (number) of blocks\\
			\hline
			$\epsilon$ & Smallest unit of transaction fee\\
			\hline
			$\mathcal{Q}^t$ & Transaction pool when mining block $t$\\
			\hline
			$d$ & Delay cost per block\\
			\hline
			$u_k$ & Buyer $k$'s payoff\\
			\hline
			$v_n$ & Seller $n$'s payoff\\
			\hline
			$w_m^t$ & Miner $m$'s payoff when mining block $t$\\
			\hline
			$sw$ & Social welfare\\
			\hline
			$\mathds{1}$ & Indicator function\\
			\hline
	\end{tabular}}
\end{table}

\subsubsection{System Designer's Mechanism Design} The system designer proposes a mechanism to maximize social welfare, aiming for the most beneficial trading outcome for the group. Typical mechanism design strategies include block size design \cite{Bitcoin_XT}, transaction fee design \cite{Eth_fee}, and consensus protocol design \cite{PoS_protocol}. The ideal mechanism design should be easy to implement and compatible with the current system protocol, possibly the simplest as in \cite{Bitcoin_XT}. Our mechanism design focuses on adjusting the block size, as detailed later in Section \ref{sec:general}.

\subsubsection{Buyers' and Sellers' Fee Decision}  We consider there are $K$ buyers and $N$ sellers, denoted by the sets $\mathcal{K} = \{1, 2, \cdots, K\}$ and $\mathcal{N} = \{1, 2, \cdots, N\}$, respectively. Each buyer $k\in \mathcal{K}$ generates a transaction aiming to purchase $b_k$ units of asset (such as cryptocurrency and non-fungible tokens (NFTs)) on the blockchain, while each seller $n \in \mathcal{N}$ proposes a transaction to sell $q_n$ units of asset, where $b_k, q_n \in [\underline{b}, \overline{b}]$ are continuous. If buyer $k$ is matched with seller $n$, the buyer receives a normalized utility of $R_k \in [0, 1]$ per unit of asset, characterizing his valuation or satisfaction level in obtaining the asset, and seller $n$ incurs a normalized cost of $C_n \in [0, 1]$ per unit of asset for selling the asset.

To facilitate these transactions, each buyer $k$ and seller $n$ \textit{decides on his fee payments $f_k^{\rm buy} \geq 0$ and $f_n^{\rm sell} \geq 0$}, respectively. The smallest fee unit is $\epsilon$ (e.g., one satoshi in Bitcoin). The generated transactions enter the transaction pool as in mid of Fig. \ref{BC_process}, which is pending for miners to match and include in the blockchain. Due to the limited number of transactions each block can contain, some transactions remain in the pending transaction pool, leading to a delay until they can be processed in subsequent blocks. Each buyer and seller must balance fee payment and transaction delay.\footnote{\color{blue}In the blockchain system, transactions typically experience some delay ranging from several minutes to hours \cite{Tx_delay}. As miners prioritize high-fee transactions \cite{wang2020measurement}, a buyer/seller generally uses the fee recommendation software to set the transaction fee, where he enters his delay requirement, and the software recommends the corresponding transaction fee \cite{Fee_per_byte}. Such a decision process is how a buyer/seller tradeoffs between transaction fee and delay.}

\subsubsection{Miners' Transaction Selection} We analyze a total of $T$ blocks, denoted as the set $\mathcal{T} \triangleq \{1, \cdots, T\}$. There are $M$ miners in the system, represented by the set $\mathcal{M} = \{1, 2, \cdots, M\}$. When mining block $t \in \mathcal{T}$, each miner $m \in \mathcal{M}$ \textit{selects a set $\mathcal{X}_m^t$ of buyer's and seller's transactions to match} and include in the blockchain, to maximize his fee revenue. The selection satisfies $\mathcal{X}_m^t \subseteq \mathcal{Q}^t$, where $\mathcal{Q}^t$ is the transaction set pending inclusion in the blockchain with initial $\mathcal{Q}^1 = \mathcal{K}\cup\mathcal{N}$. 

The block size $A$ allows each block to hold up to $A$ pairs of buyer's and seller's transactions. The block size constraint is formulated as follows:
\begin{equation}\label{block_size_con}
	|\mathcal{X}_m^t| \leq 2A.
\end{equation}
We assume the block size constraint is an even number without loss of generality. If the constraint is $|\mathcal{X}_m^t| \leq \phi$, where $\phi >0$ is odd, it is effectively $|\mathcal{X}_m^t| \leq \phi - 1$, as the remaining unpaired transaction would violate the matching constraints in (\ref{match_con1})-(\ref{match_con2}).

\textit{Matching constraint}: We respectively denote $\mathcal{X}_m^{t,\text{buy}}$ and $\mathcal{X}_m^{t,\text{sell}}$ as miner $m$'s selected buyer and seller sets, where $\mathcal{X}_m^t = \mathcal{X}_m^{t,\text{buy}} \cup\mathcal{X}_m^{t,\text{sell}}$. We use the index $\{x_{m,kn}^t| k\in \mathcal{X}_m^{t,\text{buy}}, n\in \mathcal{X}_m^{t,\text{sell}}\}$ to denote whether miner $m$ matches buyer $k$ with seller $n$ or not when mining block $t$:
\begin{equation}
	\hspace{-3mm}x_{m,kn}^t = \begin{cases}
		1, &\hspace{-2mm}\text{if miner $m$ matches buyer $k$ with seller $n$} \\
		&\hspace{-2mm}\text{when mining block $t$,}\\
		0, &\hspace{-2mm}\text{otherwise.}
	\end{cases}
\end{equation}
Within miner $m$'s selection, each buyer's transaction must be matched with one seller's in a typical setting (e.g., \cite{Chainlink_intro}\cite{Serum_intro}), ensuring the completion of the matching process. The matching conditions are formulated as:
\begin{align}
	&\sum_{n \in \mathcal{X}_m^{t,\text{sell}}} x_{m,kn}^t = 1, \quad \forall k \in \mathcal{X}_m^{t,\text{buy}},\label{match_con1}\\
	&\sum_{k \in \mathcal{X}_m^{t,\text{buy}}} x_{m,kn}^t = 1, \quad \forall n \in \mathcal{X}_m^{t,\text{sell}}.\label{match_con2}
\end{align}

Moreover, the buyer's utility must be higher than the seller's cost for matching profitable transactions. This constraint is formulated as:
\begin{equation}\label{match_con3}
	R_k \geq C_n \quad \text{if } x_{m,kn}^t = 1, \quad \forall k \in \mathcal{X}_m^{t,\text{buy}}.
\end{equation}

As in \cite{Mining_to_market}\cite{liu2022storage}, we denote the probability of miner $m$ mining a block as $\alpha_m$. For example, in proof-of-work systems, this probability is proportional to the miner's computational power relative to the total. In proof-of-stake systems, it reflects the amount of cryptocurrency the miner has staked. Therefore, each miner's probability of mining a block varies due to their computational or staking power differences. 

Miners' strategies follow the following principles based on existing blockchain systems in practice. 
\begin{itemize}
	\item \textit{Myopic selection}: According to \cite{Bitcoin_protocol} and \cite{messias2021selfish}, miners myopically select transactions to maximize their payoff for each block, prioritizing immediate transaction fee rewards. The rationale behind the myopic strategy is as follows: To benefit from a non-myopic strategy, a miner needs to mine at least two consecutive blocks. However, in typical blockchain systems like Bitcoin, each miner's mining power generally satisfies $\alpha_m \leq 0.2$ \cite{Mining_dist}. Consequently, the probability of a miner successfully mining two consecutive blocks, $\alpha_m^2 \leq 0.04$, is very small. Thus, they tend to be myopic in matching seller and buyer pairs. 
	
	\item \textit{Tie breaking}: When each miner $m$ faces two matching choices with the same total fee, he randomly selects one option. Moreover, miner $m$ rejects zero-fee transactions, as confirmed by blockchain protocols \cite{Eth_fee}.
\end{itemize}

\textit{Pending transaction set update}: 
After the miners' selection, one miner successfully generates a block (e.g., by solving a puzzle in a proof-of-work system or by being selected based on their stake in a proof-of-stake system). The selected transactions are then appended to the blockchain and removed from the pending transaction set $\mathcal{Q}^t$. This update process to the next block $t+1$ is formulated as follows:
\begin{equation}
	\mathcal{Q}^{t+1}(\boldsymbol{\mathcal{X}}) = \begin{cases}
		\mathcal{Q}^t \setminus \mathcal{X}_1^t, & \text{with probability (w.p.) } \alpha_1,\\
		\cdots \\
		\mathcal{Q}^t \setminus \mathcal{X}_M^t, & \text{w.p. } \alpha_M,
	\end{cases}
\end{equation}
where $\boldsymbol{\mathcal{X}} = (\mathcal{X}_m^t,\forall m\in\mathcal{M},\forall t\in\mathcal{T})$ is a function of all miners' strategies over time.


Next, we formulate the dynamic game-theoretical interactions among buyers, sellers, and miners.

\subsection{Dynamic Game Formulation among Buyers, Sellers and Miners}\label{sub:payoff}
This subsection models the dynamic interactions among miners, buyers, and sellers as a two-stage game. The buyers and sellers decide the transaction fees in Stage I, and miners select and match buyers' and sellers' transactions in Stage II. We will first introduce the payoffs for miners, buyers, and sellers. Then we introduce the dynamic game formulation.

\subsubsection{Each Miner $m$'s Payoff} 
When miner $m$ mines block $t$, he receives transaction fees from his selected transactions as the payoff. Otherwise, he receives nothing. We denote $\mathcal{X}_m^{t,\text{buy}}$ and $\mathcal{X}_m^{t,\text{sell}}$ as miner $m$'s selected buying and sell transaction sets, where $\mathcal{X}_m^t = \mathcal{X}_m^{t,\text{buy}} \cup\mathcal{X}_m^{t,\text{sell}}$. Miner $m$'s expected payoff during the mining of block $t$ is formulated as:
\begin{equation}\label{Miner:payoff}
	w_m^t(\mathcal{X}_m^t,A,\boldsymbol{f}) = \alpha_m (\sum_{k\in\mathcal{X}_m^{t,\text{buy}} } f_{k}^{\rm buy} + \sum_{n\in\mathcal{X}_m^{t,\text{sell}} } f_{n}^{\rm sell}),
\end{equation}
where $\boldsymbol{f} = (f_k^{\rm buy}, f_n^{\rm sell},\forall k \in \mathcal{K}, \forall n \in \mathcal{N})$ denotes all buyers' and sellers' fee decisions. Note that the fee does not vary based on the quantity of the matched order (e.g., in Ethereum, the fee for a transaction is determined by its gas consumption, which is a measure of the computational resources required to process the transaction, and the gas price, which is set by the user).

\subsubsection{Each Buyer $k$'s Payoff} Each buyer $k$'s payoff function comprises matching surplus and delay cost.

When buyer $k$ is matched with seller $n$, the trading quantity is the minimum of purchasing quantity $b_k$ and selling quantity $q_n$, denoted as $\min\{b_k, q_n\}$, which follows the typical setting for reduce-only orders \cite{Order_type}. Buyer $k$ receives a utility of $R_k$ and seller $n$ incurs a cost of $C_n$ per unit of the asset. Such a linear utility/cost model is widely used in cryptocurrency trading settings \cite{Linear_utility}, as each unit of cryptocurrency has a uniform value. To be fair, the buyer pays the widely-used mid price (e.g., \cite{mid_price,mid_price2,mid_price3}) to the seller, which is $\frac{R_k + C_n}{2}$. Thus, buyer $k$'s surplus from matching with seller $n$ is:
\begin{equation}
	\beta_{kn}^{\rm buy} = \min\{b_k,  q_n\}\big(R_k - \frac{R_k + C_n}{2}\big).
\end{equation}
When miner $m$ matches buyer $k$ with seller $n$ in mining round $t$ (i.e., $x_{m,kn}^t = 1$) and successfully mines a block (w.p., $\alpha_m$), buyer $k$ receives the surplus $\beta_{kn}^{\rm buy}$. Additionally, buyer $k$ pays a fee $f_k^{\rm buy}$ to miner $m$. Therefore, the total matching surplus for buyer $k$ is:
\begin{equation}
	u_k^{\rm Sur}(\boldsymbol{f},\boldsymbol{\mathcal{X}}) = \sum_{t \in \mathcal{T}}\sum_{m \in \mathcal{M}} \alpha_m \sum_{n \in \mathcal{X}_m^{t,\text{sell}}} x_{m,kn}^t \big(\beta_{kn}^{\rm buy} - f_k^{\rm buy}\big).
\end{equation}


When buyer $k$'s transaction is included in block $l_k^{\rm buy} \geq 1$, he bears the delay cost of $(l_k^{\rm buy} - 1)d$, where $d \geq 0$ is the delay cost per block. Buyer $k$'s delay cost is
\begin{equation}
	u_k^{\rm Delay}(\boldsymbol{\mathcal{X}}) = \big(l_k^{\rm buy}(\boldsymbol{\mathcal{X}}) - 1\big)d,
\end{equation}
where $l_k^{\rm buy}$ is a function of all miners' strategies over time $\boldsymbol{\mathcal{X}}$.

To sum up, each buyer $k$'s payoff function is formulated as follows:
\begin{equation}\label{eq:buyer_payoff}
	u_k(\boldsymbol{f}, \boldsymbol{\mathcal{X}}) = u_k^{\rm Sur}(\boldsymbol{f},\boldsymbol{\mathcal{X}}) - u_k^{\rm Delay}(\boldsymbol{\mathcal{X}}).
\end{equation}

\subsubsection{Each Seller $n$'s Payoff} Each seller $n$'s payoff function also comprises matching surplus and delay cost.

When buyer $k$ is matched with seller $n$, seller $n$ collects the price payment from buyer $k$. Seller $n$'s surplus from matching with buyer $n$ is:
\begin{equation}
	\beta_{nk}^{\rm sell} = \min\{b_k,  q_n\}\big(\frac{R_k + C_n}{2} - C_n\big).
\end{equation}
The total matching surplus for seller $n$ is:
\begin{equation}
	v_n^{\rm Sur}(\boldsymbol{f},\boldsymbol{\mathcal{X}}) = \sum_{t \in \mathcal{T}}\sum_{m \in \mathcal{M}} \alpha_m \sum_{k \in \mathcal{X}_m^{t,\text{buy}}} x_{m,kn}^t \big(\beta_{nk}^{\rm sell} - f_n^{\rm sell}\big).
\end{equation}

When seller $n$'s transaction is recorded in block $l_n^{\rm sell}\geq 1$, he bears the delay cost of:
\begin{equation}
	v_n^{\rm Delay}(\boldsymbol{\mathcal{X}}) = \big(l_n^{\rm sell}(\boldsymbol{\mathcal{X}}) - 1\big)d.
\end{equation}

To sum up, each seller $n$'s payoff function is formulated as follows:
\begin{equation}\label{eq:seller_payoff}
	v_n(\boldsymbol{f}, \boldsymbol{\mathcal{X}}) = v_n^{\rm Sur}(\boldsymbol{f},\boldsymbol{\mathcal{X}})  - v_n^{\rm Delay}(\boldsymbol{\mathcal{X}}).
\end{equation}

Based on the payoffs, we model the strategic interactions among the buyers, sellers, and miners as a two-stage game. 

\subsubsection{Stage II: Miner's Transaction Selection} During the mining of block $t\in\mathcal{T}$,\footnote{We assume a typical scenario that $T\geq \min\{N,K\}/A$, ensuring that all transactions can be recorded on the blockchain if miners choose to do so as in \cite{liu2022storage}\cite{roughgarden2020transaction}\cite{sun2023}.} each miner $m$ selects the transaction set $\mathcal{X}_m^t$ to maximize his payoff as follows:
\begin{equation}\label{eq:miner_payoff}
	\begin{aligned}
		\max &\hspace{1mm} 	w_m^t(\mathcal{X}_m^t,A,\boldsymbol{f}) \text{ in }(\ref{Miner:payoff})  \\
		\text{s.t.} &\hspace{1mm} (\ref{block_size_con}) \text{ and } (\ref{match_con1})-(\ref{match_con3}),\\
		\text{var.} &\hspace{1mm} \mathcal{X}_m^t\subseteq \mathcal{Q}^t.
	\end{aligned}
\end{equation}
In problem (\ref{eq:miner_payoff}), each miner myopically selects transactions from the pending transaction pool to maximize the fees he can earn. Note that this objective differs from centralized matching, which generally aims to maximize social welfare. This divergence is a key reason BBOB may result in lower social welfare, which will be further analyzed in the next section.

\subsubsection{Stage I: Buyers' and Sellers' Fee Setting} We formulate buyers' and sellers' fee setting as a non-cooperative game by foreseeing the miners' myopic behavior $\boldsymbol{\mathcal{X}}$, where all buyers and seller set their transaction fees simultaneously to maximize their individual payoffs.
\begin{gam}[Stage I: Transaction Fee Decisions Game]\label{Stage_II_game} In Stage I, the Fee Setting Game is a tuple $\Gamma = (\mathcal{L}, \mathcal{F}, \boldsymbol{W})$ defined by:
	\begin{itemize}
		\item Players: The set of buyers and sellers $\mathcal{L} = \mathcal{K} \cup \mathcal{N}$. 
		\item Strategies: Each buyer $k \in \mathcal{K}$ and seller $n \in \mathcal{N}$ decides on his fee payment $f_k^{\rm buy} \in \mathcal{F}_k^{\rm buy} \triangleq \{f_k^{\rm buy} \mid f_k^{\rm buy} \geq 0\}$ and $f_n^{\rm sell} \in \mathcal{F}_n^{\rm sell} \triangleq \{f_n^{\rm sell} \mid f_n^{\rm sell} \geq 0\}$, respectively. The set of feasible strategy profiles is $\mathcal{F} = (\prod_{k \in \mathcal{K}}\times \mathcal{F}_k^{\rm buy})\times(\prod_{n \in \mathcal{N}}\times \mathcal{F}_n^{\rm sell})$.
		\item Payoffs: The vector $\boldsymbol{W} = (u_k, v_n, \forall k \in \mathcal{K}, \forall n \in \mathcal{N})$ contains all buyers' and sellers' payoffs as defined in (\ref{eq:buyer_payoff}) and (\ref{eq:seller_payoff}), respectively.
	\end{itemize}
\end{gam}
In Game \ref{Stage_II_game}, each buyer or seller balances the fee payment with matching surplus and delay. By paying a higher fee, he is more likely to be matched with other higher-fee transactions and experience a lower delay. The analysis of Game \ref{Stage_II_game} is challenging because a pure strategy Nash equilibrium does not always exist, while buyers and sellers interact to decide their mixed-strategy distributions.

For modeling simplicity, we first assume that buying utilities and quantities, as well as selling costs and quantities, are static and publicly available information. Market participants can estimate these values for each participant by analyzing public historical transaction data. In Section \ref{subsec:incomplete}, we will relax this assumption by considering these values as unknown and following general distributions.

\subsubsection{PoA Definition} 
We use the PoA to evaluate the performance of the BBOB system, defined as the ratio between the social optimum and the worst social welfare under the equilibrium of the BBOB system. This measure quantifies the maximal efficiency loss due to the self-interested behaviors of all participants, providing a quantitative answer to Key Questions 1 and 2 without and with our mechanism, respectively.

Formally, the social welfare is defined as the sum of all buyers', sellers', and miners' payoffs:
\begin{equation}\label{SW}
	sw(\boldsymbol{f}, \boldsymbol{\mathcal{X}}, A) = \sum_{k \in \mathcal{K}} u_k(\boldsymbol{f}, \boldsymbol{\mathcal{X}}) + \sum_{n \in \mathcal{N}} v_n(\boldsymbol{f}, \boldsymbol{\mathcal{X}})
	+ \sum_{t \in \mathcal{T}} \sum_{m \in \mathcal{M}}w_m^t(\mathcal{X}_m^t,A,\boldsymbol{f}).
\end{equation}
The social optimum represents the maximum achievable social welfare when the system designer has full control over the decisions of buyers, sellers, and miners, and is expressed as:
\begin{equation}\label{SW_opt}
	\begin{aligned}
		sw^{\rm opt}\triangleq\max &\hspace{1mm} sw(\boldsymbol{f}, \boldsymbol{\mathcal{X}}, A)\\
		\text{var.}\hspace{0.2mm} &\hspace{1mm} \boldsymbol{f}, \boldsymbol{\mathcal{X}},A \in \mathbb{Z}^+,
	\end{aligned}
\end{equation}
where $\mathbb{Z}^+$ represents the set of positive integers. PoA is defined as:
\begin{equation}
	\text{PoA} = \max\limits_{A,d,\boldsymbol{R},\boldsymbol{C},\boldsymbol{b},\boldsymbol{q}}\frac{sw^{\rm opt}}{sw(\boldsymbol{f}^{\rm NE}, \boldsymbol{\mathcal{X}}^*, A)},
\end{equation}
where $\boldsymbol{f}^{\rm NE}$ and $\boldsymbol{\mathcal{X}}^*$ denotes the equilibrium strategies of buyers, sellers, and miners, which will be derived in the next section. As the BBOB system \textit{cannot} control the decisions of buyers, sellers, and miners, the social welfare at equilibrium is lower than the social optimum, and the PoA is always greater than 1.

\section{Game Theoretic Analysis for Poor PoA}\label{sec:analysis}
In this section, we analyze how miners' self-interested actions negatively affect the BBOB system. We will use the backward induction to derive the equilibrium of Stages II and I in Sections \ref{subsec:stage3NE} and Section \ref{subsec:stage2NE}, and then analyze the system PoA in Section \ref{subsec:stage1NE}.

\subsection{Stage II: Miners' Selfish Transaction Selection}\label{subsec:stage3NE}
In this subsection, we solve problem (\ref{eq:miner_payoff}) to derive miners' optimal strategies.

When mining block $t$, we denote that there are $K^t$ buying transactions and $N^t$ selling transactions within the pending transaction set $\mathcal{Q}^t$. Without loss of generality, we arrange the transactions in decreasing order of fees: $f_{1^t}^{\rm buy} > f_{2^t}^{\rm buy} > \cdots > f_{K^t}^{\rm buy}$ and $f_{1^t}^{\rm sell} > f_{2^t}^{\rm sell} > \cdots > f_{N^t}^{\rm sell}$. We define the set of top $i$ fee buying and selling transactions as $\mathcal{Q}^t(i)$:
\begin{align}
	&\mathcal{Q}^t(i) \triangleq \mathcal{Q}^{t,\text{buy}}(i)\cup \mathcal{Q}^{t,\text{sell}}(i),\\
	&\mathcal{Q}^{t,\text{buy}}(i) \triangleq \{1^t,2^t,\cdots,\min\{i^t,K^t\}\},\\
	&\mathcal{Q}^{t,\text{sell}}(i) \triangleq \{1^t,2^t,\cdots,\min\{i^t,N^t\}\}.
\end{align}

Theorem \ref{Stage_III_result} summarizes the optimal strategy for miners.
\begin{thm}[Miner's Optimal Strategy in Stage II]\label{Stage_III_result}
	During mining block $t$, miner $m$'s optimal strategy is to choose the top $i^{t*}$ fee buying and selling transactions in set:
	\begin{equation}\label{Miner_NE}
		\mathcal{X}_m^{t*} = \mathcal{Q}^t(i^{t*}),
	\end{equation}
	where 
	\begin{equation}
		i^{t*} = \argmax_{\substack{i^t\leq\min\{A,K^t,N^t\},\\ \mathcal{X}_m^{t} = \mathcal{Q}^t(i^t) \emph{ satisfying } (\ref{match_con1})-(\ref{match_con3})}} (\sum_{j = 1}^{i^t} f_{j}^{\rm buy} + \sum_{j = 1}^{i^t} f_{j}^{\rm sell}).
	\end{equation}
\end{thm}
The proof of propositions and theorems is in Section VII.

Despite having different probabilities of mining a block, each miner selects all the highest-fee transactions that can be matched to maximize his fee revenue. The BOBB protocol operates in the way that there is no direct competition among miners to lower the fee choice threshold, as the mining probability $\alpha_m$ is a linear factor of fee revenue in (\ref{Miner:payoff}).

\subsection{Stage I: Buyers' and Sellers' Fee Decision Game}\label{subsec:stage2NE}
In this subsection, we solve the buyers' and sellers' equilibrium strategies in Game \ref{Stage_II_game} by taking the miners' payoff-maximizing strategy in (\ref{Miner_NE}) into account.

We first define the pure strategy Nash equilibrium (PSNE). At the PSNE, no buyer or seller can increase his payoff by unilaterally changing his deterministic strategy. 
\begin{defin}[PSNE]\label{NE}
	A strategy profile $\boldsymbol{f}^{\rm NE} = (f_k^{\rm buy,NE}, f_n^{\rm sell,NE}, \forall k \in \mathcal{K}, \forall n \in \mathcal{N})$ constitutes a \emph{pure strategy Nash equilibrium} of Game \ref{Stage_II_game} in Stage I if
	\begin{align}
		&\hspace{0mm}u_k(f_k^{\rm buy,NE},\boldsymbol{f}_{-k}^{\rm NE}, \boldsymbol{\mathcal{X}}^*) \geq u_k(f_k^{\rm buy},\boldsymbol{f}_{-k}^{\rm NE}, \boldsymbol{\mathcal{X}}^*),\nonumber\\
		&\hspace{50mm}\forall f_k^{\rm sell} \geq 0, k \in \mathcal{K},\\
		&\hspace{0mm}v_n(f_n^{\rm sell,NE},\boldsymbol{f}_{-n}^{\rm NE}, \boldsymbol{\mathcal{X}}^*) \geq v_n(f_n^{\rm sell},\boldsymbol{f}_{-n}^{\rm NE}, \boldsymbol{\mathcal{X}}^*),\nonumber\\
		&\hspace{50mm}\forall f_n^{\rm sell} \geq 0, n \in \mathcal{N},
	\end{align}
	where $\boldsymbol{f}_{-k}^{\rm NE} = (f_j^{\rm buy,NE},f_n^{\rm sell,NE},\forall j\in\mathcal{K}\setminus\{k\},\forall n \in \mathcal{N})$ and $\boldsymbol{f}_{-n}^{\rm NE} = (f_k^{\rm buy,NE},f_j^{\rm sell,NE},\forall k\in\mathcal{K},\forall j \in \mathcal{N}\setminus\{n\})$ denote the equilibrium strategies of all buyers and sellers other than buyer $k$ or seller $n$, respectively.
\end{defin}

Without loss of generality, we arrange the buying utilities in decreasing order $R_1 > R_2 > \cdots > R_K$ and the selling costs in increasing order $C_1 < C_2 < \cdots < C_N$. We assume $R_1 > C_1$ to avoid the trivial situation where no transactions are matched. We then define the threshold $A_{th}$ to help establish condition of PSNE:
\begin{equation}\label{A_th}
	\hspace{-1mm}A_{th} \triangleq \begin{cases}
		i, &\hspace{-2.5mm}\text{if $R_i \geq C_i$ and $R_{i+1} < C_{i+1}$,}\\
		\min\{K,N\}, &\hspace{-2.5mm}\text{otherwise.}
	\end{cases}
\end{equation}

\newcounter{myeqcounter}
\setcounter{myeqcounter}{\theequation}
\addtocounter{myeqcounter}{1}
\newcounter{myeqcounterr}
\setcounter{myeqcounterr}{\theequation}
\addtocounter{myeqcounterr}{2}

Proposition \ref{prop:no_NE} summarizes the PSNE in Stage I.
\begin{prop}[Buyers' and Sellers' PSNE in Stage I]\label{prop:no_NE} The buyers' and sellers' PSNE $(f_k^{\rm buy,NE}, f_n^{\rm sell,NE},$ $ \forall k \in \mathcal{K}, \forall n \in \mathcal{N})$ is as follows:
	\begin{enumerate}
		\item If $A <A_{th}$, then there does not exist a PSNE in Stage I.
		\item If $A \geq A_{th}$, then $f_k^{\rm buy,NE}$ and $f_n^{\rm sell,NE}$ are:
		\begin{align}
			\hspace{-2.2mm}f_k^{\rm buy,NE} &= \begin{cases}
				\sigma_{th}^{\rm buy}(A_{th},A)+\epsilon,&\hspace{-2mm}\text{if $k\leq \min\{A, N\}$,}\hspace{0mm}\emph{ (\themyeqcounter a)}\\
				\sigma_{th}^{\rm buy}(A_{th},A), &\hspace{-2mm}\text{if $k> \min\{A, N\}$,}\hspace{0mm}\emph{ (\themyeqcounter b)}
			\end{cases}\nonumber\\
			\hspace{-2.2mm}f_n^{\rm sell,NE} &= \begin{cases}
				\sigma_{th}^{\rm sell}(A_{th},A)+\epsilon,&\hspace{-2mm}\text{if $n\leq \min\{A, K\}$,}\hspace{1mm}\emph{ (\themyeqcounterr a)}\\
				\sigma_{th}^{\rm sell}(A_{th},A),&\hspace{-2mm}\text{if $n> \min\{A, K\}$,}\hspace{1mm}\emph{ (\themyeqcounterr b)}
			\end{cases}\nonumber
		\end{align}
	
	\begin{figure*}
		\addtocounter{equation}{2}
		\begin{align}
			\label{sigma_buy} \sigma_{th}^{\rm buy}(A_{th},A) &= \begin{cases}
				\max\Big\{\frac{\sum_{n = 1}^{\min\{\lceil\frac{A_{th}}{A}\rceil A, N\}}\mathds{1}(R_{\min\{\lceil\frac{A_{th}}{A}\rceil A, N\}+1}\geq C_n)\min\{b_{\min\{\lceil\frac{A_{th}}{A}\rceil A, N\}+1},q_n\}(R_{\min\{\lceil\frac{A_{th}}{A}\rceil A, N\}+1} - C_n)}{2\sum_{n = 1}^{\min\{\lceil\frac{A_{th}}{A}\rceil A, N\}}\mathds{1}(R_{\min\{\lceil\frac{A_{th}}{A}\rceil A, N\}+1}\geq C_n)} \\
				- (\lceil\frac{\min\{\lceil\frac{A_{th}}{A}\rceil A, N\}+1}{A}\rceil A -1)d,0\Big\},\hspace{11mm}\text{if $\min\{\lceil\frac{A_{th}}{A}\rceil A, N\}<K$ and $R_{\min\{\lceil\frac{A_{th}}{A}\rceil A, N\}+1} \geq C_1$,}\\
				0,\hspace{59mm}\text{otherwise.}
			\end{cases}\\  
			\label{sigma_sell}\sigma_{th}^{\rm sell}(A_{th},A) &= \begin{cases}
				\max\Big\{\frac{\sum_{k = 1}^{\min\{\lceil\frac{A_{th}}{A}\rceil A, K\}}\mathds{1}(R_k\geq C_{\min\{\lceil\frac{A_{th}}{A}\rceil A, K\}+1})\min\{b_k,q_{\min\{\lceil\frac{A_{th}}{A}\rceil A, K\}+1}\}(R_k - C_{\min\{\lceil\frac{A_{th}}{A}\rceil A, K\}+1})}{2\sum_{k = 1}^{\min\{\lceil\frac{A_{th}}{A}\rceil A, K\}}\mathds{1}(R_k\geq C_{\min\{\lceil\frac{A_{th}}{A}\rceil A, K\}+1})} \\
				- (\lceil\frac{\min\{\lceil\frac{A_{th}}{A}\rceil A, K\}+1}{A}\rceil A -1)d,0\Big\},\hspace{11mm}\text{if $\min\{\lceil\frac{A_{th}}{A}\rceil A, K\}<N$ and $C_{\min\{\lceil\frac{A_{th}}{A}\rceil A, K\}+1} \leq R_1$,}\\
				0,\hspace{59mm}\text{otherwise.}
			\end{cases}
		\end{align}
		\vspace{-5mm}\rule[0pt]{17.9cm}{0.05em}
	\end{figure*}
		where $\sigma_{th}^{\rm buy}$ and $\sigma_{th}^{\rm sell}$ are in (\ref{sigma_buy}) and (\ref{sigma_sell}).
		\addtocounter{equation}{0}
	\end{enumerate}
\end{prop}
Here we elaborate the insights of Proposition \ref{prop:no_NE}: 
\begin{enumerate}
	\item If $A < A_{th}$: The block size is small, requiring multiple blocks to record all matched transactions. We illustrate the absence of a PSNE with a two-buyer-two-seller example and a block size of 1. When one buyer (or seller) pays a fixed fee, the other can take advantage by either paying an $\epsilon$ higher fee to reduce delay costs or paying the minimum fee of $\epsilon$ to minimize the payment. Consequently, the first buyer (or seller) will also respond by paying an $\epsilon$ higher fee or paying $\epsilon$. This back-and-forth strategy implies no PSNE.
	\item If $A \geq A_{th}$: 
	The block size is large enough that any leftover transactions cannot be matched after the first block. At PSNE, low-utility buyers in (\themyeqcounter b) (or high-cost sellers in (\themyeqcounterr b)) pay the maximum fee they can afford to get matched, resulting in zero transaction benefit, but their transactions remain unmatched as others pay higher fees. Meanwhile, high-utility buyers in (\themyeqcounter a) (or low-cost sellers in (\themyeqcounterr a)) pay a fee $\epsilon$ higher than the low-utility buyers (or high-cost sellers), ensuring their transactions get matched.
\end{enumerate}
To analyze buyers' and sellers' equilibrium strategies when $A < A_{th}$, we must refer to the mixed strategy Nash equilibrium (MSNE), where each buyer's or seller's fees follow a randomized distribution function. This makes the analysis challenging, as they interact to decide their mixed-strategy strategy functions against each other. Nevertheless, we are able to derive MSNE utilizing its property: within the support of the fee distribution, each buyer or seller receives the same expected payoff regardless of his specific fee choice within that range.

\begin{defin}[MSNE]\label{Mix_NE}
	A vector of probability measure $(\mu_k^{\rm buy,NE}(f_k^{\rm buy}),\mu_n^{\rm sell,NE}(f_k^{\rm sell}),\forall k \in \mathcal{K}, $ $\forall n \in \mathcal{N})$ constitutes a \emph{mixed strategy Nash equilibrium} if the following inequalities hold for any $k \in \mathcal{I},n \in \mathcal{N},\mu_k^{\rm buy}$, and $\mu_k^{\rm sell}$.		
	\begin{multline}\label{Def_mix_NE1}
		\int\limits_{\mathcal{F}} 
		{\mathbb{E}[u_k(f_k^{\rm buy},\boldsymbol{f}_{-k}, \boldsymbol{\mathcal{X}}^*)]}d\big(\mu_k^{\rm buy,NE}(f_k^{\rm buy})\times\boldsymbol{\mu}_{-k}^{\rm NE}(\boldsymbol{f}_{-k})\big)
		\\ \geq \hspace{-2mm}\int\limits_{\mathcal{F}} 
		{\mathbb{E}[u_k(f_k^{\rm buy},\boldsymbol{f}_{-k}, \boldsymbol{\mathcal{X}}^*)]}d\big(\mu_k^{\rm buy}(f_k^{\rm buy})\times\boldsymbol{\mu}_{-k}^{\rm NE}(\boldsymbol{f}_{-k})\big),
		\vspace{-3mm}
	\end{multline}
	\vspace{-3mm}
	\begin{multline}\label{Def_mix_NE2}
		\int\limits_{\mathcal{F}} 
		{\mathbb{E}[v_n(f_n^{\rm sell},\boldsymbol{f}_{-n}, \boldsymbol{\mathcal{X}}^*)]}d\big(\mu_n^{\rm sell,NE}(f_n^{\rm sell})\times\boldsymbol{\mu}_{-n}^{\rm NE}(\boldsymbol{f}_{-n})\big)
		\\
		\geq \hspace{-2mm}\int\limits_{\mathcal{F}} 
		{\mathbb{E}[v_n(f_n^{\rm sell},\boldsymbol{f}_{-n}, \boldsymbol{\mathcal{X}}^*)]}d\big(\mu_n^{\rm sell}(f_n^{\rm sell})\times\boldsymbol{\mu}_{-n}^{\rm NE}(\boldsymbol{f}_{-n})\big).
	\end{multline}
\end{defin}

\setcounter{myeqcounter}{\theequation}
\addtocounter{myeqcounter}{2}
\setcounter{myeqcounterr}{\theequation}
\addtocounter{myeqcounterr}{3}

\newcounter{myeqcounterrr}
\setcounter{myeqcounterrr}{\theequation}
\addtocounter{myeqcounterrr}{4}
\newcounter{myeqcounterrrr}
\setcounter{myeqcounterrrr}{\theequation}
\addtocounter{myeqcounterrrr}{5}

We then define the intermediate function $g$ to help establish the equilibrium strategy:
\begin{equation}
	g(p,f,I,A) \triangleq \sum\limits_{n=0}^{I-1}\binom{I-1}{n}p^n(1-p)^{I-1-n}(f+\lceil\frac{n+1}{A}\rceil d).
\end{equation}
Equations (\themyeqcounter) and (\themyeqcounterr) characterize the buyers' and sellers' equilibrium strategies in terms of cumulative distribution function (CDF).

\begin{figure*}
	\addtocounter{equation}{4}
	\vspace{-2.5mm}
	\begin{equation*}
		\begin{aligned}
			F^{\rm buy,NE}(f) \hspace{-0.5mm}&=\hspace{-0.5mm} \begin{cases}
				0, &\hspace{-131.25mm}\text{if $f < \sigma_{th}^{\rm buy}(A_{th},A)+\epsilon$,}\hspace{83.4mm} \text{(\themyeqcounter a)}\\
				\text{solution to }g(1-F^{\rm buy,NE}(f),f,\min\{\lceil\frac{A_{th}}{A}\rceil A, K,N\},A) \hspace{-0.5mm}= \hspace{-0.5mm}g(1,\sigma_{th}^{\rm buy}(A_{th},A)+\epsilon,\min\{\lceil\frac{A_{th}}{A}\rceil A, K,N\},A), \\
				&\hspace{-131.25mm}\text{if $\sigma_{th}^{\rm buy}(A_{th},A)+\epsilon\leq f$ and $f\leq \sigma_{th}^{\rm buy}(A_{th},A)+\epsilon + (\lceil\frac{\min\{\lceil\frac{A_{th}}{A}\rceil A, N, K\}}{A}\rceil-1)d$, \hspace{0mm}(\themyeqcounter b)}\\
				1, &\hspace{-131.25mm}\text{if $f> \sigma_{th}^{\rm buy}(A_{th},A)+\epsilon + (\lceil\frac{\min\{\lceil\frac{A_{th}}{A}\rceil A, N, K\}}{A}\rceil-1)d$.}\hspace{41mm} \text{(\themyeqcounter c)}
			\end{cases}\\   \noalign{\vskip-0pt}
			F^{\rm sell,NE}(f) \hspace{-0.5mm}&= \hspace{-0.5mm}\begin{cases}
				0, &\hspace{-130mm}\text{if $f < \sigma_{th}^{\rm sell}(A_{th},A)+\epsilon$},\hspace{84mm} \text{(\themyeqcounterr a)}\\
				\text{solution to }g(1-F^{\rm sell,NE}(f),f,\min\{\lceil\frac{A_{th}}{A}\rceil A, N, K\},A) \hspace{-0.5mm}= \hspace{-0.5mm}g(1,\sigma_{th}^{\rm sell}(A_{th},A)+\epsilon,\min\{\lceil\frac{A_{th}}{A}\rceil A, N, K\},A),\\
				&\hspace{-130mm}\text{if $\sigma_{th}^{\rm sell}(A_{th},A)+\epsilon\leq f$ and $f\leq \sigma_{th}^{\rm sell}(A_{th},A)+\epsilon + (\lceil\frac{\min\{\lceil\frac{A_{th}}{A}\rceil A, N, K\}}{A}\rceil-1)d$, \hspace{1.7mm}(\themyeqcounterr b)}\\
				1, &\hspace{-130mm}\text{if $f> \sigma_{th}^{\rm sell}(A_{th},A)+\epsilon + (\lceil\frac{\min\{\lceil\frac{A_{th}}{A}\rceil A, N, K\}}{A}\rceil-1)d$.}\hspace{42mm} \text{(\themyeqcounterr c)}
			\end{cases}
		\end{aligned}
	\end{equation*}
	\vspace{-0mm}
	\vspace{-5mm}\rule[0pt]{17.9cm}{0.05em}
\end{figure*}

Proposition \ref{stage_II_result} summarizes the buyers' and sellers' MSNE.

\begin{prop}[Buyers' and Sellers' MSNE in Stage I]\label{stage_II_result} 
	When $A< A_{th}$, the buyers' and sellers' MSNE $(\mu_k^{\rm buy,NE}(f_k^{\rm buy}),$ $\mu_n^{\rm sell,NE}(f_k^{\rm sell}),\forall k \in \mathcal{K}, \forall n \in \mathcal{N})$ is as follows:
	\begin{align}
		\hspace{0mm}& \begin{cases}
			\mu_k^{\rm buy,NE}\sim\emph{CDF in (\themyeqcounter)},&\hspace{-1mm}\text{if $k\leq \min\{\lceil\frac{A_{th}}{A}\rceil A, N\}$,}\hspace{0mm}\emph{ (\themyeqcounterrr a)}\\
			f_k^{\rm buy,NE}=\sigma_{th}^{\rm buy}(A_{th},A) , &\hspace{-1mm}\text{if $k> \min\{\lceil\frac{A_{th}}{A}\rceil A, N\}$,}\hspace{0mm}\emph{ (\themyeqcounterrr b)}
		\end{cases}\nonumber\\
		\hspace{0mm} & \begin{cases}
			\mu_n^{\rm sell,NE}\sim\emph{CDF in (\themyeqcounterr)},&\hspace{-1mm}\text{if $n\leq \min\{\lceil\frac{A_{th}}{A}\rceil A, K\}$,}\hspace{1.5mm}\emph{ (\themyeqcounterrrr a)}\\
			f_n^{\rm sell,NE}=\sigma_{th}^{\rm sell}(A_{th},A),&\hspace{-1mm}\text{if $n> \min\{\lceil\frac{A_{th}}{A}\rceil A, K\}$.}\hspace{1.5mm}\emph{ (\themyeqcounterrrr b)} 
		\end{cases}\nonumber
	\end{align}
\end{prop}

Here we elaborate on the insights of Proposition \ref{stage_II_result}. If $A < A_{th}$, buyers and sellers adopt the mixed strategy. Low-utility buyers in (\themyeqcounterrr b) pay the maximum fee they can afford in an attempt to get matched but fail. High-utility buyers in (\themyeqcounterrr a) pay fees that follow a distribution due to the nonexistence of a PSNE stated in Proposition \ref{prop:no_NE}. Their fee distribution's lower bound in (\themyeqcounter a) is $\epsilon$ higher than those with low utilities to ensure to get matched. The same insights apply to the sellers.

\subsection{PoA to Compare NE With Social Optimum}\label{subsec:stage1NE}
In this subsection, we derive the PoA of the state-of-the-art BBOB equilibrium in Theorem \ref{thm:PoA1}.
\begin{thm}[PoA of current system]\label{thm:PoA1}
	The PoA is unbounded, i.e., $\text{PoA} = \infty$, which means NE's arbitrarily large social welfare loss compared to the social optimum.
\end{thm}
We explain the insights of Theorem \ref{thm:PoA1} by a two-buyer-two-seller example in Figs. \ref{Fig:utility1}-\ref{Fig:utility2}, highlighting that setting the block size $A$ too high or too low can lead to arbitrarily large social welfare loss. Each buyer and seller has a unit of buying and selling quantity, respectively.
\begin{itemize}
	\item \begin{figure}[h]
		\centering
		{\includegraphics[width=5cm]{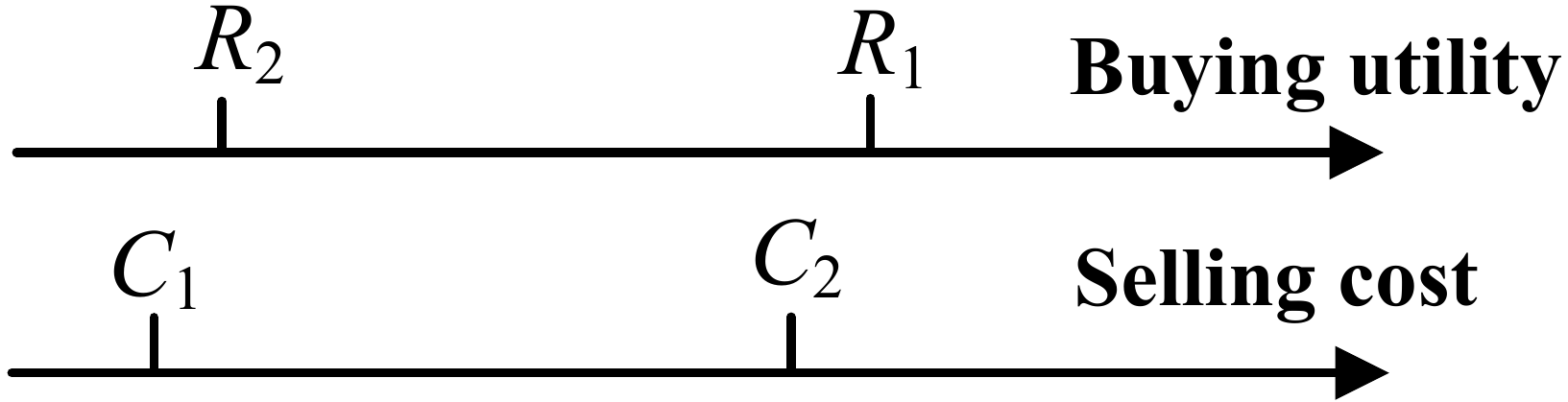}}
		\vspace{-2mm}
		\caption{Buying utility and selling cost for high block size.}\label{Fig:utility1}
		\vspace{-2mm}
	\end{figure}
	\textit{High block size with $A\geq 2$ example}: As shown in Fig. \ref{Fig:utility1}, with $C_1 < R_2 < C_2 < R_1$, the social optimum is to only match $R_1$ with $C_1$ to maximize the difference between buying utility and selling cost. However, if block size $A \geq 2$ and each buyer and seller set an $\epsilon$ fee according to Proposition \ref{prop:no_NE}, miners will match $R_2$ with $C_1$ and $R_1$ with $C_2$ to maximize their fee revenue. Note that $R_2$ must be matched with $C_1$ since a valid match requires the buying utility to be higher than the selling cost. Consequently, miners' matching can collect fees from four transactions instead of the two under the social optimum. Therefore, the ratio between social optimum and actual social welfare is:
	\begin{equation*}
		\frac{sw^{\rm opt}}{sw(\boldsymbol{f}^{\rm NE}, \boldsymbol{\mathcal{X}}^*, A)} = \frac{R_1-C_1}{R_1+R_2-C_1-C_2}.
	\end{equation*}
	As $R_1 - C_1$ can be made arbitrarily large while keeping $R_2 - C_1$ and $R_1 - C_2$ constant, the PoA as the maximum ratio becomes unbounded.
	\item \begin{figure}[h]
		\centering
		{\includegraphics[width=5cm]{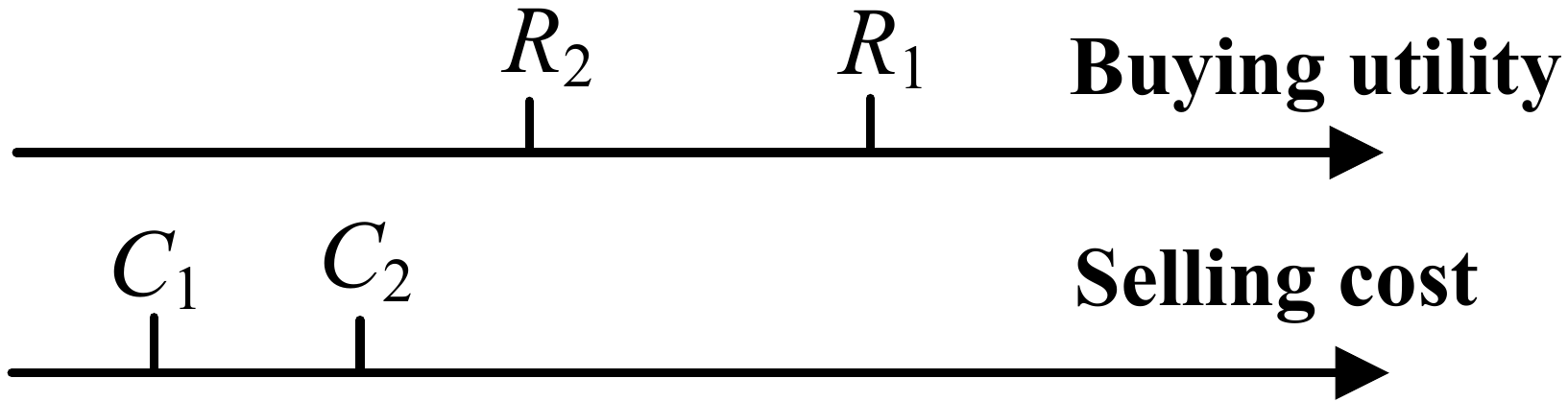}}
		\vspace{-2mm}
		\caption{Buying utility and selling cost for low block size.}\label{Fig:utility2}
		\vspace{-2mm}
	\end{figure}
	\textit{Low block size with $A = 1$ example}: As shown in Fig. \ref{Fig:utility2}, with $C_1 < C_2 < R_2 < R_1$, the social optimum is to set block size as $A = 2$ and include all transactions in one block to maximize the difference between buying utility and selling cost without delay cost. However, if block size $A = 1$ and each buyer and seller sets fees according to the mixed strategy in Proposition \ref{stage_II_result}, miners need two blocks to include all transactions, resulting in two transactions bearing delay cost $d$. Therefore, the performance ratio with optimum is:
	\begin{equation*}
		\frac{sw^{\rm opt}}{sw(\boldsymbol{f}^{\rm NE}, \boldsymbol{\mathcal{X}}^*, A)} = \frac{R_1+R_2-C_1-C_2}{R_1+R_2-C_1-C_2 - 2d}.
	\end{equation*}
	As $d$ can be made arbitrarily large or buyers and sellers are impatient to trade in practice while keeping $R_1, R_2, C_1,$ and $C_2$ constant, the PoA as the maximum ratio becomes unbounded.
\end{itemize}

\textit{Theorem \ref{thm:PoA1} answers Key Question 1: Setting the block size too high or too low can lead to arbitrarily low social welfare. When the block size is too high, miners' self-interested behavior can cause them to ignore desirable matches, resulting in significant social welfare loss. Conversely, when the block size is too low, the system incurs substantial delay costs. These observations underscore the motivation for our mechanism design.}

\section{Our Adjustable Block Size Mechanism}\label{sec:general}
In this section, we present the adjustable block size mechanism. First, we analyze the scenario where bid and ask prices and quantities are public information in Section \ref{subsec:com}. Then, we extend our mechanism and analysis to the case where these parameters follow random distributions in Section \ref{subsec:incomplete}. 

\subsection{Adjustable Block Size Mechanism for Complete Information}\label{subsec:com}
Motivated by the Bitcoin protocol that optimizes block size \cite{Bitcoin_XT}, the system designer \textit{decides the block size $A \in \mathbb{Z}^+$} in Stage 0 in Fig. \ref{BC_process}, before buyers and sellers decide the transaction fees in Stage I. The system may impose a hard block size limit due the limited network bandwidth, which will be analyzed in Section \ref{sec:per}-B. Our Adjustable Block Size (ABS) mechanism design problem is formulated as follows:
\begin{equation}\label{Mech_design}
	\begin{aligned}
		\max &\hspace{2mm} sw(\boldsymbol{f}^{\rm NE}(R_k,b_k,C_n,q_n,\forall k\in\mathcal{K}, \forall n \in\mathcal{N}), \boldsymbol{\mathcal{X}}^*, A) \\
		\text{var.} &\hspace{2mm} A \in \mathbb{Z}^+.
	\end{aligned}
\end{equation}
Problem (\ref{Mech_design}) is challenging as it needs to consider the mixed-strategy distribution interactions among all buyers and sellers under any block size design, which may involve trading over multiple rounds due to the limited block size. Note that our mechanism design incorporates the tradeoff between block size and delay, as delay is part of buyers' and sellers' payoff function in the social welfare.

We address this challenge and the following algorithm outline the procedure for determining the optimal block size in the complete information setting:
\begin{algorithm}  
	{\color{blue}
		\begin{algorithmic}[1]
			\STATE Arrange the buying utilities in decreasing order $R_1 > R_2 > \cdots > R_K$ and the selling costs in increasing order $C_1 < C_2 < \cdots < C_N$.
			\STATE Set the optimal block size as $A_{th}$ in (\ref{A_th}). 
		\end{algorithmic}  
		\caption{\color{blue}Optimal Block Size Under Complete Information}  \label{alg:1} 
	}
\end{algorithm} 

{\color{blue} The computational complexity of Algorithm \ref{alg:1} is $O(\max\{K,N\}\log \max\{K,N\})$ by the merge sort algorithm,\footnote{https://www.javatpoint.com/merge-sort} as the primary task is to sort the buying utilities and selling costs. This level of complexity is manageable even in systems with a large number of buyers and sellers.}
\begin{thm}[PoA under ABZ]\label{thm:stage_I_result} 
	The PoA under our ABS mechanism with $A = A_{th}$ is upper bounded as follows:
	\begin{equation}
		\text{PoA} \leq \frac{\text{ }\overline{b}\text{ }}{\text{ }\underline{b}\text{ }}.
	\end{equation}
	When considering homogeneous-quantity trading (e.g., NFT trading) in BBOB with $\underline{b} = \overline{b}$, setting the block size as $A = A_{th}$ achieves social optimum and PoA = 1.
\end{thm}
Here we explain the insights of Theorem \ref{thm:stage_I_result}. Note that $R_{A_{th}} \geq C_{A_{th}}$ and $R_{A_{th}+1} < C_{A_{th}+1}$ according to (\ref{A_th}). Therefore, including transactions with $R_i$ and $C_i$ for top $i \leq A_{th}$ pairs enhances social welfare, while including $R_j$ and $C_j$ for bottom $j > A_{th}$ reduces social welfare. Hence, we set the block size as $A_{th}$. 
\textit{The ABS in Theorem \ref{thm:stage_I_result} achieves the social optimum under homogeneous-quantity trading. This is particularly useful and efficient in BBOB for NFT trading, where the trading quantity for buyers and sellers is generally one.} 

The PoA is larger than 1 when $\underline{b} \neq \overline{b}$ for heterogeneous-quantity trading, because miners may randomly match transactions as long as such a random match does not affect their total fee revenue. This can result in transactions with large buying quantities matched with transactions with small selling quantities, causing small trading quantities and social welfare loss.\footnote{To mitigate this issue, we could enforce a rule requiring transactions with large buying quantities to be matched with transactions with high selling quantities. However, this approach may fundamentally alter the core principles of blockchain and compel agents to adopt an entirely new protocol.} 
\subsection{ABS Mechanism for Incomplete Information}\label{subsec:incomplete}
In practice, the system designer may not know the exact values of buying utilities and quantities from buyers, as well as selling costs and quantities from sellers \cite{gao2023average,gao2024dynamic}. Here, these parameters' values are drawn from random distributions.

As a typical setting, we suppose the buying utilities and selling costs follow i.i.d. random distributions in the normalized range $[0,1]$, with general CDFs of $R(\cdot)$ and $C(\cdot)$. The buying and selling quantities follow i.i.d. distributions between $[\underline{b},\overline{b}]$ with general CDFs of $B(\cdot)$ and $Q(\cdot)$. 

The ABS mechanism design problem under such incomplete information is formulated as follows:
\begin{equation}\label{Mech_design2}
	\begin{aligned}
		\max &\hspace{2mm} \mathbb{E}[sw(\boldsymbol{f}^{\rm NE}(R_k,b_k,C_n,q_n,\forall k\in\mathcal{K}, \forall n \in\mathcal{N}), \boldsymbol{\mathcal{X}}^*, A)] \\
		\text{s.t.} &\hspace{2mm} R_k \sim R(r), b_k\sim B(b), \forall k\in\mathcal{K}, \\
		&\hspace{2mm} C_n \sim C(c), q_n\sim Q(q), \forall n\in\mathcal{N}, \\
		\text{var.} &\hspace{2mm} A \in \mathbb{Z}^+.
	\end{aligned}
\end{equation}
Problem (\ref{Mech_design2}) is analytically challenging to solve due to the adoption of mixed strategies by buyers and sellers, which are randomized and interactive distributions of others' strategies. Additionally, the general distributions of trading information for buyers and sellers further complicates the analysis. Despite these difficulties, we are able to derive the PoA bounds.

{\color{blue}The following algorithm outlines the procedure for determining the optimal block size in the incomplete information setting:}
\begin{algorithm}  
	{\color{blue}
		\begin{algorithmic}[1]
			\STATE Solve for $\eta$, the unique solution to the following equation:
			\begin{equation}
				N C(\eta) = K \big(1-R(\eta)\big).
			\end{equation}
			\STATE Given any $\psi \in(0,1)$, the optimal block size is:
			\begin{equation}\label{A_dist}
				A^{*,\text{dist}} = \lfloor N\big(C(\eta) + N^{-\psi}\big) \rfloor,
			\end{equation} 
		\end{algorithmic}  
		\caption{\color{blue}Optimal Block Size Under Incomplete Information} 
		\label{alg:2}  
	}	 
\end{algorithm} 

{\color{blue}Note that the computational complexity of Algorithm \ref{alg:2} is independent the number of buyers $K$ and sellers $N$, indicating that the mechanism is scalable.}
\begin{thm}\label{thm:dist}
	When $N\rightarrow \infty$, the PoA under the ABS mechanism with block size of $A^{*,\text{dist}}$ in (\ref{A_dist}) satisfies:
	\begin{equation}\label{eq:approx2}
		\lim\limits_{N\rightarrow \infty} \text{PoA}  \leq \frac{\text{ }\overline{b}\text{ }}{\text{ }\underline{b}\text{ }}.
	\end{equation} 
	When considering homogeneous-quantity trading (e.g., NFT trading) in BBOB with $\underline{b} = \overline{b}$, setting the block size as $A = A^{*,\text{dist}}$ achieves asymptotically social optimum and the PoA satisfies
	\begin{equation}
		\lim\limits_{N\rightarrow \infty} \emph{PoA} = 1.
	\end{equation}
\end{thm}
Here we explain the insights of Theorem \ref{thm:dist}. Note that $\mathbb{E}[R_{\lfloor NC(\eta)\rfloor}] \geq \mathbb{E}[C_{\lfloor NC(\eta)\rfloor}]$ and $\mathbb{E}[R_{\lfloor NC(\eta)\rfloor+1}] $

\hspace{-3mm}$< \mathbb{E}[C_{\lfloor NC(\eta)\rfloor+1}]$ as $N\rightarrow \infty$. Therefore, we set the block size as $A^{*,\text{dist}}$ in (\ref{A_dist}), where the parameter $N^{-\psi}$ provides block size redundancy to ensure that the mechanism includes all transaction pairs with actual $R_i \geq C_i$ for all $i = 1, 2, \cdots, \min\{K, N\}$ to maximize social welfare. \textit{Moreover, the block size in Theorem \ref{thm:dist} achieves the asymptotically social optimum under homogeneous-quantity trading, relying only on the knowledge of the distributions for setting the block size of ABS.}

\textit{Theorems \ref{thm:stage_I_result} and \ref{thm:dist} answer Key Question 2: the optimal block size achieves asymptotically social optimum under homogeneous-quantity trading (e.g., NFTs) and provides a bounded PoA under heterogeneous-quantity trading, even without specific matching information, which is easy to implement in practice.}

\section{Performance Evaluations with Real Dataset}\label{sec:per}
We conduct experiments to evaluate the mechanism in terms of social welfare, the performance ratio between the mechanism's social welfare and social optimum, and the robustness of the system under time-varying numbers of buyers and sellers and different block size limits. 
\subsection{Experiment Setting} Our experiment considers homogeneous-quantity and heterogeneous-quantity trading, with parameters set based on practical datasets. The homogeneous-quantity trading experiment is based on NFT trading data, including bid and ask prices \cite{NFT_trading_data}. The heterogeneous-quantity trading experiment utilizes the Bitcoin trading dataset \cite{Dataset}, which includes bid prices, ask prices, bid quantities, ask quantities, and trading times.

\begin{figure}[t]
	\centering
	\vspace{-2mm}
	\subfigure[NFT trading (homogeneous quantity) about bid/ask price distributions from \cite{NFT_trading_data}. The trading quantity per matching pair is constantly 1. ]{
		\includegraphics[width=0.47\linewidth]{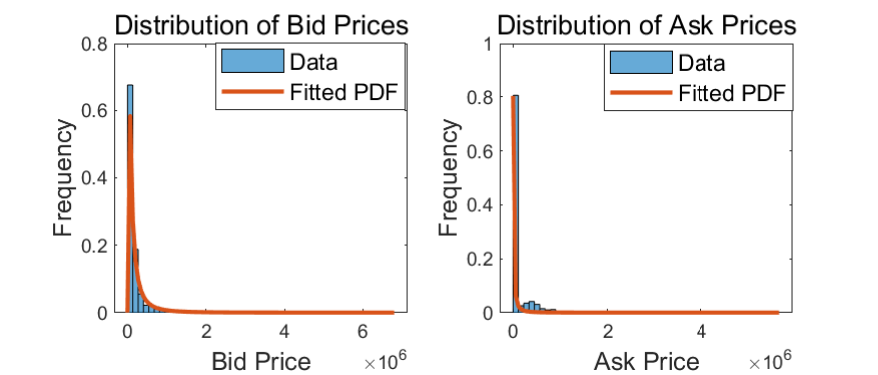}}
	\hspace{2mm}
	\subfigure[Bitcoin trading (heterogeneous quantity) about bid/ask price and bid/ask quantity distributions from \cite{Dataset}.]{
		\includegraphics[width=0.47\linewidth]{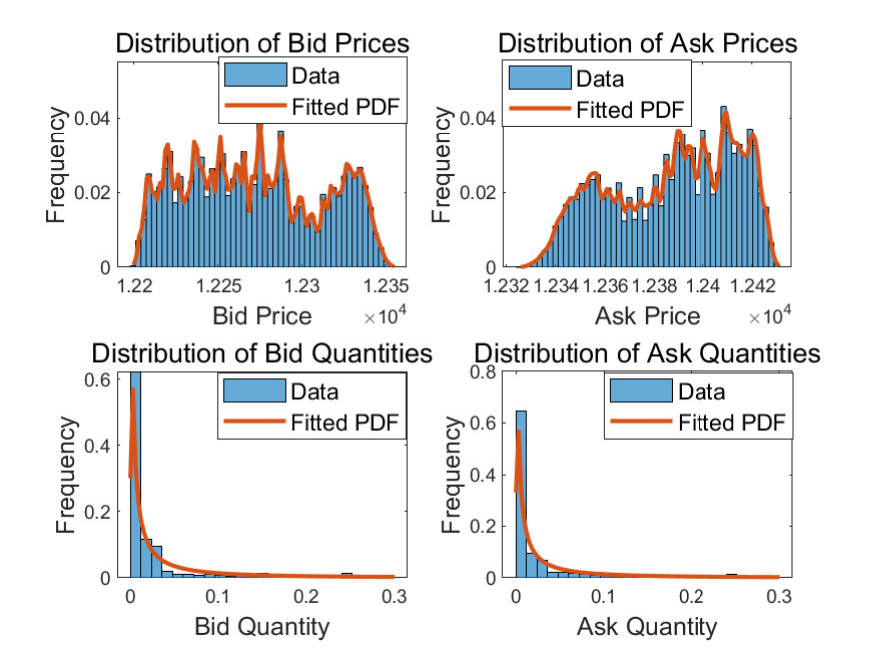}}
	\vspace{-2mm}
	\caption{The histogram of real dataset and fitted PDF.}\label{Fig:dataset}
	\vspace{-2mm}
\end{figure}
Fig. \ref{Fig:dataset}(a) illustrates the bid price and ask price of NFT trading, where the trading quantities are constantly 1. Fig. \ref{Fig:dataset}(b) illustrates the bid price, ask price, bid quantity, and ask quantity for Bitcoin trading.\footnote{For Bitcoin trading, the smallest unit of bid and ask quantities is not one but $10^{-8}$ Bitcoin tokens. These trading quantities are relatively small because of the high price of each Bitcoin token.}  We fit each attribute of the data to a distribution. For homogeneous-quantity trading, we only need to generate random bid and ask prices based on fitted distributions from the NFT trading dataset \cite{NFT_trading_data}. For heterogeneous-quantity trading, we generate bid prices, ask prices, bid quantities, and ask quantities based on the Bitcoin trading dataset \cite{Dataset}.

In reality, a buyer's utility (or a seller's selling cost) can differ from their bid (or ask) price. In the experiment, we follow \cite{yu2012best} to set the ratio of 1.05 between buyers' bid prices and their utilities, as well as between sellers' ask prices and their costs. The remaining parameters' values are summarized as follows: the number of miners is 250k,\footnote{\url{https://www.statista.com/statistics/1334722/ethereum-held-by-miners/}} the average ratio of buyers to sellers is $K/N = 1$, the delay cost is $d = \$0.3$ per block, and the block size factor $\psi = 0.85$.

In the experiment, we compare our mechanism with the following benchmarks:
\begin{enumerate}
	\item \textit{Benchmark with BBOB matching algorithm \cite{Matching_protocol}}: The benchmark mechanism is the state-of-the-art BBOB matching algorithm, which sets the block size to be maximal. This mechanism further recommends transaction matching for miners, but miners are rational and do not necessarily follow the recommendation. This issue is prevalent in other matching algorithms for BBOB.\footnote{\url{https://dydx.exchange/}}
	\item \textit{Our ABS facing 20\% non-selfish miners}: In practice, some miners follow the protocol recommendations.\footnote{\url{https://thenextweb.com/news/a-brief-history-of-bitcoin-mining-hardware}} Therefore, we evaluate our ABS mechanism assuming 20\% of miners are cooperative and follow the recommendations of matching algorithm \cite{Matching_protocol}. 
\end{enumerate}
\subsection{Implementation of ABS Mechanism}

We implement the ABS mechanism on a local instance of the Ethereum blockchain by modifying the ``gas limit,'' which refers to a system parameter that restricts the number of computational steps (or ``gas'') a block can contain. This Ethereum-based testbed demonstrates that block size adjustments can be effectively realized within a real-world blockchain environment.
\begin{figure}[h]
	\centering
	\includegraphics[width=0.99\linewidth]{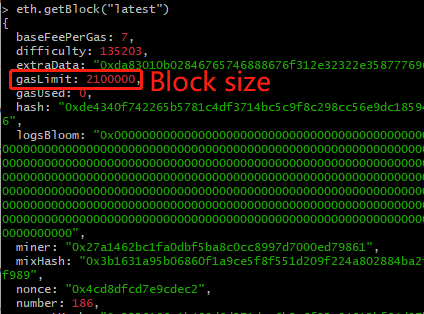}
	\caption{ABS implementation in Ethereum blockchain testbed.}
	\label{fig:block_info}
\end{figure}

We use Geth version 1.11.2 \cite{Eth_client} as the Ethereum client. We modified the Geth source code's ``block\_validator.go'' file, setting the gas limit to adjust the block size. After modifying the client, we ran a local Ethereum instance with the updated gas limit. Fig. \ref{fig:block_info} illustrates the modification as setting the gas limit to 2,100,000 to accommodate approximately 100 transactions per block (each transaction typically consumes 21,000 gas).

By leveraging this implementation, we are able to conduct our experiments and evaluate the performance of our ABS mechanism. Section VIII provides a detailed explanation of the implementation steps.

\subsection{Experiment Results} We compare our ABS mechanism with the two benchmarks and evaluate the impacts of the randomness of buyer/seller number and the block size limit.
\subsubsection{Mechanism Comparison}
\begin{figure}[h]
	\centering
	\includegraphics[width=0.475\linewidth]{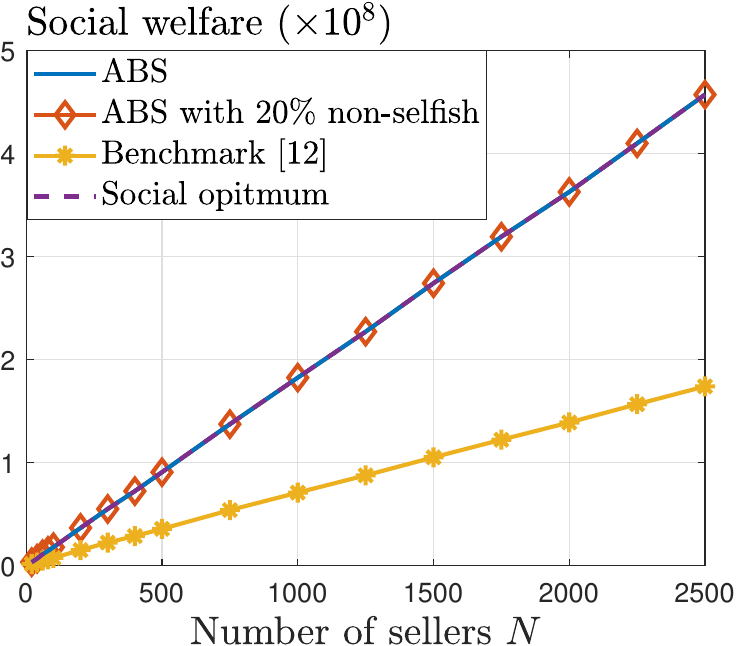}
	\caption{Mechanism comparison for homogeneous-quantity trading.}
	\label{fig:single_unit1}
\end{figure}
\begin{figure}[h]
	\centering
	\subfigure[Social welfare.]{
		\includegraphics[width=0.475\linewidth]{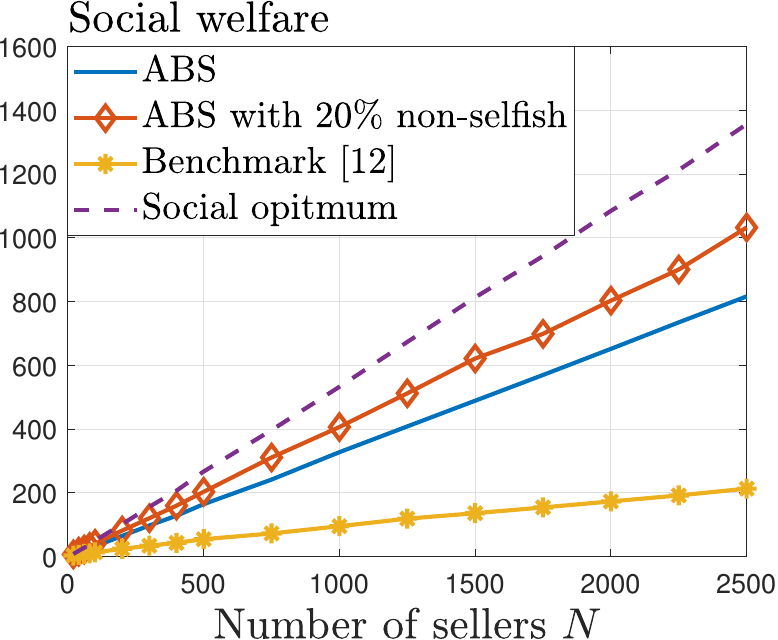}}
	\subfigure[Performance ratio.]{
		\includegraphics[width=0.475\linewidth]{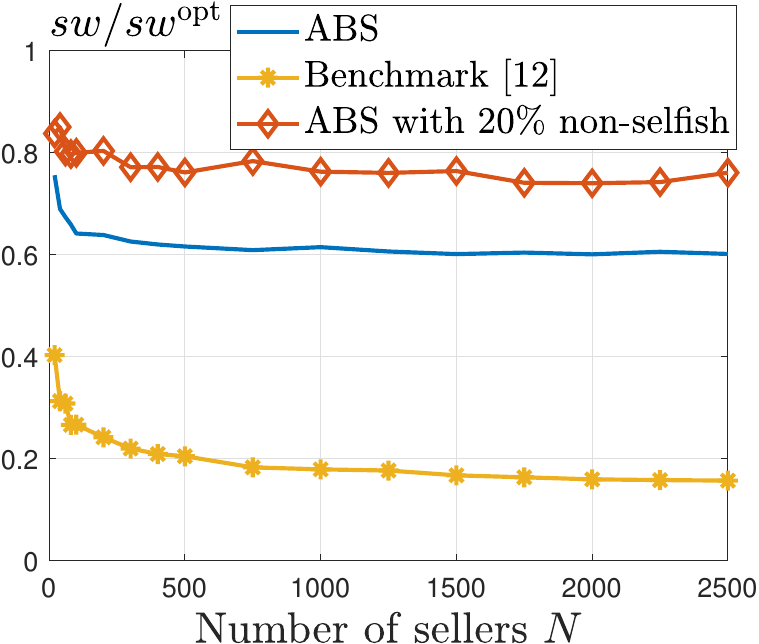}}
	\vspace{-2mm}
	\caption{Mechanism comparison for heterogeneous-quantity trading.}
	\label{Fig:sw}
\end{figure}


We establish the homogeneous-quantity trading and heterogeneous-quantity trading settings by using the fitted PDFs from the NFT trading dataset \cite{NFT_trading_data} and the Bitcoin trading dataset \cite{Dataset}, respectively.

For homogeneous-quantity trading, Fig. \ref{fig:single_unit1} compares different mechanisms in terms of social welfare. The social welfare of all mechanisms increases with the number of sellers due to more matching opportunities. Notably, our ABS mechanism, ABS with 20\% non-selfish miners, and the social optimum overlap in the figure. This confirms the result in Theorem \ref{thm:dist} for our ABS mechanism's asymptotic optimum and demonstrates the close-to-optimal performance even with non-large seller number cases. Moreover, the benchmark mechanism only achieves about 40\% of our mechanism's social welfare.

For heterogeneous-quantity trading, Fig. \ref{Fig:sw}(a) shows that our mechanism achieves 3.7 times higher social welfare than the benchmark mechanism by strategically adjusting the block size to prevent social welfare loss. Additionally, the presence of non-selfish miners who follow the recommendation further enhances the social welfare of our mechanism. This highlights that our mechanism performs well even when the system deviates from our rational-miner assumption.

In Fig. \ref{Fig:sw}(b), the performance ratio of our mechanism converges to 0.6, indicating that it achieves approximately 60\% of the social optimum. With 20\% non-selfish miners, the performance ratio can be improved to 78\%. 

Overall, our ABS mechanism's performance is close to optimum for practical homogeneous-quantity trading like NFT trading. Moreover, it dominates the benchmark mechanism in heterogeneous-quantity trading settings.
	\subsubsection{Impact of Buyer/Seller Number Randomness}
	\begin{figure}[h]
		\centering
		\subfigure[Homogeneous-quantity trading.]{
			\includegraphics[width=0.475\linewidth]{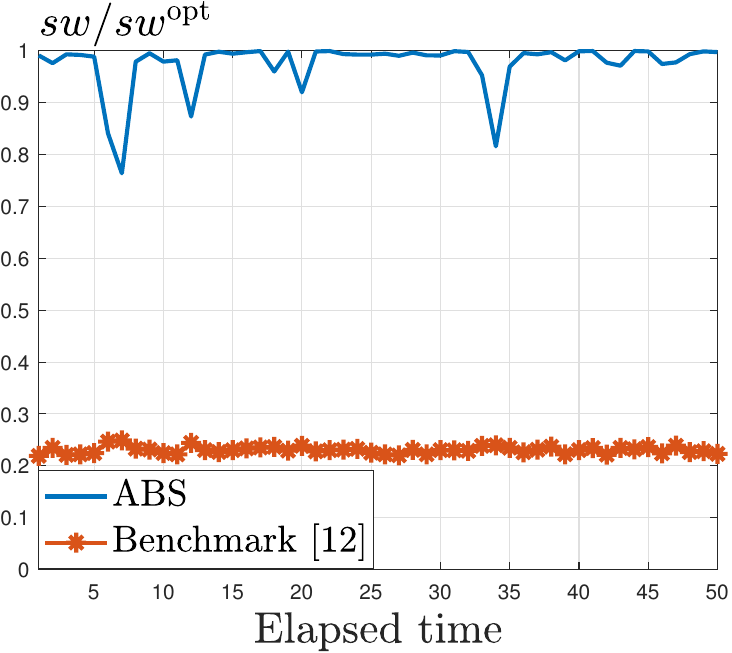}}
		\subfigure[Heterogeneous-quantity trading.]{
			\includegraphics[width=0.475\linewidth]{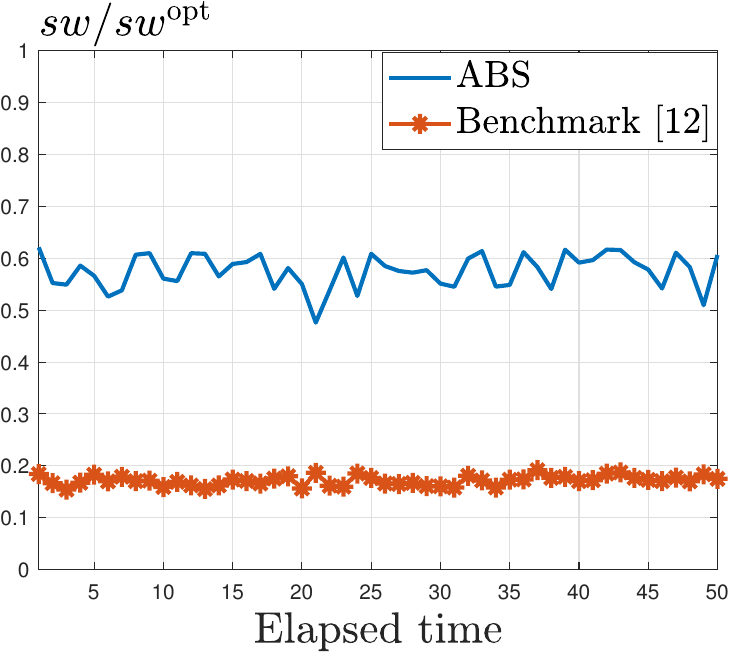}}
		\vspace{-2mm}
		\caption{Impact of the random number of users on performance ratio.}
		\label{Fig:sw5}
		\vspace{-3mm}
	\end{figure} 
	In practice, the number of buyers and sellers can vary randomly over time. Here we analyze cases where the number of sellers $N$ follows the number of buyers/sellers in the Bitcoin trading dataset \cite{Dataset}. Our mechanism sets the block size based on (\ref{A_dist}), using the average value of $N$. Fig. \ref{Fig:sw5} illustrates how the performance ratio changes over time.
	
	Fig. \ref{Fig:sw5}(a) illustrates the homogeneous-quantity trading. The performance ratio fluctuates with time as the number of sellers changes. Our ABS mechanism achieves an average performance ratio of 96\%, while the benchmark mechanism achieves an average of 40\%. 
	
	Fig. \ref{Fig:sw5}(b) illustrates heterogeneous-quantity trading. Our ABS and benchmark mechanisms achieve lower performance ratios than homogeneous-quantity trading, as miners can match high-buying-quantity transactions with low-selling-quantity transactions, causing social welfare loss. Nevertheless, our mechanism achieves an average performance ratio of 58\%, which is 3.3 times higher than the benchmark mechanism.
	
	To sum up, Fig. \ref{Fig:sw5} shows that our ABS mechanism still outperforms the benchmark even when the number of buyers and sellers is random, showcasing the practicality of our mechanism in real-world implementations. Additionally, our benchmark mechanism highlights the shortcomings of existing matching algorithms in blockchain that neglect miners' incentives, indicating that our mechanism achieves universal dominance over existing matching algorithms.
	
	\subsubsection{Impact of Block Size Limit on ABS}
	\begin{figure}
		\centering
		\subfigure[Optimal block size.]{
			\includegraphics[width=0.47\linewidth]{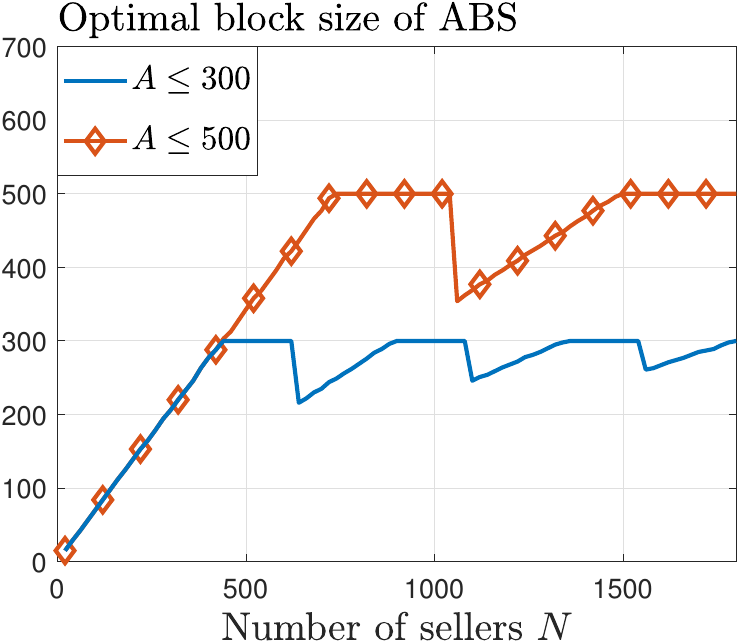}}
		\subfigure[Social welfare comparison.]{
			\includegraphics[width=0.475\linewidth]{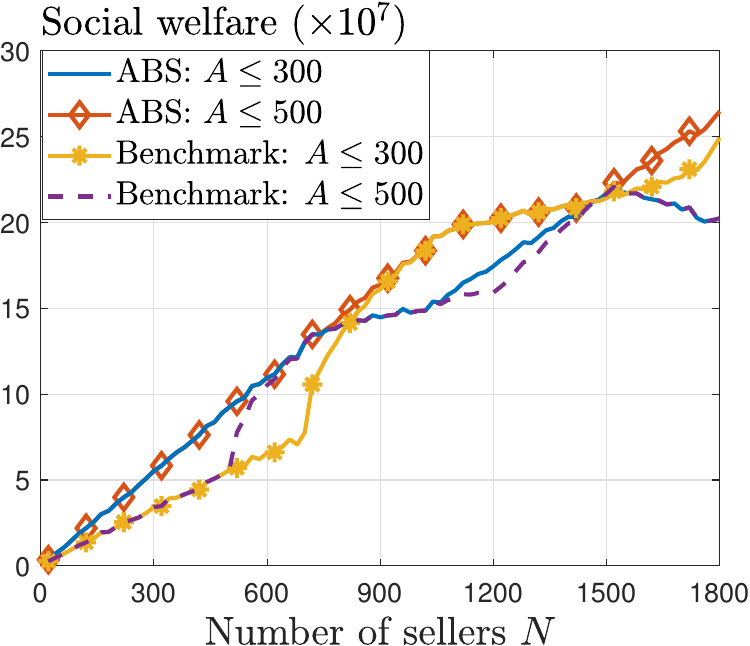}}
		\vspace{-2mm}
		\caption{Impact of block size limit on homogeneous-quantity trading.}
		\vspace{-3mm}
		\label{Fig:sw6}
	\end{figure}
	
	\begin{figure}
		\centering
		\subfigure[Optimal block size.]{
			\includegraphics[width=0.47\linewidth]{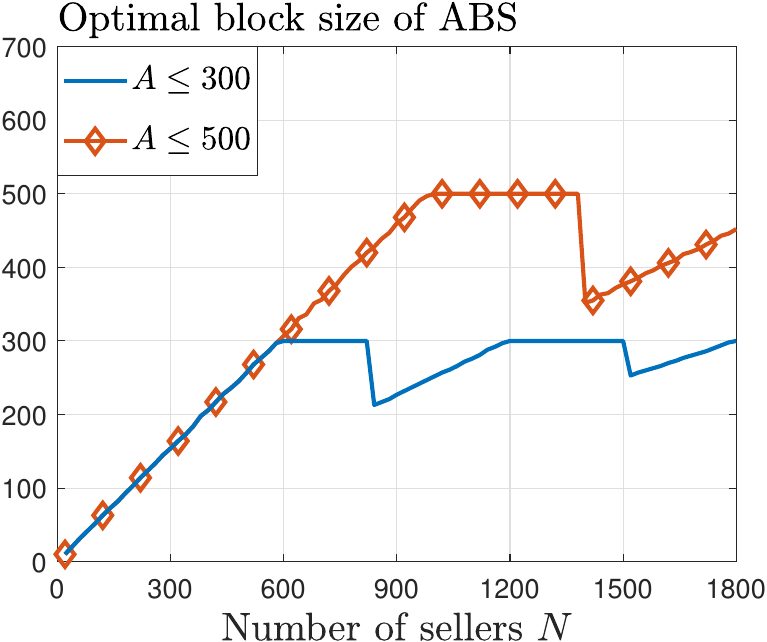}}
		\subfigure[Social welfare comparison.]{
			\includegraphics[width=0.475\linewidth]{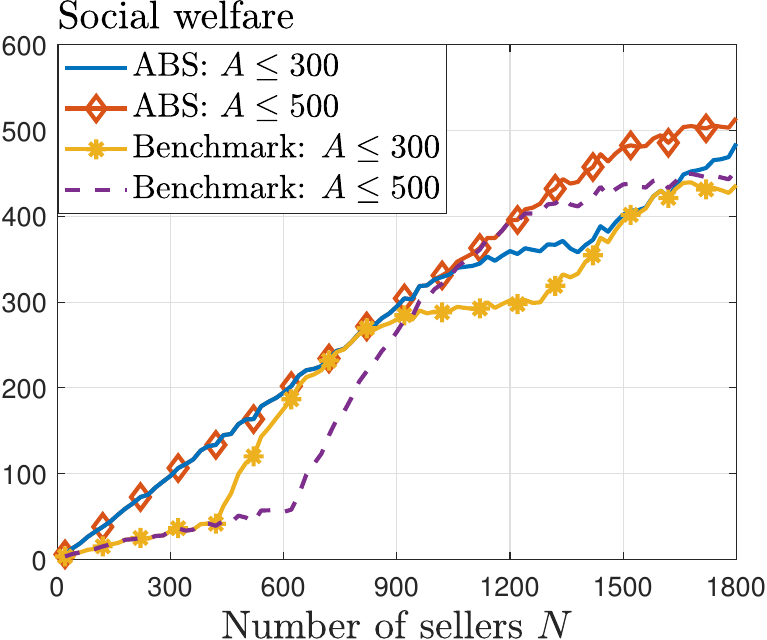}}
		\vspace{-2mm}
		\caption{Impact of block size limit on heterogeneous-quantity trading.}
		\vspace{-3mm}
		\label{Fig:sw7}
	\end{figure}
	
	The blockchain system generally has a block size limit due to the limited network bandwidth \cite{SIG1}\cite{sun2021time}. Here, we examine the impact of block size limits on the performance of our mechanism. 
	
	We consider block size limits of $A\leq 300$ and $A \leq 500$ and use a brute-force search to optimally solve problem (\ref{Mech_design2}) with the additional block size limit constraint. 
	
	Fig. \ref{Fig:sw6} illustrates the optimal block size for our ABS mechanism and the corresponding social welfare for homogeneous-quantity trading. In Fig. \ref{Fig:sw6}(a), when the number of sellers $N$ is small, the optimal block size of our ABS mechanism increases with $N$. However, as $N$ increases, our ABS mechanism sometimes strategically reduces the block size. This is because larger block sizes can lead to a higher tendency to ignore desirable matches, as explained in Theorem~\ref{thm:PoA1}, which decreases social welfare. Consequently, our ABS mechanism may decrease the block size and bear a higher delay cost to mitigate this issue. This finding underscores the importance of our ABS mechanism even under block size limits.
	
	Fig. \ref{Fig:sw6}(b) compares the social welfare of our ABS mechanism with the benchmark. Even with the block size limit, our mechanism achieves higher social welfare than the benchmark, highlighting its practical necessity. The improvement is more significant when the block size limit is large (i.e., $A\leq 500$) and the number of sellers is small (i.e., $N\leq 600$). Such an improvement is even more pronounced for heterogeneous-quantity trading in Fig. \ref{Fig:sw7}(b), with an average improvement of 65\%.
	
	Fig. \ref{Fig:sw7} illustrates the optimal block size and the corresponding social welfare for heterogeneous-quantity trading. The insights are consistent with those observed in the homogeneous-quantity trading setting.
	
	Hence, our ABS mechanism can still improve social welfare under the block size limit constraint. The improvement is particularly significant when the number of buyers and sellers is small.
	\section{Conclusion}\label{sec:conclusion}
	\vspace{-1pt}
	In this paper, we proposed a block size optimization mechanism to address the inefficiencies in blockchain-based order book (BBOB) systems caused by miners' self-interested actions. Our dynamic game-theoretical model highlighted how these self-interested actions lead to suboptimal transaction matches, significantly reducing social welfare. To mitigate this issue, we designed a mechanism where the system designer optimizes the block size to balance the tendency to ignore desirable matches with maintaining high throughput. Our analysis revealed that the optimized block size mechanism could achieve the social optimum under homogeneous-quantity matching and provide a bounded approximation ratio under heterogeneous-quantity matching. We implemented our proposed mechanism on an Ethereum blockchain testbed, validating its practical applicability. Through extensive experiments using actual BBOB data, we demonstrated that our mechanism achieved social optimum for homogeneous-quantity matching like NFT. For heterogeneous-quantity matching, our mechanism enhances social welfare by up to 3.7 times compared to the existing benchmark and achieves at least 60\% of the social optimum. Furthermore, our mechanism demonstrated robustness in scenarios with random variations in the number of buyers and sellers. 
	
	In future work, we will consider the impact of different miner incentive structures that can adapt to varying levels of miner rationality. Furthermore, we will consider participants making strategic decisions on bid and ask prices.

\newpage
\section{Model Analysis}

\subsection{Stage II of Miners' Transaction Selection Analysis}
	
\textbf{Proof of Theorem \ref{Stage_III_result}.} 

Theorem \ref{Stage_III_result} follows naturally from the objective of miners to maximize their payoff. Miners will always select the highest-fee transaction set that can be matched. Propositions 1 and 2 later show that buyers (or sellers) who cannot be matched will pay lower transaction fees than those who can be matched. Therefore, the strategy of selecting only the top-fee transactions ensures that each miner maximizes their payoff.

This completes the proof of Theorem 1.

\newpage

\subsection{Stage I Analysis}

\textbf{Proof of Proposition \ref{prop:no_NE}.} 

We will first prove the case of $A < A_{th}$, and then prove the case of $A\geq A_{th}$.

\subsubsection{Proof of Case $A< A_{th}$} We use the contradiction method to prove the non-existence of pure strategy Nash equilibrium. We assume there exists a pure strategy Nash equilibrium $\boldsymbol{f}^{\rm NE} = (f_k^{\rm buy,NE}, f_n^{\rm sell,NE}, \forall k \in \mathcal{K}, \forall n \in \mathcal{N})$. Then we arrange the buyers' and sellers' fees in a decreasing order: $f_1^{\rm buy,NE} \geq f_2^{\rm buy,NE}\geq \cdots \geq f_K^{\rm buy,NE}$ and $ f_1^{\rm sell,NE} \geq  f_2^{\rm sell,NE} \geq \cdots  f_N^{\rm sell,NE}$. Then based on the assumption that there exists a pure strategy Nash equilibrium, we will prove that $ f_1^{\rm buy,NE} =  f_2^{\rm buy,NE} = \cdots =  f_A^{\rm buy,NE}$ and $ f_1^{\rm sell,NE} =  f_2^{\rm sell,NE} = \cdots =  f_A^{\rm sell,NE}$ by contradiction.

As we arrange the fee in a decreasing order, we assume that buyer $i$'s fee satisfies $f_i^{\rm buy,NE} >  f_{i+1}^{\rm buy,NE}$, where $1 \leq i \leq A-1$. Then buyer $i$ can deviate to set the transaction fee as $f_{i+1}^{\rm buy,NE}$, where he pay strictly less fee and do not change the miners' matching result and delay cost. This contradicts the pure strategy Nash equilibrium. Hence, we have proved that $ f_1^{\rm buy,NE} =  f_2^{\rm buy,NE} = \cdots =  f_A^{\rm buy,NE}$. Following the same proof logic, we can also prove that  $ f_1^{\rm sell,NE} =  f_2^{\rm sell,NE} = \cdots =  f_A^{\rm sell,NE}$.

Next, we prove that $ f_{A+1}^{\rm buy,NE} <  f_A^{\rm buy,NE}$. If buyer $A+1$ sets the fee as $f_{A+1}^{\rm buy,NE} = f_A^{\rm buy,NE}$, then buyer $A$ has $\frac{1}{A+1}$ probability that transaction is included in the second block and bear the delay cost. He can deviate to pay fee $f_A^{\rm buy,NE} + \epsilon$ to bear no delay cost, strictly increasing his payoff. This contradicts the pure strategy Nash equilibrium. Hence, we must have $f_{A+1}^{\rm buy,NE} < f_A^{\rm buy,NE}$. 

Then, we prove that $ f_{A+1}^{\rm buy,NE} =  f_A^{\rm buy,NE}-\epsilon$. If buyer $A+1$ sets the fee as $f_{A+1}^{\rm buy,NE} < f_A^{\rm buy,NE}-\epsilon$, then buyer $A$ can deviate to set the transaction fee as $f_{A+1}^{\rm buy,NE} + \epsilon$, where he pay strictly less fee and do not change the miners' matching result and delay cost, strictly increasing his payoff. This contradicts the pure strategy Nash equilibrium. Hence, we must have $f_{A+1}^{\rm buy,NE} = f_A^{\rm buy,NE} - \epsilon$.

However, when  $ f_{A+1}^{\rm buy,NE} =  f_A^{\rm buy,NE}-\epsilon$, then buyer $A+1$ can deviate to set the fee as $f_A^{\rm buy,NE} + 2\epsilon$, where his transaction is included in blockchain in the first block. He strictly decreases his delay cost by $d$ while paying slightly higher fee. Hence, this contradicts the pure strategy Nash equilibrium.

To sum up, we have proved the non-existence of pure strategy Nash equilibrium for Case $A< A_{th}$.

\subsubsection{Proof of Case $A\geq A_{th}$} For this case, we just need to verify that no buyer or seller can increase his payoff by deviating from the pure strategy Nash equilibrium. 

\textbf{We first analyze buyers with $k> \min\{A,N\}$}. Under the equilibrium strategies characterized by (10) and (11), miners will select buyers with $k\leq \min\{A,N\}$ and sellers with $n\leq \min\{A,K\}$ to included in the blockchain. For buyers with $k> \min\{A,N\}$ and sellers with $n\leq \min\{A,K\}$, their transactions will not be matched and get zero payoff. For a buyer with $k> \min\{A,N\}$, if he deviates to set lower transaction fee, his transaction is still not included in the blockchain, and his payoff is unchanged. If buyer $k$ deviates to increase transaction fee to $f_k^{\rm buy} \geq \sigma_{th}^{\rm sell}(A_{th},A)+\epsilon$, the matching surplus of his transaction is:
\begin{equation}
	u_k^{\rm Sur}(f_k^{\rm buy}) = \frac{\sum_{n = 1}^{\min\{\lceil\frac{A_{th}}{A}\rceil A, N\}}\mathbbm{1}(R_k\geq C_n)\min\{b_k,q_n\}(R_k - C_n)}{2\sum_{n = 1}^{\min\{\lceil\frac{A_{th}}{A}\rceil A, N\}}\mathbbm{1}(R_k\geq C_n)} -f_k^{\rm buy}.
\end{equation}
The delay cost is:
\begin{equation}
	u_k^{\rm Delay}(f_k^{\rm buy}) = (\lceil\frac{\min\{\lceil\frac{A_{th}}{A}\rceil A, N\}}{A}\rceil A -1)d.
\end{equation}
Hence, the payoff of setting transaction fee as $f_k^{\rm buy}$ is
\begin{equation}\label{payoff1}
	u_k = u_k^{\rm Sur}(f_k^{\rm buy})  - u_k^{\rm Delay}(f_k^{\rm buy}).
\end{equation}

Note that for buyer $k = \min\{A, N\}+1$, the above (\ref{payoff1}) achieves the highest payoff of:
\begin{equation}
	\begin{aligned}
		u_k &\leq u_{\min\{\lceil\frac{A_{th}}{A}\rceil A, N\}+1} \\
		&= \frac{\sum_{n = 1}^{\min\{\lceil\frac{A_{th}}{A}\rceil A, N\}}\mathbbm{1}(R_{\min\{\lceil\frac{A_{th}}{A}\rceil A, N\}+1}\geq C_n)\min\{b_{\min\{\lceil\frac{A_{th}}{A}\rceil A, N\}+1},q_n\}(R_{\min\{\lceil\frac{A_{th}}{A}\rceil A, N\}+1} - C_n)}{2\sum_{n = 1}^{\min\{\lceil\frac{A_{th}}{A}\rceil A, N\}}\mathbbm{1}(R_{\min\{\lceil\frac{A_{th}}{A}\rceil A, N\}+1}\geq C_n)} \\
		&\hspace{5mm}- (\lceil\frac{\min\{\lceil\frac{A_{th}}{A}\rceil A, N\}}{A}\rceil A -1)d - \sigma_{th}^{\rm sell}(A_{th},A)-\epsilon\\
		& \leq -\epsilon.
	\end{aligned}
\end{equation}

To sum up, for buyers with $k> \min\{A,N\}$, increasing or decreasing the transaction fee cannot increase his payoff. Following the same analysis process, we can also prove that for sellers with $n> \min\{K,N\}$, he cannot increase his payoff by deviating from the pure strategy Nash equilibrium.

\textbf{We then analyze buyers with $k\leq  \min\{A,N\}$}. We derive his payoff and then compare it with he deviating to set higher or lower transaction fee. For buyer $k\leq  \min\{A,N\}$ setting the transaction fee as $f_k^{\rm buy,NE}$, his matching surplus is
\begin{equation}
	u_k^{\rm Sur}(f_k^{\rm buy,NE}) = \frac{\sum_{n = 1}^{\min\{\lceil\frac{A_{th}}{A}\rceil A, N\}}\mathbbm{1}(R_k\geq C_n)\min\{b_k,q_n\}(R_k - C_n)}{2\sum_{n = 1}^{\min\{\lceil\frac{A_{th}}{A}\rceil A, N\}}\mathbbm{1}(R_k\geq C_n)} -f_k^{\rm buy,NE}.
\end{equation}
The delay cost is:
\begin{equation}
	u_k^{\rm Delay}(f_k^{\rm buy,NE}) = (\lceil\frac{\min\{\lceil\frac{A_{th}}{A}\rceil A, N\}}{A}\rceil A -1)d.
\end{equation}
Hence, the payoff of setting transaction fee as $f_k^{\rm buy,NE}$ is
\begin{equation}\label{payoff2}
	u_k = u_k^{\rm Sur}(f_k^{\rm buy,NE})  - u_k^{\rm Delay}(f_k^{\rm buy,NE}).
\end{equation}
Note that for buyer $k = \min\{A, N\}$, the above (\ref{payoff2}) achieves the lowest payoff of:
\begin{equation}
	\begin{aligned}
		u_k &\geq u_{\min\{\lceil\frac{A_{th}}{A}\rceil A, N\}} \\
		&= \frac{\sum_{n = 1}^{\min\{\lceil\frac{A_{th}}{A}\rceil A, N\}}\mathbbm{1}(R_{\min\{\lceil\frac{A_{th}}{A}\rceil A, N\}+1}\geq C_n)\min\{b_{\min\{\lceil\frac{A_{th}}{A}\rceil A, N\}+1},q_n\}(R_{\min\{\lceil\frac{A_{th}}{A}\rceil A, N\}+1} - C_n)}{2\sum_{n = 1}^{\min\{\lceil\frac{A_{th}}{A}\rceil A, N\}}\mathbbm{1}(R_{\min\{\lceil\frac{A_{th}}{A}\rceil A, N\}+1}\geq C_n)} \\
		&\hspace{5mm}- (\lceil\frac{\min\{\lceil\frac{A_{th}}{A}\rceil A, N\}}{A}\rceil A -1)d - \sigma_{th}^{\rm sell}(A_{th},A)-\epsilon\\
		& \geq 0.
	\end{aligned}
\end{equation}

When buyer $k$ deviates to pay a higher transaction fee, he cannot change the matching results and his delay cost. Hence, he has no incentive to increase his transaction fee. When buyer $k\leq  \min\{A,N\}$ deviates to decrease transaction fee to $f_k^{\rm buy} \leq \sigma_{th}^{\rm sell}(A_{th},A)$, then he cannot get matched and getting the payoff of zero.

To sum up, for buyers with $k\leq \min\{A,N\}$, increasing or decreasing the transaction fee cannot increase his payoff. Following the same analysis process, we can also prove that for sellers with $n\leq \min\{K,N\}$, he cannot increase his payoff by deviating from the pure strategy Nash equilibrium.

This completes the proof of Proposition \ref{prop:no_NE}.

\newpage

\textbf{Proof of Proposition \ref{stage_II_result}}

\textbf{In this proof, we only present the analysis of buyers, as the proof sketch is exactly the same as for sellers. Here we first show that buyers who adopt pure strategy in (\themyeqcounterrr b) cannot increase his payoff by setting different transaction fee. Next we prove the mixed strategy in (\themyeqcounterrr a).}

\vspace{10mm}
\textbf{Buyers with $k> \min\{\lceil\frac{A_{th}}{A}\rceil A, N\}$}: For any buyer $k> \min\{\lceil\frac{A_{th}}{A}\rceil A, N\}$, his transaction cannot get matched and he gets zero payoff. If he deviates to set lower transaction fee, his transaction still cannot get matched, which cannot increase his payoff. If buyer $k$ deviates to set higher transaction fee, there are two situations:
\begin{enumerate}
	\item If $f_k^{\rm buy} > \sigma_{th}^{\rm buy}(A_{th},A)+\epsilon + (\lceil\frac{\min\{\lceil\frac{A_{th}}{A}\rceil A, N, K\}}{A}\rceil-1)d$, then the matching surplus of his transaction is:
	\begin{equation}
		u_k^{\rm Sur}(f_k^{\rm buy}) = \frac{\sum_{n = 1}^{\min\{\lceil\frac{A_{th}}{A}\rceil A, N\}}\mathbbm{1}(R_k\geq C_n)\min\{b_k,q_n\}(R_k - C_n)}{2\sum_{n = 1}^{\min\{\lceil\frac{A_{th}}{A}\rceil A, N\}}\mathbbm{1}(R_k\geq C_n)} -f_k^{\rm buy}.
	\end{equation}
	The delay cost is:
	\begin{equation}
		u_k^{\rm Delay}(f_k^{\rm buy}) = 0.
	\end{equation}
	Hence, the payoff of setting transaction fee as $f_k^{\rm buy} \geq \sigma_{th}^{\rm buy}(A_{th},A)+\epsilon + (\lceil\frac{\min\{\lceil\frac{A_{th}}{A}\rceil A, N, K\}}{A}\rceil-1)d$ is
	\begin{equation}
		u_k(f_k^{\rm buy}) = \frac{\sum_{n = 1}^{\min\{\lceil\frac{A_{th}}{A}\rceil A, N\}}\mathbbm{1}(R_k\geq C_n)\min\{b_k,q_n\}(R_k - C_n)}{2\sum_{n = 1}^{\min\{\lceil\frac{A_{th}}{A}\rceil A, N\}}\mathbbm{1}(R_k\geq C_n)} -f_k^{\rm buy} <0 .
	\end{equation}
    Hence, for buyer $k$, setting the transaction fee as $f_k^{\rm buy} > \sigma_{th}^{\rm buy}(A_{th},A)+\epsilon + (\lceil\frac{\min\{\lceil\frac{A_{th}}{A}\rceil A, N, K\}}{A}\rceil-1)d$ cannot increase his payoff.
	\item Setting transaction fee as $f_k^{\rm buy}$ satisfying $\sigma_{th}^{\rm buy}(A_{th},A)+\epsilon \leq f_k^{\rm buy} \leq \sigma_{th}^{\rm buy}(A_{th},A)+\epsilon + (\lceil\frac{\min\{\lceil\frac{A_{th}}{A}\rceil A, N, K\}}{A}\rceil-1)d$, then the matching utility of his transaction is:
	\begin{equation}
		\frac{\sum_{n = 1}^{\min\{\lceil\frac{A_{th}}{A}\rceil A, N\}}\mathbbm{1}(R_k\geq C_n)\min\{b_k,q_n\}(R_k - C_n)}{2\sum_{n = 1}^{\min\{\lceil\frac{A_{th}}{A}\rceil A, N\}}\mathbbm{1}(R_k\geq C_n)}.
	\end{equation}
	The sum of fee payment and delay cost is:
	\begin{equation}
		g(1-F^{\rm sell,NE}(f_k^{\rm buy}),f_k^{\rm buy},\min\{\lceil\frac{A_{th}}{A}\rceil A, N, K\},A).
	\end{equation}
	Hence, the payoff of setting transaction fee as $f_k^{\rm buy}$ is
	\begin{equation}
		\begin{aligned}
			&u_k(f_k^{\rm buy}) \\
			= &\frac{\sum_{n = 1}^{\min\{\lceil\frac{A_{th}}{A}\rceil A, N\}}\mathbbm{1}(R_k\geq C_n)\min\{b_k,q_n\}(R_k - C_n)}{2\sum_{n = 1}^{\min\{\lceil\frac{A_{th}}{A}\rceil A, N\}}\mathbbm{1}(R_k\geq C_n)} - g(1-F^{\rm sell,NE}(f_k^{\rm buy}),f_k^{\rm buy},\min\{\lceil\frac{A_{th}}{A}\rceil A, N, K\},A)\\
			= &\frac{\sum_{n = 1}^{\min\{\lceil\frac{A_{th}}{A}\rceil A, N\}}\mathbbm{1}(R_k\geq C_n)\min\{b_k,q_n\}(R_k - C_n)}{2\sum_{n = 1}^{\min\{\lceil\frac{A_{th}}{A}\rceil A, N\}}\mathbbm{1}(R_k\geq C_n)} - g(1,\sigma_{th}^{\rm sell}(A_{th},A)+\epsilon,\min\{\lceil\frac{A_{th}}{A}\rceil A, N, K\},A)\\
			= &\frac{\sum_{n = 1}^{\min\{\lceil\frac{A_{th}}{A}\rceil A, N\}}\mathbbm{1}(R_k\geq C_n)\min\{b_k,q_n\}(R_k - C_n)}{2\sum_{n = 1}^{\min\{\lceil\frac{A_{th}}{A}\rceil A, N\}}\mathbbm{1}(R_k\geq C_n)}  - (\min\{\lceil\frac{A_{th}}{A}\rceil A, N, K\}-1)d- \sigma_{th}^{\rm sell}(A_{th},A)-\epsilon\\
			= &-\epsilon \\
			< & 0.
		\end{aligned}
	\end{equation}
    Hence, for buyer $k$, setting the transaction fee as $\sigma_{th}^{\rm buy}(A_{th},A)+\epsilon \leq f_k^{\rm buy} \leq \sigma_{th}^{\rm buy}(A_{th},A)+\epsilon + (\lceil\frac{\min\{\lceil\frac{A_{th}}{A}\rceil A, N, K\}}{A}\rceil-1)d$ cannot increase his payoff.
\end{enumerate}
To sum up, for any buyer $k> \min\{\lceil\frac{A_{th}}{A}\rceil A, N\}$, setting different transaction fee cannot increase his payoff.

\vspace{10mm}
\textbf{Next we prove the mixed strategy for buyers with $k\leq \min\{\lceil\frac{A_{th}}{A}\rceil A, N\}$. Here we first present Lemma 2.1-2.4 to analyze the property of mixed-strategy distribution, then we derive the mixed strategy Nash equilibrium.}

For all buyer (or seller) who adopt mixed strategy, we consider their strategies follows the same probability distribution. Note that the distributions of buyers and sellers are different. Moreover, as some transactions are not matched, (e.g., if $N > K$, then some buyers' transactions cannot be matched), the strategies of matched and unmatched ones are different. For buyer $k$ who adopt the mixed strategy, we define his strategy is $\mu_{k}^{\rm NE}$, which is a probability measure over $[0,\infty)$. We define the corresponding CDF as $G_k^{\rm NE}(f_k^{\rm buy})$.

We use $L$ and $U$ to represent the strategy of transaction fee's \textit{lower support} and \textit{upper support}, respectively, which are defined as follows:
\begin{equation} 
	\begin{aligned} 
		L &\triangleq \sup \left\{\underline{f}_k : \mu_{k}^{\rm NE}(\{f_k^{\rm buy} \geq \underline{f}_k\})=1\right\}, \\ 
		U &\triangleq\inf \left\{\overline{f}_k : \mu_{k}^{\rm NE}(\{f_k^{\rm buy} \leq \overline{f}_k\})=1\right\}. \end{aligned}
\end{equation}

\textbf{Lemma 2.1.     }\emph{We define $L$ and $U$ as the lower and upper support of transaction fee-per-byte $\mu_{k}^{\rm NE}$ at mixed strategy Nash equilibrium, respectively. Then $\sigma_{th}^{\rm buy}(A_{th},A)+\epsilon \leq L,U \leq  \sigma_{th}^{\rm buy}(A_{th},A)+\epsilon + (\lceil\frac{\min\{\lceil\frac{A_{th}}{A}\rceil A, N, K\}}{A}\rceil-1)d$.}
\vspace{5mm}

\emph{Proof of Lemma 2.1.} 

\textbf{We will prove the range of $L$ and $U$ by contradiction. Notice that as we will focus on a particular buyer $k$'s payoff, we will denote his payoff function as $u_k(f_k^{\rm buy})$ to simplify the notation.}

Assume $U>\sigma_{th}^{\rm buy}(A_{th},A)+\epsilon + (\lceil\frac{\min\{\lceil\frac{A_{th}}{A}\rceil A, N, K\}}{A}\rceil-1)d$: then $\mu_{k}^{\rm NE}(\{f_k^{\rm buy}>\sigma_{th}^{\rm buy}(A_{th},A)+\epsilon + (\lceil\frac{\min\{\lceil\frac{A_{th}}{A}\rceil A, N, K\}}{A}\rceil-1)d\})>0$ at equilibrium. When buyer $k$ sets fee $f_k^{\rm buy}> \sigma_{th}^{\rm buy}(A_{th},A)+\epsilon + (\lceil\frac{\min\{\lceil\frac{A_{th}}{A}\rceil A, N, K\}}{A}\rceil-1)d$, the matching surplus of his transaction is:
\begin{equation}
	u_k^{\rm Sur}(f_k^{\rm buy}) = \frac{\sum_{n = 1}^{\min\{\lceil\frac{A_{th}}{A}\rceil A, N\}}\mathbbm{1}(R_k\geq C_n)\min\{b_k,q_n\}(R_k - C_n)}{2\sum_{n = 1}^{\min\{\lceil\frac{A_{th}}{A}\rceil A, N\}}\mathbbm{1}(R_k\geq C_n)} -f_k^{\rm buy}.
\end{equation}
The delay cost is:
\begin{equation}
	u_k^{\rm Delay}(f_k^{\rm buy}) = 0.
\end{equation}
Hence, the payoff of setting transaction fee as $f_k^{\rm buy}$ is
\begin{equation}
	u_k(f_k^{\rm buy}) = \frac{\sum_{n = 1}^{\min\{\lceil\frac{A_{th}}{A}\rceil A, N\}}\mathbbm{1}(R_k\geq C_n)\min\{b_k,q_n\}(R_k - C_n)}{2\sum_{n = 1}^{\min\{\lceil\frac{A_{th}}{A}\rceil A, N\}}\mathbbm{1}(R_k\geq C_n)} -f_k^{\rm buy} .
\end{equation}

Note that $f_k^{\rm buy}> \sigma_{th}^{\rm buy}(A_{th},A)+\epsilon + (\lceil\frac{\min\{\lceil\frac{A_{th}}{A}\rceil A, N, K\}}{A}\rceil-1)d$. Hence, buyer $k$ can increase his payoff by setting the transaction fee as $\sigma_{th}^{\rm buy}(A_{th},A)+\epsilon$, the corresponding payoff is
\begin{equation}
	\begin{aligned}
		u_k(\sigma_{th}^{\rm buy}(A_{th},A)+\epsilon) = & \frac{\sum_{n = 1}^{\min\{\lceil\frac{A_{th}}{A}\rceil A, N\}}\mathbbm{1}(R_k\geq C_n)\min\{b_k,q_n\}(R_k - C_n)}{2\sum_{n = 1}^{\min\{\lceil\frac{A_{th}}{A}\rceil A, N\}}\mathbbm{1}(R_k\geq C_n)} \\
	 	&- (\lceil\frac{\min\{\lceil\frac{A_{th}}{A}\rceil A, N, K\}}{A}\rceil-1)d-\big(\sigma_{th}^{\rm buy}(A_{th},A)+\epsilon\big)\\
	 	>&u_k(f_k^{\rm buy}).
	\end{aligned}
\end{equation}
This contradicts the Nash equilibrium, where no buyer or seller can increase his payoff by deviating. Such a contradiction shows that the upper bound $U$ satisfies $U\leq \sigma_{th}^{\rm buy}(A_{th},A)+\epsilon + (\lceil\frac{\min\{\lceil\frac{A_{th}}{A}\rceil A, N, K\}}{A}\rceil-1)d $.

Assume $L<\sigma_{th}^{\rm buy}(A_{th},A)+\epsilon $: then $\mu_{k}^{\rm NE}(\{f_k^{\rm buy}<\sigma_{th}^{\rm buy}(A_{th},A)+\epsilon\})>0$ at equilibrium. When buyer $k$ sets fee $f_k^{\rm buy}< \sigma_{th}^{\rm buy}(A_{th},A)+\epsilon $, his transaction cannot get matched and buyer $k$ gets zero payoff. Then buyer $k$ can increase his payoff by setting the transaction fee as $f_k^{\rm buy} =\sigma_{th}^{\rm buy}(A_{th},A)+\epsilon$, where his payoff is 
\begin{equation}
	\begin{aligned}
		u_k(\sigma_{th}^{\rm buy}(A_{th},A)+\epsilon) = & \frac{\sum_{n = 1}^{\min\{\lceil\frac{A_{th}}{A}\rceil A, N\}}\mathbbm{1}(R_k\geq C_n)\min\{b_k,q_n\}(R_k - C_n)}{2\sum_{n = 1}^{\min\{\lceil\frac{A_{th}}{A}\rceil A, N\}}\mathbbm{1}(R_k\geq C_n)} \\
		&- (\lceil\frac{\min\{\lceil\frac{A_{th}}{A}\rceil A, N, K\}}{A}\rceil-1)d-\big(\sigma_{th}^{\rm buy}(A_{th},A)+\epsilon\big)\\
		>&0.
	\end{aligned}
\end{equation}

This contradicts the Nash equilibrium, where no buyer or seller can increase his payoff by deviating. Such a contradiction shows that the lower bound $L$ satisfies $L\geq \sigma_{th}^{\rm buy}(A_{th},A)+\epsilon$.

\textbf{Thus, all possible cases leads to contradiction, we have $\sigma_{th}^{\rm buy}(A_{th},A)+\epsilon \leq L,U \leq  \sigma_{th}^{\rm buy}(A_{th},A)+\epsilon + (\lceil\frac{\min\{\lceil\frac{A_{th}}{A}\rceil A, N, K\}}{A}\rceil-1)d$.} \textbf{This completes the proof of Lemma 2.1.}

\vspace{10mm}
We define an \textit{atom} as a point where the probability of a random variable equals to the value of this point is strictly positive. For example, $f_k^{\rm buy*}$ is an atom if $\mu_k^{\rm NE}(\{f_k^{\rm buy} = f_k^{\rm buy*}\}) > 0$.

\textbf{Lemma 2.2.     }\emph{For buyer's strategy at mixed-strategy Nash equilibrium $\mu_k^{\rm NE}$, it doesn't have any atom over $[L,U]$, i.e.,  $\forall f_k^{\rm buy \prime}\in[L,U], \mu_k^{\rm NE}(\{f_k^{\rm buy} = f_k^{\rm buy \prime}\}) = 0$.}

\vspace{5mm}
\textbf{\emph{Proof of Lemma 2.2.}}

\textbf{We prove the result by contradiction method.} Assuming there exists an atom in $\mu_k^{\rm NE}$ (i.e., $\exists f_k^{\rm buy \prime}\in[L,U], \mu_k^{\rm NE}(\{f_k^{\rm buy} = f_k^{\rm buy \prime}\}) > 0$), we will construct a strategy that achieves higher payoff than equilibrium strategy. Notice that as we will focus on a particular buyer $k$'s payoff, we will denote his payoff function as $u_k(f_k^{\rm buy})$ to simplify the notation.

\textbf{Assume $\mu_k^{\rm NE}$ has an atom $x_1$ within $[L,U]$, i.e., $\mu_k^{\rm NE}(\{f_k^{\rm buy}=x_1\})>0$.} Then we define following probabilities:
\begin{equation}
	\begin{cases}
		\mu_k^{\rm NE}(\{f_k^{\rm buy} > x_1\})= \zeta_1\geq 0,\\
		\mu_k^{\rm NE}(\{f_k^{\rm buy} = x_1\})= \zeta_2>0,\\
		\mu_k^{\rm NE}(\{f_k^{\rm buy} < x_1\})= \zeta_3\geq 0,\\
		\zeta_1+\zeta_2+\zeta_3 = 1.
	\end{cases}
\end{equation}

At mixed strategy Nash equilibrium, any strategy support yields same expected payoff. Thus, buyer $k$'s expected payoff can be calculated at transaction fee equaling $x_1$. The matching utility of his transaction is:
\begin{equation}\label{matching_utility}
	\frac{\sum_{n = 1}^{\min\{\lceil\frac{A_{th}}{A}\rceil A, N\}}\mathbbm{1}(R_k\geq C_n)\min\{b_k,q_n\}(R_k - C_n)}{2\sum_{n = 1}^{\min\{\lceil\frac{A_{th}}{A}\rceil A, N\}}\mathbbm{1}(R_k\geq C_n)}.
\end{equation}
The sum of fee payment and delay cost is:
\begin{equation}\label{Atom_deviation1}
	\begin{aligned}
		& \sum_{n=0}^{\min\{\lceil\frac{A_{th}}{A}\rceil A,K, N\}-1}\binom{\min\{\lceil\frac{A_{th}}{A}\rceil A,K, N\}-1}{n}(\zeta_1+\zeta_2)^n(1-\zeta_1-\zeta_2)^{\min\{\lceil\frac{A_{th}}{A}\rceil A, K, N\}-1-n}\\
		&\sum_{j=0}^{n}\binom{n}{j}(\frac{\zeta_2}{\zeta_1+\zeta_2})^j(\frac{\zeta_1}{\zeta_1+\zeta_2})^{n-j}(x_1+\sum_{l=1}^{j+1}\lceil\frac{n-j+l}{A}\rceil\frac{d}{j+1})\\
		<& \sum_{n=0}^{\min\{\lceil\frac{A_{th}}{A}\rceil A, K, N\}-1}\binom{\min\{\lceil\frac{A_{th}}{A}\rceil A, K, N\}-1}{n}\zeta_1^n(1-\zeta_1)^{\min\{\lceil\frac{A_{th}}{A}\rceil A, K, N\}-1-n}(x_1+\lceil\frac{n+1}{A}\rceil d)\\
		=& g(\zeta_1,x_1,\min\{\lceil\frac{A_{th}}{A}\rceil A, N, K\},A).
	\end{aligned}
\end{equation}

Hence, the payoff of setting transaction fee as $x_1$ satisfies
\begin{equation}
	\begin{aligned}
		&u_k(f_k^{\rm buy}) \\
		= &\frac{\sum_{n = 1}^{\min\{\lceil\frac{A_{th}}{A}\rceil A, N\}}\mathbbm{1}(R_k\geq C_n)\min\{b_k,q_n\}(R_k - C_n)}{2\sum_{n = 1}^{\min\{\lceil\frac{A_{th}}{A}\rceil A, N\}}\mathbbm{1}(R_k\geq C_n)} - g(\zeta_1,x_1,\min\{\lceil\frac{A_{th}}{A}\rceil A, N, K\},A).\\
	\end{aligned}
\end{equation}

Define 
\begin{equation}
	\begin{aligned}
		\phi = &\sum_{n=0}^{\min\{\lceil\frac{A_{th}}{A}\rceil A,K, N\}-1}\binom{\min\{\lceil\frac{A_{th}}{A}\rceil A,K, N\}-1}{n}(\zeta_1+\zeta_2)^n(1-\zeta_1-\zeta_2)^{\min\{\lceil\frac{A_{th}}{A}\rceil A, K, N\}-1-n}\\
		&\sum_{j=0}^{n}\binom{n}{j}(\frac{\zeta_2}{\zeta_1+\zeta_2})^j(\frac{\zeta_1}{\zeta_1+\zeta_2})^{n-j}\sum_{l=1}^{j+1}\lceil\frac{n-j+l}{A}\rceil\frac{d}{j+1}\\
		&-\sum_{n=0}^{\min\{\lceil\frac{A_{th}}{A}\rceil A,K, N\}-1}\binom{\min\{\lceil\frac{A_{th}}{A}\rceil A,K, N\}-1}{n}\zeta_1^n(1-\zeta_1)^{\min\{\lceil\frac{A_{th}}{A}\rceil A,K, N\}-1-n}\lceil\frac{n+1}{A}\rceil d.\\
	\end{aligned}
\end{equation}

We set a sufficiently small $\delta_2\in(0,\frac{\phi}{2})$, such that there is no atom between $(x_1, x_1+\delta_2]$ and satisfies following condition:  
$$\begin{aligned}
	&\sum_{n=0}^{\min\{\lceil\frac{A_{th}}{A}\rceil A,K, N\}-1}\binom{\min\{\lceil\frac{A_{th}}{A}\rceil A,K, N\}-1}{n}\zeta_1^n(1-\zeta_1)^{\min\{\lceil\frac{A_{th}}{A}\rceil A,K, N\}-1-n}\lceil\frac{n+1}{A}\rceil d\\
	&- \sum_{n=0}^{\min\{\lceil\frac{A_{th}}{A}\rceil A,K, N\}-1}\binom{\min\{\lceil\frac{A_{th}}{A}\rceil A,K, N\}-1}{n}[1-G_k^{\rm NE}(x_1+\delta_2)]^n[G_k^{\rm NE}(x_1+\delta_2)]^{\min\{\lceil\frac{A_{th}}{A}\rceil A,K, N\}-1-n}\lceil\frac{n+1}{A}\rceil d\geq 0.
\end{aligned}$$ 
(there must exist $\delta_2$ satisfy the above condition, as $x_1+\delta_2$ can be arbitrarily close to $x_1$ and atom cannot continuously exists). We consider buyer $k$ sets the trasnaction fee as $x_1+\delta_2$, the sum of buyer $k$' fee payment and delay cost is
\begin{equation}\label{Atom_deviation2}
	\begin{aligned}
		& \sum_{n=0}^{\min\{\lceil\frac{A_{th}}{A}\rceil A,K, N\}-1}\binom{\min\{\lceil\frac{A_{th}}{A}\rceil A,K, N\}-1}{n}[1-G_k^{\rm NE}(x_1+\delta_2)]^n[G_k^{\rm NE}(x_1+\delta_2)]^{\min\{\lceil\frac{A_{th}}{A}\rceil A,K, N\}-1-n}\\
		&[(x_1+\delta_2)+\lceil\frac{n+1}{A}\rceil d]\\
		\leq & \sum_{n=0}^{\min\{\lceil\frac{A_{th}}{A}\rceil A,K, N\}-1}\binom{\min\{\lceil\frac{A_{th}}{A}\rceil A,K, N\}-1}{n}\zeta_1^n(1-\zeta_1)^{\min\{\lceil\frac{A_{th}}{A}\rceil A,K, N\}-1-n}(x_1+\delta_2+\lceil\frac{n+1}{A}\rceil d)\\
		= &  \sum_{n=0}^{\min\{\lceil\frac{A_{th}}{A}\rceil A,K, N\}-1}\binom{\min\{\lceil\frac{A_{th}}{A}\rceil A,K, N\}-1}{n}\zeta_1^n(1-\zeta_1)^{\min\{\lceil\frac{A_{th}}{A}\rceil A,K, N\}-1-n}(x_1+\lceil\frac{n+1}{A}\rceil d) + \delta_2\\
		<& \sum_{n=0}^{\min\{\lceil\frac{A_{th}}{A}\rceil A,K, N\}-1}\binom{\min\{\lceil\frac{A_{th}}{A}\rceil A,K, N\}-1}{n}\zeta_1^n(1-\zeta_1)^{\min\{\lceil\frac{A_{th}}{A}\rceil A,K, N\}-1-n}(x_1+\lceil\frac{n+1}{A}\rceil d) +\frac{\phi}{2}\\
		<& \sum_{n=0}^{\min\{\lceil\frac{A_{th}}{A}\rceil A,K, N\}-1}\binom{\min\{\lceil\frac{A_{th}}{A}\rceil A,K, N\}-1}{n}(\zeta_1+\zeta_2)^n(1-\zeta_1-\zeta_2)^{\min\{\lceil\frac{A_{th}}{A}\rceil A, K, N\}-1-n}\\
		&\sum_{j=0}^{n}\binom{n}{j}(\frac{\zeta_2}{\zeta_1+\zeta_2})^j(\frac{\zeta_1}{\zeta_1+\zeta_2})^{n-j}(x_1+\sum_{l=1}^{j+1}\lceil\frac{n-j+l}{A}\rceil\frac{d}{j+1})\\
	\end{aligned}
\end{equation}
Note that the matching utilities are the same when buyer $k$ set the transaction fee as $x_1+\delta_2$ and $x_1$, which is in equation (\ref{matching_utility}).
\textbf{Thus buyer $k$ can increase his payoff by setting the transaction fee as $x_1+\delta_2$, which is a contradiction to Nash equilibrium. Thus, we have proved $\mu_k^{\rm NE}$ doesn't have any atom. This completes the proof of Lemma 2.2.}

\vspace{20mm}

\textbf{Lemma 2.3.     }\emph{We use $\mu_k^{\rm NE}$ to represent buyer $k$'s strategy at mixed strategy Nash equilibrium, $L$ and $U$ to represent the strategy of transaction fee-per-byte's lower and upper support, and $G_k^{\rm NE}(f_k^{\rm buy})$ to represent the CDF of the strategy of transaction fee-per-byte. Then $G_k^{\rm NE}(f_k^{\rm buy})$ is strictly increasing over $[L,U]$.}

\vspace{5mm}
\emph{Proof of Lemma 2.3.}

\textbf{We will prove that the CDF $G_k^{\rm NE}(f_k^{\rm buy})$ is strictly increasing over $[L,U]$ by contradiction. Assuming the CDF is not strictly increasing, we will construct a strategy that achieves higher payoff than equilibrium strategy. Notice that as we will focus on a particular buyer $k$'s payoff, we will denote his payoff function as $u_k(f_k^{\rm buy})$ to simplify the notation.} 

\textbf{Assume $\mu_k^{\rm NE}$'s CDF $G_k^{\rm NE}(f_k^{\rm buy})$ is constant over $[x_1,x_1^{'}]$ (i.e, $\mu_k^{\rm NE}(\{x_1\leq f_k^{\rm buy}\leq x_1^{'}\}) = 0$), $x_1<x_1^{'}$, and $x_1,x_1^{'} \in [L,U]$, such that $G_k^{\rm NE}(f_k^{\rm buy})$ is not strictly increasing over $[x_1,x_1^{'}]$.} Then, there is a positive probability of buyer $k$'s transaction fee-per-byte is between $[x_1^{'}, U]$, i.e., $\mu_k^{\rm NE}(\{f_k^{\rm buy}\geq x_1^{'}\})>0$. (Otherwise the upper support should be $x_1$, which is defined as strictly smaller than $U$.) Then there are two possible cases:
\begin{enumerate}
	\item Strategy support includes point $x_1^{'}$. Notice that there is no atom at $x_1^{'}$ based on Lemma 2.2. At mixed strategy Nash equilibrium, any strategy support yields same expected payoff. Thus, buyer $k$'s expected payoff can be calculated at fee-per-byte equaling $x_1^{\prime}$. The matching utility is
	\begin{equation}\label{matching_utility2}
		\frac{\sum_{n = 1}^{\min\{\lceil\frac{A_{th}}{A}\rceil A, N\}}\mathbbm{1}(R_k\geq C_n)\min\{b_k,q_n\}(R_k - C_n)}{2\sum_{n = 1}^{\min\{\lceil\frac{A_{th}}{A}\rceil A, N\}}\mathbbm{1}(R_k\geq C_n)}.
	\end{equation}
	The sum of fee and delay cost is
	\begin{equation}
		\begin{aligned}
			\sum_{n=0}^{\min\{\lceil\frac{A_{th}}{A}\rceil A,K, N\}-1}\binom{\min\{\lceil\frac{A_{th}}{A}\rceil A,K, N\}-1}{n}[1-G_k^{\rm NE}(x_1^{'})]^n[G_k^{\rm NE}(x_1^{'})]^{\min\{\lceil\frac{A_{th}}{A}\rceil A,K, N\}-1-n}(x_1^{'}+\lceil\frac{n+1}{A}\rceil d).\\
		\end{aligned}
	\end{equation}
	\textbf{We consider buyer $k$ sets the transaction fee as $\frac{x_1^{'}+x_1}{2}$}. His matching utility is the same as in (\ref{matching_utility2}). However, the sum of fee and delay cost is
	\begin{equation}
		\begin{aligned}
			& \sum_{n=0}^{\min\{\lceil\frac{A_{th}}{A}\rceil A,K, N\}-1}\binom{\min\{\lceil\frac{A_{th}}{A}\rceil A,K, N\}-1}{n}[1-G_k^{\rm NE}(\frac{x_1^{'}+x_1}{2})]^n [G_k^{\rm NE}(\frac{x_1^{'}+x_1}{2})]^{\min\{\lceil\frac{A_{th}}{A}\rceil A,K, N\}-1-n}\\
			&(\frac{x_1^{'}+x_1}{2} +\lceil\frac{n+1}{A}\rceil d)\\
			=& \sum_{n=0}^{\min\{\lceil\frac{A_{th}}{A}\rceil A,K, N\}-1}\binom{\min\{\lceil\frac{A_{th}}{A}\rceil A,K, N\}-1}{n}[1-G_k^{\rm NE}(x_1^{'})]^n[G_k^{\rm NE}(x_1^{'})]^{\min\{\lceil\frac{A_{th}}{A}\rceil A,K, N\}-1-n}\\
			&(x_1^{'}+\lceil\frac{n+1}{A}\rceil d) - \frac{x_1^{'}-x_1}{2}.
		\end{aligned}
	\end{equation}
	We compare the payoff of setting the transaction fee as $\frac{x_1^{'}+x_1}{2}$ and the mixed strategy at equilibrium, we have
	$$\begin{aligned}
		u_k(\frac{x_1^{'}+x_1}{2})-u_k(x_1^{'}) = \frac{x_1^{'}-x_1}{2}>0.
	\end{aligned}$$
Note that the matching utilities are the same when buyer $k$ set the transaction fee as $x_1+\delta_2$ and $x_1$, which is in equation (\ref{matching_utility2}).
	\textbf{Thus buyer $k$ can increase his payoff by setting the transaction fee as $\frac{x_1^{'}+x_1}{2}$, which is a contradiction to Nash equilibrium.}
	\item Users' strategy support does not include point $x_1^{'}$. There must exist a sufficiently small $\delta_3>0$, such that $x_1^{'}+\delta_3$ belongs to the support of users' strategy and satisfies following condition:  $$\begin{aligned}
		&\sum_{n=0}^{\min\{\lceil\frac{A_{th}}{A}\rceil A,K, N\}-1}\binom{\min\{\lceil\frac{A_{th}}{A}\rceil A,K, N\}-1}{n}\Bigg\{[1-G_k^{\rm NE}(x_1^{'})]^n[G_k^{\rm NE}(x_1^{'})]^{\min\{\lceil\frac{A_{th}}{A}\rceil A,K, N\}-1-n}\\
		&-[1-G_k^{\rm NE}(x_1^{'}+\delta_3)]^n[G_k^{\rm NE}(x_1^{'}+\delta_3)]^{\min\{\lceil\frac{A_{th}}{A}\rceil A,K, N\}-1-n}\Bigg\}\lceil\frac{n+1}{A}\rceil d<\frac{x_1^{'}-x_1}{2}.
	\end{aligned}$$ 
	(there must exist $\delta_3$ satisfy the above condition, as $x_1^{'}+\delta_3$ can be arbitrarily close to $x_1^{'}$, otherwise we can always consider a larger $x_1^{'}$ where the CDF is constant over $[x_1,x_1^{'}]$) 
	At mixed strategy Nash equilibrium, any strategy support yields same expected payoff. Thus, buyer $k$'s expected payoff can be calculated at fee-per-byte equaling $x_1^{\prime}+\delta_3$. The matching utility is
	\begin{equation}\label{matching_utility3}
		\frac{\sum_{n = 1}^{\min\{\lceil\frac{A_{th}}{A}\rceil A, N\}}\mathbbm{1}(R_k\geq C_n)\min\{b_k,q_n\}(R_k - C_n)}{2\sum_{n = 1}^{\min\{\lceil\frac{A_{th}}{A}\rceil A, N\}}\mathbbm{1}(R_k\geq C_n)}.
	\end{equation}
	The sum of fee and delay cost is
	$$\begin{aligned}
		& \sum_{n=0}^{\min\{\lceil\frac{A_{th}}{A}\rceil A,K, N\}-1}\binom{\min\{\lceil\frac{A_{th}}{A}\rceil A,K, N\}-1}{n}[1-G_k^{\rm NE}(x_1^{'}+\delta_3)]^n[G_k^{\rm NE}(x_1^{'}+\delta_3)]^{\min\{\lceil\frac{A_{th}}{A}\rceil A,K, N\}-1-n}\\
		&[(x_1^{'}+\delta_3)+\lceil\frac{n+1}{A}\rceil d].\\
	\end{aligned}$$
	\textbf{We consider buyer $k$ sets the transaction fee as $\frac{x_1^{'}+x_1}{2}$.} The matching surplus is the same as in (\ref{matching_utility3}). However, the sum of fee and delay cost is
	$$\begin{aligned}
		& \sum_{n=0}^{\min\{\lceil\frac{A_{th}}{A}\rceil A,K, N\}-1}\binom{\min\{\lceil\frac{A_{th}}{A}\rceil A,K, N\}-1}{n}[1-G_k^{\rm NE}(\frac{x_1^{'}+x_1}{2})]^n[G_k^{\rm NE}(\frac{x_1^{'}+x_1}{2})]^{\min\{\lceil\frac{A_{th}}{A}\rceil A,K, N\}-1-n}\\
		&(\frac{x_1^{'}+x_1}{2}+\lceil\frac{n+1}{A}\rceil d)\\
		=& \sum_{n=0}^{\min\{\lceil\frac{A_{th}}{A}\rceil A,K, N\}-1}\binom{\min\{\lceil\frac{A_{th}}{A}\rceil A,K, N\}-1}{n}[1-G_k^{\rm NE}(x_1^{'})]^n[G_k^{\rm NE}(x_1^{'})]^{\min\{\lceil\frac{A_{th}}{A}\rceil A,K, N\}-1-n}\\
		&(\frac{x_1^{'}+x_1}{2}+\lceil\frac{n+1}{A}\rceil d).
	\end{aligned}$$
	We compare the payoff of setting the transaction fee as $\frac{x_1^{'}+x_1}{2}$ and the mixed strategy $ \mu_k^{\rm NE}$, we have
	$$\begin{aligned}
		&u_k(\frac{x_1^{'}+x_1}{2})-u_k(x_1^{'}+\delta_3)\\
	  = &(\frac{x_1^{'}-x_1}{2}+\delta_3)-\sum_{n=0}^{\min\{\lceil\frac{A_{th}}{A}\rceil A,K, N\}-1}\binom{\min\{\lceil\frac{A_{th}}{A}\rceil A,K, N\}-1}{n}\Bigg\{[1-G_k^{\rm NE}(x_1^{'})]^n[G_k^{\rm NE}(x_1^{'})]^{\min\{\lceil\frac{A_{th}}{A}\rceil A,K, N\}-1-n}\\
		&-[1-G_k^{\rm NE}(x_1^{'}+\delta_3)]^n[G_k^{\rm NE}(x_1^{'}+\delta_3)]^{\min\{\lceil\frac{A_{th}}{A}\rceil A,K, N\}-1-n}\Bigg\}\lceil\frac{n+1}{A}\rceil d>0.
	\end{aligned}$$
Note that the matching utilities are the same when buyer $k$ set the transaction fee as $x_1+\delta_2$ and $x_1$, which is in equation (\ref{matching_utility3}).
	\textbf{Thus buyer $k$ can increase his payoff by setting the transaction fee as $\frac{x_1^{'}+x_1}{2}$, which is a contradiction to Nash equilibrium.}
\end{enumerate}
\textbf{Thus, we have proved that the CDF $G_k^{\rm NE}$ is strictly increasing over $[L,U]$. This completes the proof of Lemma 2.3.}

\vspace{50mm}

\textbf{Lemma 2.4.     }\emph{We use $ \mu_k^{\rm NE}$ to represent buyer $k$'s strategy at mixed strategy Nash equilibrium, $L$ and $U$ to represent the strategy of transaction fee's lower and upper support, and $G_k^{\rm NE}(f_k^{rm buy})$ to represent the CDF of the strategy of transaction fee-per-byte. Then $L = \sigma_{th}^{\rm buy}(A_{th},A)+\epsilon$ and $U = \sigma_{th}^{\rm buy}(A_{th},A)+\epsilon+\lceil\frac{\min\{\lceil\frac{A_{th}}{A}\rceil A,K, N\}}{A}\rceil d$.}

\vspace{5mm}
\textbf{\emph{Proof of Lemma 2.4.}}

\textbf{We prove by contradiction. Note that Lemma 2.1 has proved that $L \geq \sigma_{th}^{\rm buy}(A_{th},A)+\epsilon$. Hence, we assume $L > \sigma_{th}^{\rm buy}(A_{th},A)+\epsilon$, we will construct a strategy that achieves higher payoff than equilibrium strategy. Notice that as we will focus on a particular buyer $k$'s payoff, we will denote his payoff function as $u_k(f_k^{\rm buy})$ to simplify the notation.} 

When $L > \sigma_{th}^{\rm buy}(A_{th},A)+\epsilon$. There are two cases:
\begin{itemize}
	\item User $i$'s transaction fee strategy support at mixed strategy Nash equilibrium includes point $L$. At mixed strategy Nash equilibrium, any strategy support yields same expected payoff. Thus, buyer $k$'s expected payoff can be calculated at fee-per-byte equaling $L$. The matching utility is
	\begin{equation}\label{matching_utility4}
		\frac{\sum_{n = 1}^{\min\{\lceil\frac{A_{th}}{A}\rceil A, N\}}\mathbbm{1}(R_k\geq C_n)\min\{b_k,q_n\}(R_k - C_n)}{2\sum_{n = 1}^{\min\{\lceil\frac{A_{th}}{A}\rceil A, N\}}\mathbbm{1}(R_k\geq C_n)}.
	\end{equation}
	The sum of fee and delay cost is
	$$\begin{aligned}
		& \sum_{n=0}^{\min\{\lceil\frac{A_{th}}{A}\rceil A,K, N\}-1}\binom{\min\{\lceil\frac{A_{th}}{A}\rceil A,K, N\}-1}{n}[1-G_k^{\rm NE}(L)]^n[G_k^{\rm NE}(L)]^{\min\{\lceil\frac{A_{th}}{A}\rceil A,K, N\}-1-n}(L+\lceil\frac{n+1}{A}\rceil d)\\
		=& L+\lceil\frac{\min\{\lceil\frac{A_{th}}{A}\rceil A,K, N\}}{A}\rceil d.
	\end{aligned}$$
	\textbf{We consider buyer $k$ sets the transaction fee as $\sigma_{th}^{\rm buy}(A_{th},A)+\epsilon$}. The matching surplus is the same as in (\ref{matching_utility4}). However, the sum of fee and delay cost is
	$$\begin{aligned}
		& \sum_{n=0}^{\min\{\lceil\frac{A_{th}}{A}\rceil A,K, N\}-1}\binom{\min\{\lceil\frac{A_{th}}{A}\rceil A,K, N\}-1}{n}[1-G_k^{\rm NE}(L)]^n[G_k^{\rm NE}(L)]^{\min\{\lceil\frac{A_{th}}{A}\rceil A,K, N\}-1-n}\\
		&(\sigma_{th}^{\rm buy}(A_{th},A)+\epsilon+\lceil\frac{n+1}{A}\rceil d)\\
		=& (\sigma_{th}^{\rm buy}(A_{th},A)+\epsilon+\lceil\frac{n+1}{A}\rceil d)\\
		<&L+\lceil\frac{\min\{\lceil\frac{A_{th}}{A}\rceil A,K, N\}}{A}\rceil d.
	\end{aligned}$$
Note that the matching utilities are the same when buyer $k$ set the transaction fee as $x_1+\delta_2$ and $x_1$, which is in equation (\ref{matching_utility4}).
	\textbf{Thus buyer $k$ can increase his payoff by setting the transaction fee as $\sigma_{th}^{\rm buy}(A_{th},A)+\epsilon$, which is a contradiction to Nash equilibrium.}
	
	\item Users' strategy support does not include point $L$. Note that there exist a sufficiently small $\delta_4>0$, such that $L+\delta_4$ belongs to the support of users' strategy at mixed strategy Nash equilibrium and satisfies following condition:  
	$$
	\begin{aligned}
		&\lceil\frac{\min\{\lceil\frac{A_{th}}{A}\rceil A,K, N\}}{A}\rceil d -\sum_{n=0}^{\min\{\lceil\frac{A_{th}}{A}\rceil A,K, N\}-1}\binom{\min\{\lceil\frac{A_{th}}{A}\rceil A,K, N\}-1}{n}\\
		&[1-G_k^{\rm NE}(L+\delta_4)]^n[G_k^{\rm NE}(L+\delta_4)]^{\min\{\lceil\frac{A_{th}}{A}\rceil A,K, N\}-1-n}\lceil\frac{n+1}{A}\rceil d\\
		&<(L-\sigma_{th}^{\rm buy}(A_{th},A)-\epsilon).
	\end{aligned}
	$$ 
	(there must exist $\delta_4$ satisfy the above condition, as $L+\delta_4$ can be arbitrarily close to $L$, otherwise $L$ is not the lower bound.) 
	At mixed strategy Nash equilibrium, any strategy support yields same expected payoff. Thus, buyer $k$'s expected payoff can be calculated at fee-per-byte equaling $L+\delta_4$.  The matching utility is
	\begin{equation}\label{matching_utility5}
		\frac{\sum_{n = 1}^{\min\{\lceil\frac{A_{th}}{A}\rceil A, N\}}\mathbbm{1}(R_k\geq C_n)\min\{b_k,q_n\}(R_k - C_n)}{2\sum_{n = 1}^{\min\{\lceil\frac{A_{th}}{A}\rceil A, N\}}\mathbbm{1}(R_k\geq C_n)}.
	\end{equation}
	The sum of fee and delay cost is
	$$\begin{aligned}
		& \sum_{n=0}^{\min\{\lceil\frac{A_{th}}{A}\rceil A,K, N\}-1}\binom{\min\{\lceil\frac{A_{th}}{A}\rceil A,K, N\}-1}{n}[1-G_k^{\rm NE}(L+\delta_4)]^n[G_k^{\rm NE}(L+\delta_4)]^{\min\{\lceil\frac{A_{th}}{A}\rceil A,K, N\}-1-n}\\
		&(L+\delta_4+\lceil\frac{n+1}{A}\rceil d).\\
	\end{aligned}$$
	\textbf{We consider buyer $k$ sets the transaction fee as $\sigma_{th}^{\rm buy}(A_{th},A)+\epsilon$.} The matching surplus is the same as in (\ref{matching_utility5}). However, the sum of fee and delay cost is
	$$\begin{aligned}
		& \sum_{n=0}^{\min\{\lceil\frac{A_{th}}{A}\rceil A,K, N\}-1}\binom{\min\{\lceil\frac{A_{th}}{A}\rceil A,K, N\}-1}{n}[1-G_k^{\rm NE}(\sigma_{th}^{\rm buy}(A_{th},A)+\epsilon)]^n\\
		&[G_k^{\rm NE}(\sigma_{th}^{\rm buy}(A_{th},A)+\epsilon)]^{\min\{\lceil\frac{A_{th}}{A}\rceil A,K, N\}-1-n}(\sigma_{th}^{\rm buy}(A_{th},A)+\epsilon+\lceil\frac{n+1}{A}\rceil d)\\
		=& \sigma_{th}^{\rm buy}(A_{th},A)+\epsilon+\lceil\frac{\min\{\lceil\frac{A_{th}}{A}\rceil A,K, N\}}{A}\rceil d.
	\end{aligned}$$
	We compare the payoff of adopting pure strategy $\sigma_{th}^{\rm buy}(A_{th},A)+\epsilon$ and the mixed strategy $ \mu_k^{\rm NE}$, we have
	$$\begin{aligned}
		&u_k(\sigma_{th}^{\rm buy}(A_{th},A)+\epsilon)-u_k(L+\delta_4) = -\sigma_{th}^{\rm buy}(A_{th},A)-\epsilon - \lceil\frac{\min\{\lceil\frac{A_{th}}{A}\rceil A,K, N\}}{A}\rceil d + L+\delta_4\\
		& +\sum_{n=0}^{\min\{\lceil\frac{A_{th}}{A}\rceil A,K, N\}-1}\binom{\min\{\lceil\frac{A_{th}}{A}\rceil A,K, N\}-1}{n}\{[1-G_k^{\rm NE}(L+\delta_4)]\}^n[G_k^{\rm NE}(L+\delta_4)]^{\min\{\lceil\frac{A_{th}}{A}\rceil A,K, N\}-1-n}\lceil\frac{n+1}{A}\rceil d\\
		>&\delta_4\\
		>&0.
	\end{aligned}$$
	Thus buyer $k$ can increase his payoff by setting the transaction fee as $\sigma_{th}^{\rm buy}(A_{th},A)+\epsilon$, which is a contradiction to Nash equilibrium.
\end{itemize}

\textbf{This completes the proof of $L = \sigma_{th}^{\rm buy}(A_{th},A)+\epsilon$.}

\textbf{Next we prove $U =\sigma_{th}^{\rm buy}(A_{th},A)+\epsilon+\lceil\frac{\min\{\lceil\frac{A_{th}}{A}\rceil A,K, N\}}{A}\rceil d$ by contradiction.}

Note that Lemma 2.1 has proved that $U \leq \sigma_{th}^{\rm buy}(A_{th},A)+\epsilon+\lceil\frac{\min\{\lceil\frac{A_{th}}{A}\rceil A,K, N\}}{A}\rceil d$. Then we assume $U < \sigma_{th}^{\rm buy}(A_{th},A)+\epsilon+\lceil\frac{\min\{\lceil\frac{A_{th}}{A}\rceil A,K, N\}}{A}\rceil d$. Notice that any strategy within the strategy support yields same expected payoff and strategy support of fee-per-byte must include $L = \sigma_{th}^{\rm buy}(A_{th},A)+\epsilon$. Then we consider buyer $k$ sets transaction fee as $f_k^{\rm buy} = U$. The matching utility is
\begin{equation}\label{matching_utility6}
	\frac{\sum_{n = 1}^{\min\{\lceil\frac{A_{th}}{A}\rceil A, N\}}\mathbbm{1}(R_k\geq C_n)\min\{b_k,q_n\}(R_k - C_n)}{2\sum_{n = 1}^{\min\{\lceil\frac{A_{th}}{A}\rceil A, N\}}\mathbbm{1}(R_k\geq C_n)}.
\end{equation}
The sum of fee and delay cost is $U$.
However, for buyer $k$ sets the transaction fee as $L$, the sum of fee and delay cost is
\begin{equation}
	 \sigma_{th}^{\rm buy}(A_{th},A)+\epsilon+\lceil\frac{\min\{\lceil\frac{A_{th}}{A}\rceil A,K, N\}}{A}\rceil d > U.
\end{equation}
Thus, different strategies within the strategy support yield different expected payoffs, which is a contradiction to mixed strategy Nash equilibrium. \textbf{Thus, we have proved that $U =\sigma_{th}^{\rm buy}(A_{th},A)+\epsilon+\lceil\frac{\min\{\lceil\frac{A_{th}}{A}\rceil A,K, N\}}{A}\rceil d$.}

\textbf{This completes the proof of Lemma 2.4.}

\vspace{50mm}

\textbf{Finally we prove the CDF of transaction fee $G_k^{\rm NE}(f_k^{\rm buy})$ satisfies Equation (\themyeqcounter).} Notice that any strategy within the strategy support yields same expected payoff and strategy support of fee must include $\sigma_{th}^{\rm buy}(A_{th},A)+\epsilon$, thus for any strategy $f_k^{\rm buy} \in [\sigma_{th}^{\rm buy}(A_{th},A)+\epsilon,\sigma_{th}^{\rm buy}(A_{th},A)+\epsilon+\lceil\frac{\min\{\lceil\frac{A_{th}}{A}\rceil A,K, N\}}{A}\rceil d]$, the expected payoff of strategy $f_k^{\rm buy}$ and strategy $\sigma_{th}^{\rm buy}(A_{th},A)+\epsilon$ should be the same, which is:
\begin{equation}\label{CDF11}
	\begin{aligned}
		\sum\limits_{n=0}^{\min\{\lceil\frac{A_{th}}{A}\rceil A,K, N\}-1}\binom{\min\{\lceil\frac{A_{th}}{A}\rceil A,K, N\}-1}{n}[1-&G_k^{\rm NE}(f_k^{\rm buy})]^n[G_k^{\rm NE}(f_k^{\rm buy})]^{\min\{\lceil\frac{A_{th}}{A}\rceil A,K, N\}-1-n}(f_k^{\rm buy}+\lceil\frac{n+1}{A}\rceil d)  \\
		&= \sigma_{th}^{\rm buy}(A_{th},A)+\epsilon+\lceil\frac{\min\{\lceil\frac{A_{th}}{A}\rceil A,K, N\}}{A}\rceil d\\
		&\big\Downarrow\\
		g(1-G_k^{\rm NE}(f_k^{\rm buy}),f_k^{\rm buy},\min\{\lceil\frac{A_{th}}{A}\rceil A, K,N\},A)&= g(1,\sigma_{th}^{\rm buy}(A_{th},A)+\epsilon,\min\{\lceil\frac{A_{th}}{A}\rceil A, K,N\},A).
	\end{aligned}
\end{equation}
When $f_k^{\rm buy} \not\in [\sigma_{th}^{\rm buy}(A_{th},A)+\epsilon,\sigma_{th}^{\rm buy}(A_{th},A)+\epsilon+\lceil\frac{\min\{\lceil\frac{A_{th}}{A}\rceil A,K, N\}}{A}\rceil d]$, then $f_k^{\rm buy}$ no longer belongs to the strategy support and we have 
\begin{equation}\label{CDF12}
	G_k^{\rm NE}(f_k^{\rm buy}) = \begin{cases}
		0, &\text{if $f_k^{\rm buy} <\sigma_{th}^{\rm buy}(A_{th},A)+\epsilon$,}\\
		1, &\text{if $f_k^{\rm buy} >\sigma_{th}^{\rm buy}(A_{th},A)+\epsilon+\lceil\frac{\min\{\lceil\frac{A_{th}}{A}\rceil A,K, N\}}{A}\rceil d$.}
	\end{cases}
\end{equation}

Based on Equations (\ref{CDF11}) and (\ref{CDF12}), we complete the proof of Proposition 2.

\newpage
\subsection{Existing BBOB System PoA Analysis}

\textbf{Proof of Theorem \ref{thm:PoA1}.} 

\textbf{We prove this result by constructing two cases where price of anarchy (PoA) is infinity.}

\textbf{We first construct an example where setting the block size too high leads to infinity PoA. } We consider a 2-buyer-2-seller example illustrated in Fig. \ref{Fig:utility1}, which satisfies $C_1 + 2\epsilon = R_2 <C_2= R_1 - 2\epsilon$. Hence, it means $R_1\geq C_1$ and $R_2 < C_2$. We consider the buying quantities and selling quantities are all 1.  Under such setting, we consider the case $A \geq 2$.

\begin{figure}[h]
	\centering
	{\includegraphics[width=8cm]{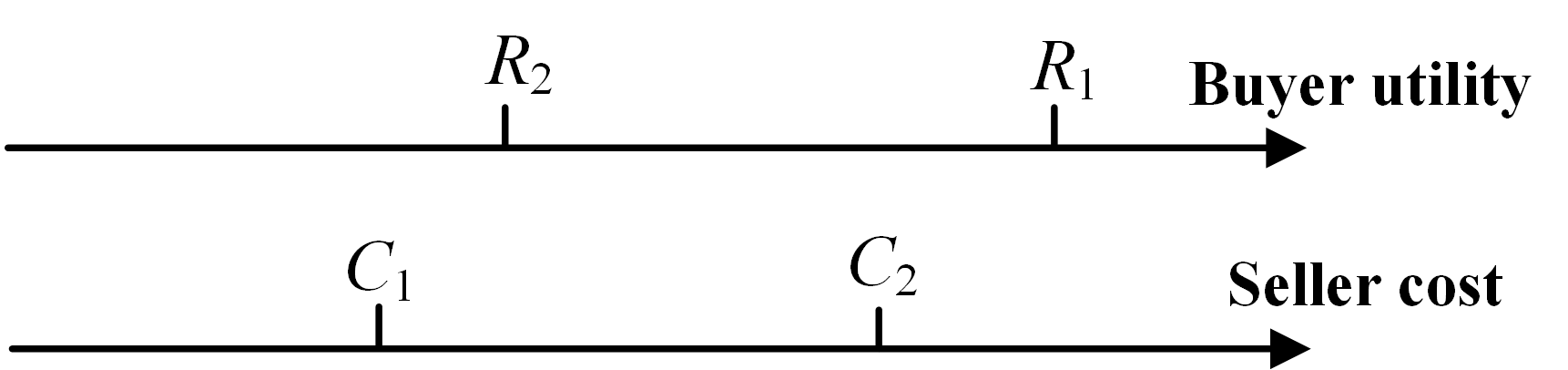}}
	\caption{Buyer's utility and seller's cost}\label{Fig:utility1}
\end{figure}
At Nash equilibrium of Stage I, users will set the transaction fee as in Table \ref{Tab:NE2}.
\begin{table}[h]
	\centering
	\caption{Stage I Equilibrium under $C_1 + 2\epsilon = R_2 <C_2= R_1 - 2\epsilon$ and $A > 1$}\label{Tab:NE2}
	\begin{tabular}{|c|c|c|}
		\hline
		& 1 & 2   \\
		\hline
		Buyer	& $\epsilon$ & $\epsilon$ \\
		\hline
		Seller & $\epsilon$ & $\epsilon$  \\
		\hline
	\end{tabular}
\end{table}

For this example, the social welfare is 
\begin{equation}
	sw = R_1 +R_2- C_1-C_2 = 4\epsilon.
\end{equation}
The social optimum is to match $R_1$ with $C_2$, which is
\begin{equation}
	sw^{\rm opt} = R_1 - C_2.
\end{equation}
The ratio between social optimum and social welfare is
\begin{equation}
	\frac{sw^{\rm opt}}{sw} = \frac{R_1 - C_2}{4\epsilon}.
\end{equation}
As long as we choose $R_1-C_2$ to be sufficiently large, PoA, which is largest possible ratio is unbounded. 

This completes the construction of the example where setting the block size too high leads to infinity PoA.

\vspace{5mm}

\textbf{We then construct an example where setting the block size too low leads to infinity PoA.}

\begin{figure}[h]
	\centering
	{\includegraphics[width=8cm]{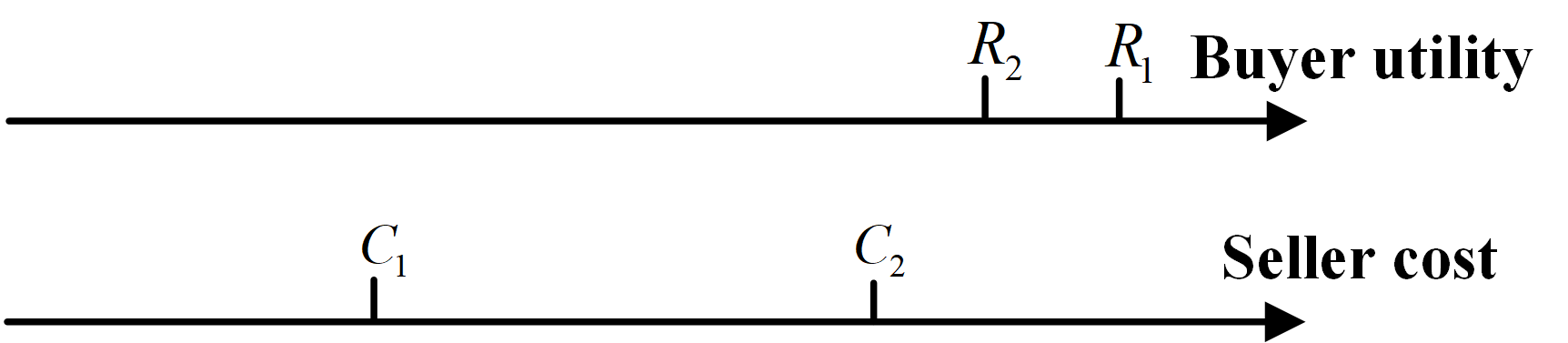}}
	\caption{Buyer's utility and seller's cost}\label{Fig:utility15}
\end{figure}

We consider a 2-buyer-2-seller example illustrated in Fig. \ref{Fig:utility15}, which satisfies $C_1 <C_2+2\epsilon< R_2 < R_1 $. We consider the buying quantities and selling quantities are all 1.  Hence, under such setting, we consider the case $A <  2$.

Based on users' fee equilibrium characterized by Propositions 1 and 2, miners must record all transactions in blockchain: 2 transactions in the first block and 2 transactions in the second block. Hence, the social welfare is 
\begin{equation}
	sw = R_1 +R_2- C_1-C_2 -2d.
\end{equation}
The social optimum is to record all transactions in the first block, which is
\begin{equation}
	sw^{\rm opt} = R_1 +R_2- C_1-C_2 >0.
\end{equation}
The ratio between social optimum and social welfare is
\begin{equation}
	\frac{sw^{\rm opt}}{sw} = \frac{ R_1 +R_2- C_1-C_2}{ R_1 +R_2- C_1-C_2 -2d}.
\end{equation}
Then we have 
\begin{equation}
	\lim\limits_{d\rightarrow(\frac{R_1+R_2-C_1-C_2}{2})^+}\frac{sw^{\rm opt}}{sw} = \lim\limits_{d\rightarrow(\frac{R_1+R_2-C_1-C_2}{2})^+}\frac{ R_1 +R_2- C_1-C_2}{ R_1 +R_2- C_1-C_2 -2d} = \infty.
\end{equation}
As long as we choose delay cost $d$ to be sufficiently close to $ \frac{R_1+R_2-C_1-C_2}{2}$, PoA, which is largest possible ratio is unbounded. 

This completes the construction of the example where setting the block size too low leads to infinity PoA.

This completes the proof of Theorem 2.

\newpage
\subsection{Stage 0 Analysis}

\subsubsection{Adjustable Block Size Mechanism for Complete Information}

\textbf{Proof of Theorem \ref{thm:stage_I_result}.}

\textbf{We first prove the result for homogeneous trading quantity $\underline{b} = \overline{b}$, by showing $A_{th}$ can achieve the social optimum under the three cases.}	

To simplify the notation, we consider $\underline{b} = \overline{b} =1$. For $\underline{b} = \overline{b} \not=1$, we can just multiply the social welfare and social optimum by a linear factor of $\underline{b}$.
\begin{enumerate}
	\item Case 1: As illustrated in Fig. \ref{Fig:case12}, buying utility and selling cost satisfy $R_1 < C_1$.
	\begin{figure}[h]
		\centering
		{\includegraphics[width=8cm]{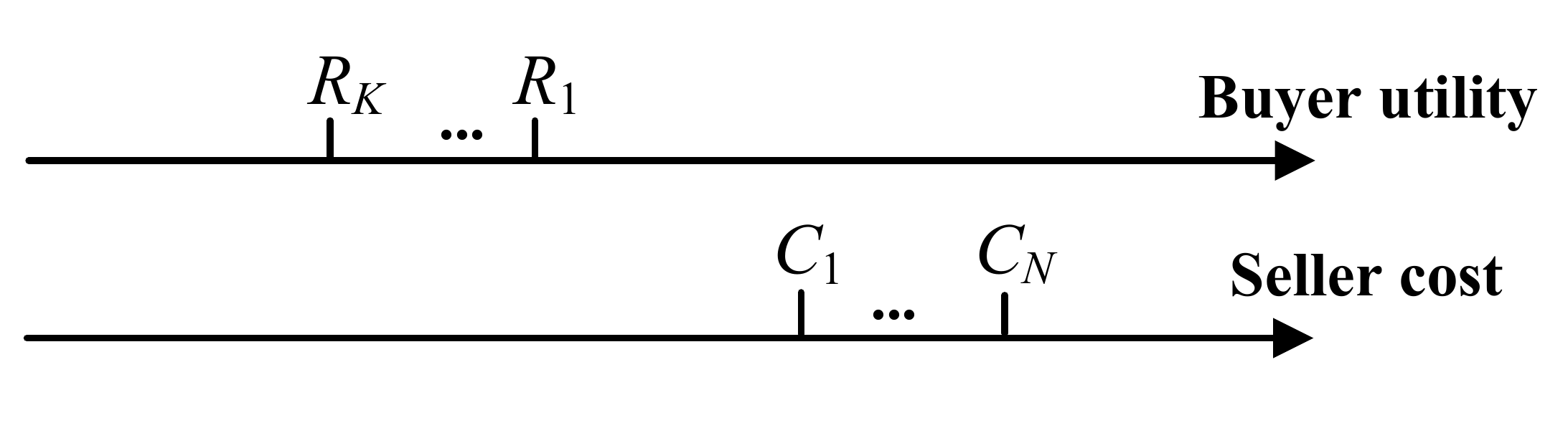}}
		\caption{Buyer's utility and seller's cost when $R_1<C_1$}\label{Fig:case12}
	\end{figure}
	 Under this case, there is no possible match. Hence, the social optimum is zero. Any block size $A$ also yields no match and corresponding social welfare is zero. Hence, any $A_{th}$ achieves social optimum.

	\item Case 2: As illustrated in Fig. \ref{Fig:case22}, buying utility and selling cost satisfy $R_K \geq  C_N$. 
	\begin{figure}[h]
		\centering
		{\includegraphics[width=8cm]{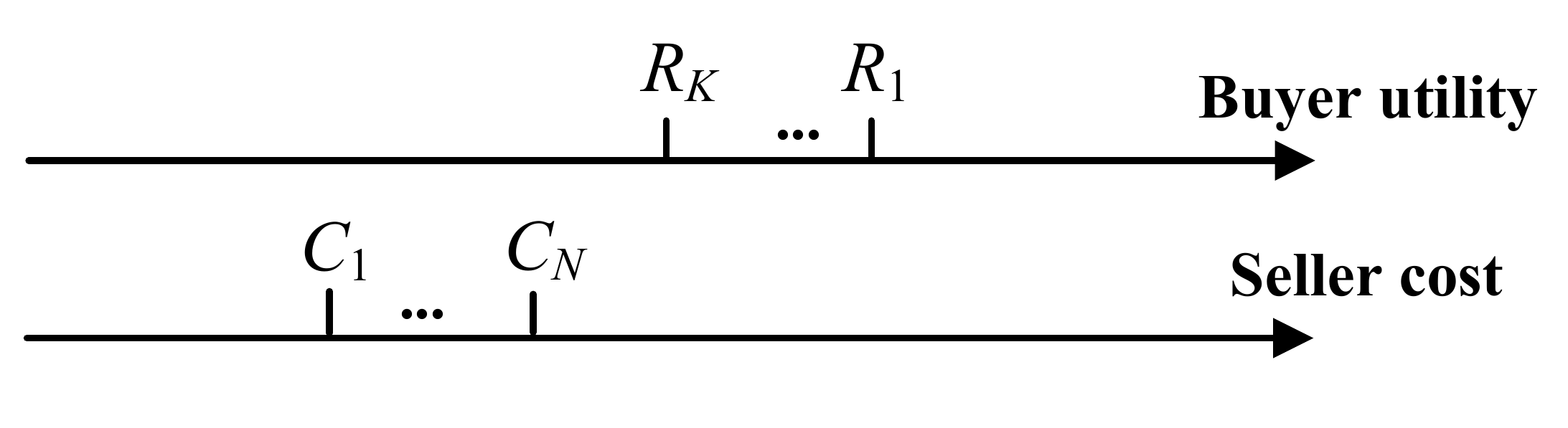}}
		\caption{Buyer's utility and seller's cost when $R_K \geq C_N$}\label{Fig:case22}
	\end{figure}

    Hence, the social optimum is as follows:
    \begin{equation}
    	sw^{\rm opt} = \sum_{n = 1}^{\min\{K,N\}} (R_n - C_n).
    \end{equation}
     
     We consider three subcases to show that $A_{th} = \min\{K,N\}$ achieves social optimum. 
	\begin{enumerate}
		\item If $K = N$, then under $A_{th}= \min\{K,N\}$, the following strategies constitute a Nash equilibrium:
		\begin{align}
			f_n^{\rm sell,NE} = f_k^{\rm buy,NE} = \epsilon, \quad\forall n\in\mathcal{N},\forall k\in\mathcal{K}.
		\end{align}
		Under such an equilibrium, miners randomly match buying and selling transactions. Even though a buyer or a seller deviates to higher fee, he is still randomly matched.
		For this subcase, the social welfare is:
		\begin{equation}
			sw  = \sum_{n\in\mathcal{N}} (R_n - C_n) = sw^{\rm opt}.
		\end{equation}
		Hence,  $A_{th}= \min\{K,N\}$ achieves social optimum.
		\item If $K > N$, then under $A_{th} = \min\{K,N\}$, the following strategies constitute a Nash equilibrium:
		\begin{align}
			&f_k^{\rm buy,NE} = \begin{cases}
				\dfrac{\sum_{n = 1}^{N}(R_{N+1} - C_n)}{2N}+\epsilon,&\text{if $k\leq N$.}\\
				\dfrac{\sum_{n = 1}^{N}(R_{N+1} - C_n)}{2N},&\text{if $k>N$.}
			\end{cases}\\
			&f_n^{\rm sell,NE} = 0, \quad\forall n\in\mathcal{N}.
		\end{align}
		Under such an equilibrium, for buyer with index $k>N$, they pay the maximal fee they can afford; while for buyer with index $k\leq N$, they pay the maximal fee $\epsilon$ higher than the buyer with index $k>N$.
		For this subcase, the social welfare is:
		\begin{equation}
			sw  = \sum_{n\in\mathcal{N}}(R_n - C_n) = sw^{\rm opt}.
		\end{equation}
		Hence,  $A_{th} = \min\{K,N\}$ achieves social optimum.
		\item If $K < N$, then under $A_{th} = \min\{K,N\}$, then the following strategies constitute a Nash equilibrium:
		\begin{align}
			& f_k^{\rm buy,NE} = 0, \quad\forall k\in\mathcal{K},\\
			&f_n^{\rm sell,NE} = \begin{cases}
				\dfrac{\sum_{k = 1}^{K}(R_k - C_{K+1})}{2K}+\epsilon,&\text{if $n\leq K$,}\\
				\dfrac{\sum_{k = 1}^{K}(R_k - C_{K+1})}{2K},&\text{if $n>K$.}
			\end{cases}
		\end{align}
		Under such an equilibrium, for seller with index $n>K$, they pay the maximal fee they can afford; while for seller with index $n\leq K$, they pay the maximal fee $\epsilon$ higher than the buyer with index $n>K$.
		For this subcase, the social welfare and social optimum are the same:
		\begin{equation}
			sw = \sum_{k\in\mathcal{K}}(R_k - C_k) = sw^{\rm opt} .
		\end{equation}
		Hence,  $A_{th} = \min\{K,N\}$ achieves social optimum.
	\end{enumerate}
	
	\item Case 3: As illustrated in Fig. \ref{Fig:case32}, there exists $j$ such that $R_j \geq C_j$ and $R_{j+1}<C_{j+1}$. 
		\begin{figure}[h]
		\centering
		{\includegraphics[width=8cm]{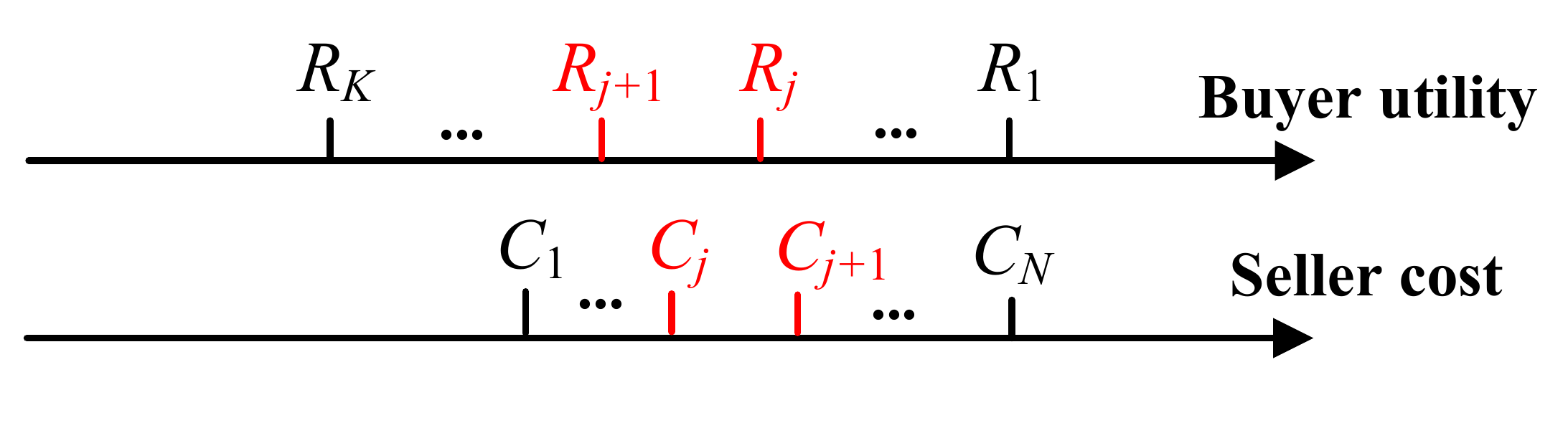}}
		\caption{Buyer's utility and seller's cost when $R_j \geq C_j$ and $R_{j+1}<C_{j+1}$}\label{Fig:case32}
	\end{figure}

    Hence, the social optimum is as follows:
    \begin{equation}
    	sw^{\rm opt} = \sum_{i = 1}^{j} (R_i - C_i).
    \end{equation}
    Under $A_{th} = j$, the following strategy constitutes a NE:
    \begin{align}
    	f_k^{\rm buy,NE} = \begin{cases}
    		0,&\text{if $1\leq k\leq j$ and $R_{j+1}<C_1$,}\\
    		\dfrac{\sum_{i = 1}^{j}\mathbbm{1}(R_{j+1}\geq C_i)[R_{j+1} - C_i]}{2\sum_{i = 1}^{j}\mathbbm{1}(R_{j+1}\geq C_i)}+\epsilon,&\text{if $1\leq k\leq j$ and $R_{j+1}\geq C_1$,}\\
    		0,&\text{if $k \geq j+1$ and $R_{j+1}<C_1$,}\\
    		\dfrac{\sum_{i = 1}^{j}\mathbbm{1}(R_{j+1}\geq C_i)[R_{j+1} - C_i]}{2\sum_{i = 1}^{j}\mathbbm{1}(R_{j+1}\geq C_i)},&\text{if $k \geq j+1$ and $R_{j+1}\geq C_1$,}
    	\end{cases}\\
    	f_n^{\rm sell,NE} = \begin{cases}
    		0,&\text{if $1\leq n\leq j$ and $C_{j+1}>R_1$,}\\
    		\dfrac{\sum_{i = 1}^{j}\mathbbm{1}(R_i\geq C_{j+1})[R_i - C_{j+1}]}{2\sum_{i = 1}^{j}\mathbbm{1}(R_i\geq C_{j+1})}+\epsilon,&\text{if $1\leq n\leq j$ and $C_{j+1}\leq R_1$.}\\
    		0,&\text{if $n \geq j+1$ and $C_{j+1}> R_1$,}\\
    		\dfrac{\sum_{i = 1}^{j}\mathbbm{1}(R_i\geq C_{j+1})[R_i - C_{j+1}]}{2\sum_{i = 1}^{j}\mathbbm{1}(R_i\geq C_{j+1})},&\text{if $n \geq j+1$ and $C_{j+1}\leq R_1$.}
    	\end{cases}
    \end{align}
	For this case, the social welfare is:
	\begin{equation}
		sw  = \sum_{i = 1}^j(R_i - C_i) = sw^{\rm opt}.
	\end{equation}
	Hence,  $A_{th} = j$ achieves social optimum.

\end{enumerate}

To sum up, for homogeneous trading quantity $\underline{b} = \overline{b}$ and setting the block size as $A = A_{th}$, the social welfare always equals social optimum. Hence, the PoA of homogeneous trading quantity $\underline{b} = \overline{b}$ is PoA = 1.

This completes the proof of for homogeneous trading quantity $\underline{b} = \overline{b}$.

%

\vspace{5mm}
\textbf{Next, we prove the case of heterogeneous trading quantities by analyzing the PoA under the three cases.}

\begin{enumerate}
	\item Case 1: As illustrated in Fig. \ref{Fig:case15}, buying utility and selling cost satisfy $R_1<C_1$. Under this case, there is no possible match. Hence, the social optimum is zero. Any block size $A$ also yields no match and corresponding social welfare is zero. Hence, any block size achieves social optimum. We have PoA $\leq \frac{\overline{b}}{\underline{b}}$.
	\begin{figure}[h]
		\centering
		{\includegraphics[width=8cm]{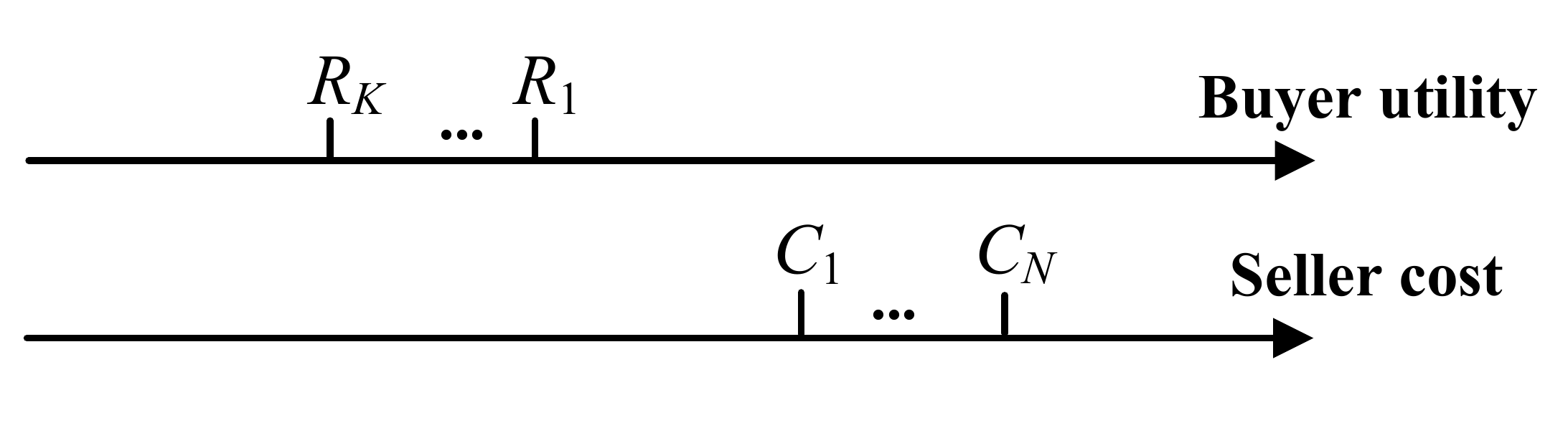}}
		\caption{Buyer's utility and seller's cost when $R_1<C_1$}\label{Fig:case15}
	\end{figure}	
	\item Case 2: As illustrated in Fig. \ref{Fig:case25}, buying utility and selling cost satisfy $R_K \geq  C_N$. 
	\begin{figure}[h]
		\centering
		{\includegraphics[width=8cm]{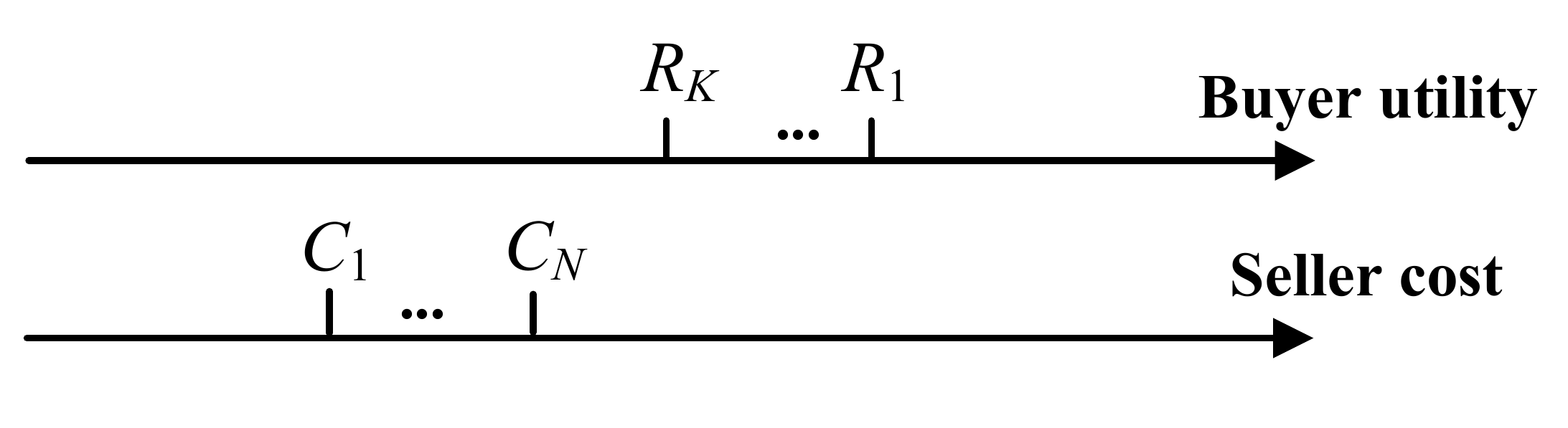}}
		\caption{Buyer's utility and seller's cost when $R_K \geq C_N$}\label{Fig:case25}
	\end{figure}
	
	Hence, the social optimum is as follows:
	\begin{equation}
		sw^{\rm opt} = \sum_{n \in\mathcal{N}} \sum_{k\in\mathcal{K}}x_{kn}\min\{b_k,q_n\}(R_k - C_n) \leq \overline{b}\sum_{n = 1}^{\min\{K,N\}} (R_n - C_n).
	\end{equation}
	
	We consider three subcases to show the social welfare of setting block size as $A_{th}= \min\{K,N\}$. 
	\begin{enumerate}
		\item If $K = N$, then under $A_{th}= \min\{K,N\}$, the following strategies constitute a Nash equilibrium:
		\begin{align}
			f_n^{\rm sell} = f_k^{\rm buy} = \epsilon, \quad\forall n\in\mathcal{N},\forall k\in\mathcal{K}.
		\end{align}
		Under such an equilibrium, miners randomly match buying and selling transactions. Even though a buyer or a seller deviates to higher fee, he is still randomly matched.
		For this subcase, the social welfare satisfies:
		\begin{equation}
			sw  \geq \underline{b}\sum_{n\in\mathcal{N}} (R_n - C_n).
		\end{equation}
		Moreover, the social optimum satisfies:
		\begin{equation}
			sw^{\rm opt} \leq \overline{b}\sum_{n\in\mathcal{N}} (R_n - C_n).
		\end{equation}
		Hence, the ratio between social optimum and social welfare satisfies
		\begin{equation}
			\frac{sw^{\rm opt}}{sw}  \leq \frac{\overline{b}}{\underline{b}}.
		\end{equation}
		\item If $K > N$, then under $A_{th}= \min\{K,N\}$, the following strategies constitute a Nash equilibrium:
		\begin{align}
			&f_k^{\rm buy} = \begin{cases}
				\dfrac{\sum_{n = 1}^{N}\min\{b_{N+1},q_n\}(R_{N+1} - C_n)}{2N}+\epsilon,&\text{if $k\leq N$.}\\
				\dfrac{\sum_{n = 1}^{N}\min\{b_{N+1},q_n\}(R_{N+1} - C_n)}{2N},&\text{if $k>N$.}
			\end{cases}\\
			&f_n^{\rm sell} = 0, \quad\forall n\in\mathcal{N}.
		\end{align}
		Under such an equilibrium, for buyer with index $k>N$, they pay the maximal fee they can afford; while for buyer with index $k\leq N$, they pay the maximal fee $\epsilon$ higher than the buyer with index $k>N$.
		For this subcase, the social welfare satisfies:
		\begin{equation}
			sw  \geq \underline{b}\sum_{n\in\mathcal{N}} (R_n - C_n).
		\end{equation}
		Moreover, the social optimum satisfies:
		\begin{equation}
			sw^{\rm opt} \leq \overline{b}\sum_{n\in\mathcal{N}} (R_n - C_n).
		\end{equation}
		Hence, the ratio between social optimum and social welfare satisfies
		\begin{equation}
			\frac{sw^{\rm opt}}{sw}  \leq \frac{\overline{b}}{\underline{b}}.
		\end{equation}
		
		\item If $k < N$, then under $A_{th}= \min\{K,N\}$, then the following strategies constitute a Nash equilibrium:
		\begin{align}
			& f_k^{\rm buy} = 0, \quad\forall k\in\mathcal{K},\\
			&f_n^{\rm sell} = \begin{cases}
				\dfrac{\sum_{k = 1}^{K}\min\{b_k,q_{K+1}\}(R_k - C_{K+1})}{2K}+\epsilon,&\text{if $n\leq K$,}\\
				\dfrac{\sum_{k = 1}^{K}\min\{b_k,q_{K+1}\}(R_k - C_{K+1})}{2K},&\text{if $n>K$.}
			\end{cases}
		\end{align}
		Under such an equilibrium, for seller with index $n>K$, they pay the maximal fee they can afford; while for seller with index $n\leq K$, they pay the maximal fee $\epsilon$ higher than the buyer with index $n>K$.
		For this subcase, the social welfare satisfies:
		\begin{equation}
			sw  \geq \underline{b}\sum_{n\in\mathcal{N}} (R_n - C_n).
		\end{equation}
		Moreover, the social optimum satisfies:
		\begin{equation}
			sw^{\rm opt} \leq \overline{b}\sum_{n\in\mathcal{N}} (R_n - C_n).
		\end{equation}
		Hence, the ratio between social optimum and social welfare satisfies
		\begin{equation}
			\frac{sw^{\rm opt}}{sw}  \leq \frac{\overline{b}}{\underline{b}}.
		\end{equation}
	\end{enumerate}

	\item Case 3: As illustrated in Fig. \ref{Fig:case35}, there exists $j$ such that $R_j \geq C_j$ and $R_{j+1}<C_{j+1}$. 
	\begin{figure}[h]
		\centering
		{\includegraphics[width=8cm]{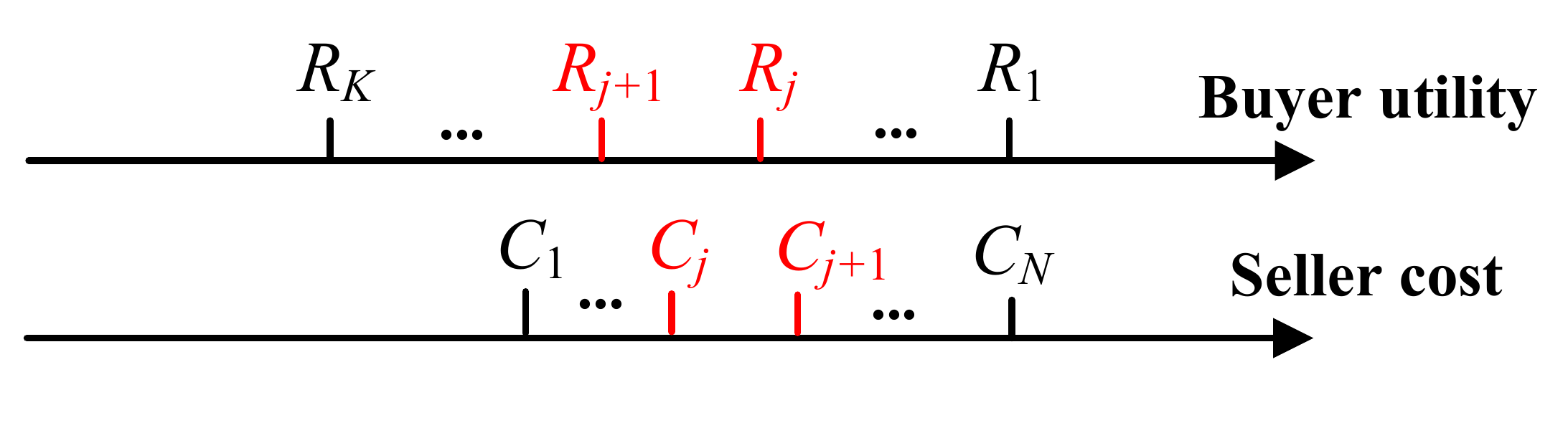}}
		\caption{Buyer's utility and seller's cost when $R_j \geq C_j$ and $R_{j+1}<C_{j+1}$}\label{Fig:case35}
	\end{figure}
	
	Hence, the social optimum is as follows:
	\begin{equation}
		sw^{\rm opt}  = \sum_{n \in\mathcal{N}} \sum_{k\in\mathcal{K}}x_{kn}\min\{b_k,q_n\}(R_k - C_n) \leq \overline{b} \sum_{i = 1}^{j} (R_i - C_i).
	\end{equation}
	Under $A_{th} = k$, the following strategy constitutes a NE:
	\begin{align}
		f_k^{\rm buy} = \begin{cases}
			0,&\text{if $1\leq k\leq j$ and $R_{j+1}<C_1$,}\\
			\dfrac{\sum_{i = 1}^{j}\mathbbm{1}(R_{j+1}\geq C_i)\min\{b_{j+1},q_i\}[R_{j+1} - C_i]}{2\sum_{i = 1}^{j}\mathbbm{1}(R_{j+1}\geq C_i)}+\epsilon,&\text{if $1\leq k\leq j$ and $R_{j+1}\geq C_1$,}\\
			0,&\text{if $k \geq j+1$ and $R_{j+1}<C_1$,}\\
			\dfrac{\sum_{i = 1}^{j}\mathbbm{1}(R_{j+1}\geq C_i)\min\{b_{j+1},q_i\}[R_{j+1} - C_i]}{2\sum_{i = 1}^{j}\mathbbm{1}(R_{j+1}\geq C_i)},&\text{if $k \geq j+1$ and $R_{j+1}\geq C_1$,}
		\end{cases}\\
		f_n^{\rm sell} = \begin{cases}
			0,&\text{if $1\leq n\leq j$ and $C_{j+1}>R_1$,}\\
			\dfrac{\sum_{i = 1}^{j}\mathbbm{1}(R_i\geq C_{j+1})\min\{b_i,q_{j+1}\}[R_i - C_{j+1}]}{2\sum_{i = 1}^{j}\mathbbm{1}(R_i\geq C_{j+1})}+\epsilon,&\text{if $1\leq n\leq j$ and $C_{j+1}\leq R_1$.}\\
			0,&\text{if $n \geq j+1$ and $C_{j+1}> R_1$,}\\
			\dfrac{\sum_{i = 1}^{j}\mathbbm{1}(R_i\geq C_{j+1})\min\{b_i,q_{j+1}\}[R_i - C_{j+1}]}{2\sum_{i = 1}^{j}\mathbbm{1}(R_i\geq C_{j+1})},&\text{if $n \geq j+1$ and $C_{j+1}\leq R_1$.}
		\end{cases}
	\end{align}
	For this case, the social welfare satisfies:
	\begin{equation}
		sw  \geq \underline{b}\sum_{n\in\mathcal{N}} (R_n - C_n).
	\end{equation}
	Moreover, the social optimum satisfies:
	\begin{equation}
		sw^{\rm opt} \leq \overline{b}\sum_{n\in\mathcal{N}} (R_n - C_n).
	\end{equation}
	Hence, the ratio between social optimum and social welfare satisfies
	\begin{equation}
		\frac{sw^{\rm opt}}{sw}  \leq \frac{\overline{b}}{\underline{b}}.
	\end{equation}

\end{enumerate}
To sum up, for heterogeneous trading quantity $\underline{b} \not= \overline{b}$ and setting the block size as $A = A_{th}$, the ratio between social optimum and social welfare always satisfies
\begin{equation}
	\frac{sw^{\rm opt}}{sw}  \leq \frac{\overline{b}}{\underline{b}}.
\end{equation}
Hence, the PoA of heterogeneous trading quantity $\underline{b} \not= \overline{b}$ satisfies
\begin{equation}
	\text{PoA}  \leq \frac{\overline{b}}{\underline{b}}.
\end{equation}

This completes the proof of for heterogeneous trading quantity $\underline{b} \not= \overline{b}$.

This completes the proof of Theorem 3.

\newpage

\newpage

\subsubsection{Adjustable Block Size Mechanism for Incomplete Information}

\textbf{Proof of Theorem \ref{thm:dist}.} 

To facilitate the proof, we define 
\begin{equation}
	\rho = \frac{K}{N},
\end{equation}
where $\rho >0$ characterizes the ratio between number of buyers and sellers. Moreover, \textbf{we first set the general block size as $A = \lfloor NC(\eta)+\delta N \rfloor$ with $0<\delta<\min\{1,\rho\} - C(\eta)$, and then we evaluate a special case where $\delta = N^{-\psi}$. Note that when $\delta = N^{-\psi}$, the block size $A$ is exactly the block size set in Theorem 4.}
\begin{figure}[h]
	\centering
	\includegraphics[width=10cm]{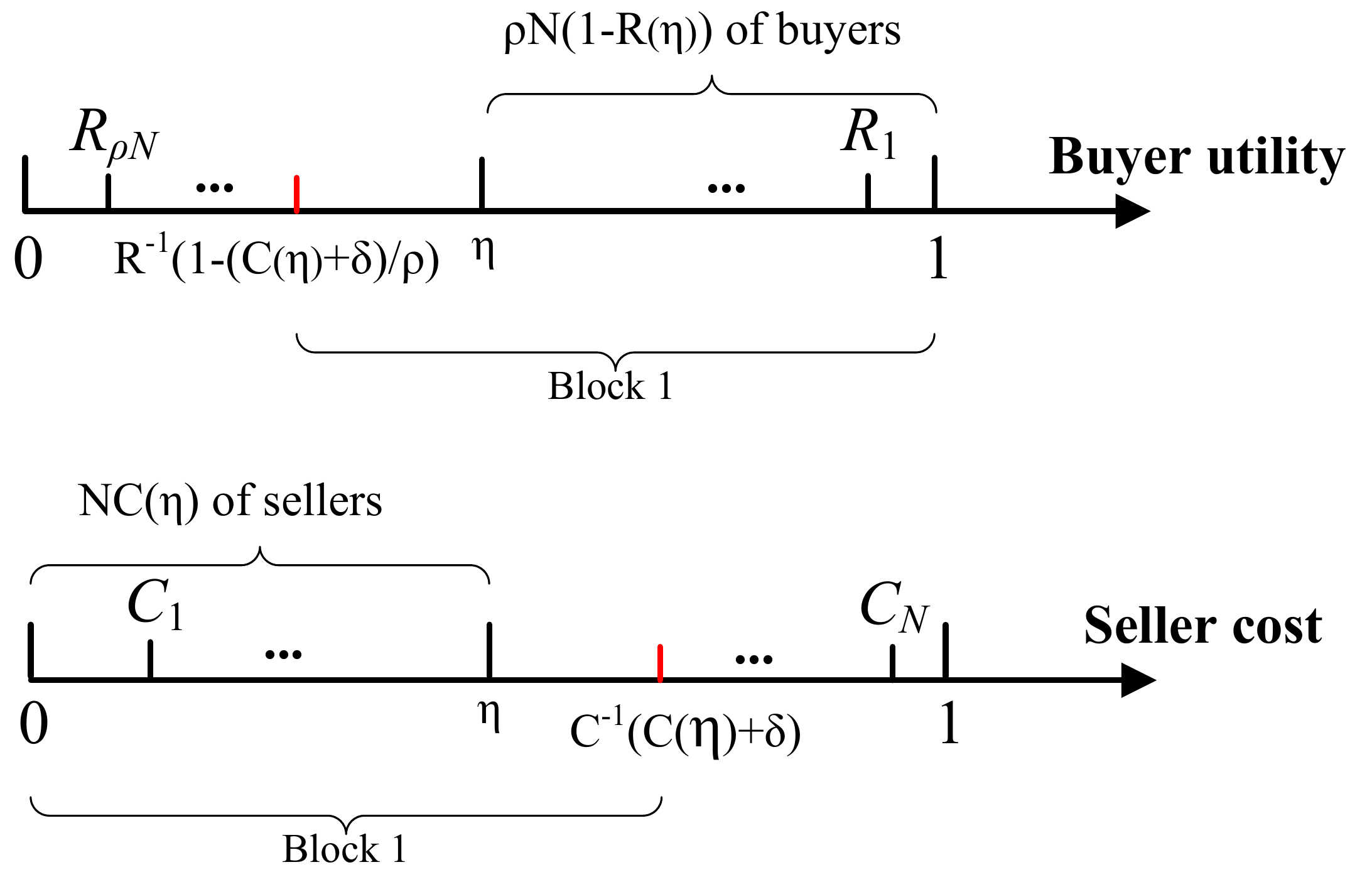}
	\caption{$A = \lfloor NC(\eta)+\delta N \rfloor$ illustration.}
	\label{Fig:large_N}
\end{figure}

We illustrate how we set block size in Fig. \ref{Fig:large_N}. We explain why we set block size as $A = \lfloor NC(\eta)+\delta N \rfloor$: We set block size as $A = \lfloor NC(\eta)+\delta N \rfloor$. So on average, block 1 includes buying transactions with utility higher than $R^{-1}(1-\frac{C(\eta)+\delta}{\rho})$ and selling transactions with cost lower than $C^{-1}(C(\eta)+\delta)$.  For $\delta = N^{-\psi}>0$, we have $C^{-1}(C(\eta)+\delta) > R^{-1}(1-\frac{C(\eta)+\delta}{\rho})$. So for the unrecorded transactions, they are unlikely to be matched (probability they can be matched exponentially decreases with $N$). Hence, for large $N$, we do not need block 2 to record the rest transactions. This avoids the delay cost $d$, achieving a constant approximation ratio.

 \vspace{10mm}
\textbf{We will first prove the homogeneous trading quantity of $\underline{b} = \overline{b}$. In the proof, we will first estimates the social optimum. Next we estimate the social welfare loss when setting $A = \lfloor NC(\eta)+\delta N \rfloor$.}

To simplify the notation, we consider $\underline{b} = \overline{b} =1$ for homogeneous trading quantity. For $\underline{b} = \overline{b} \not=1$, we can just multiply the social welfare and social optimum by a linear factor of $\underline{b}$.

\textbf{Social optimum estimation}: We denote the buying utilities in a decreasing order as $R_1 \geq R_2\geq \cdots \geq R_{\rho N}$ and denote selling costs in an increasing order as $C_1 \leq C_2 \leq \cdots \leq C_N$. Then when $R_i \geq C_i$, the social optimum case should include it in blockchain.
Hence, the social optimum can be derived as follows:
\begin{equation}\label{Eq:swopt56}
	sw^{\rm opt} = \sum\limits_{i = 1}^{\min\{\rho N,N\}} \mathbb{P}(R_i \geq C_i) \mathbb{E}[R_i-C_i|R_i\geq C_i].
\end{equation}
Note that $R_i$ is the $\rho N+1-i$-th order statistics of $\rho N$ iid random variables (i.e., buying utilities). Moreover, $C_i$ is $i$-th order statistics of $N$ iid random variables (i.e., selling costs).

\begin{figure}[h]
	\centering
	\includegraphics[width=9cm]{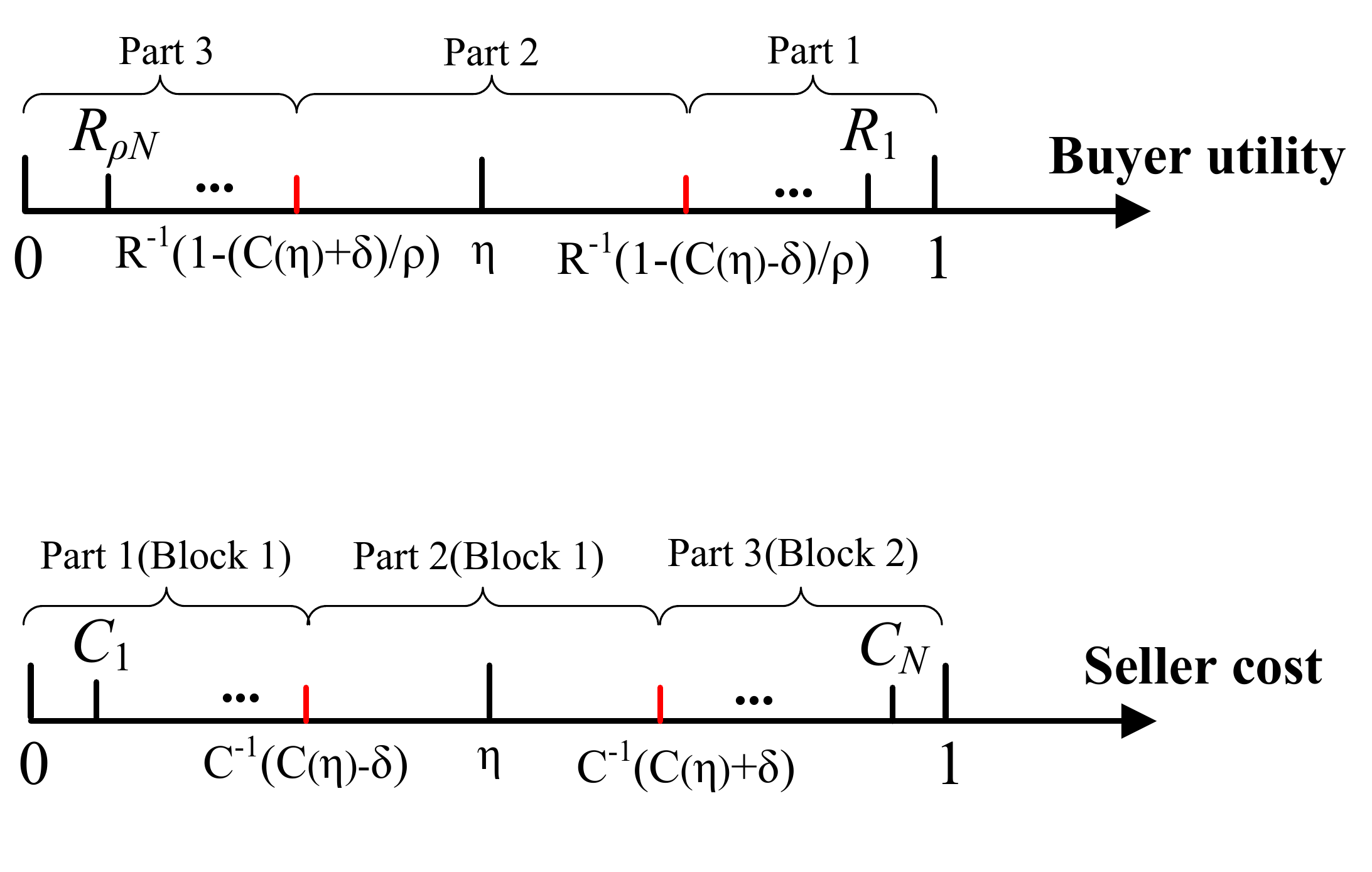}
	\caption{Three parts to estimate $sw^{\rm opt}$.}
	\label{Fig:large_N2}
\end{figure}

As illustrated in Fig. \ref{Fig:large_N2}, we estimate $sw^{\rm opt}$ in equation (\ref{Eq:swopt56}) by three parts. Our estimation is to prove that when $N$ is large, part 1 are matched with probability 1, part 2's match can be approximated, and part 3 are matched with probability of 0.

The detailed estimation is as follows:
\begin{equation}\label{Eq:swopt57}
	\begin{aligned}
		sw^{\rm opt} = &\sum\limits_{i = 1}^{\lfloor NC(\eta)-\delta N \rfloor} \mathbb{P}(R_i \geq C_i) \mathbb{E}[R_i-C_i|R_i\geq C_i] \\
		&+ \sum\limits_{i = \lfloor NC(\eta)-\delta N \rfloor+1}^{\lfloor NC(\eta)+\delta N \rfloor} \mathbb{P}(R_i \geq C_i) \mathbb{E}[R_i-C_i|R_i\geq C_i]\\
		&+ \sum\limits_{i = \lfloor NC(\eta)+\delta N \rfloor +1}^{\min\{\rho N,N\}} \mathbb{P}(R_i \geq C_i) \mathbb{E}[R_i-C_i|R_i\geq C_i].
	\end{aligned}
\end{equation}
Here we estimate three parts of equation (\ref{Eq:swopt57}) as follows:
\begin{enumerate}
	\item Estimation of $\sum\limits_{i = 1}^{\lfloor NC(\eta)-\delta N \rfloor} \mathbb{P}(R_i \geq C_i) \mathbb{E}[R_i-C_i|R_i\geq C_i]$. For $1 \leq i \leq \lfloor NC(\eta)-\delta N \rfloor$, the expectation satisfies:
	\begin{align}
		&\mathbb{E}[R_i] \approx R^{-1}\left(1-\frac{i}{\rho N+1}\right) \geq  R^{-1}\left(1-\frac{\lfloor NC(\eta)-\delta N \rfloor}{\rho N+1}\right) \approx R^{-1}\left(1 - \frac{C(\eta)-\delta}{\rho}\right).\\
		&\mathbb{E}[C_i] \approx C^{-1}\left(\frac{i}{N+1}\right) \leq C^{-1}\left(\frac{\lfloor NC(\eta)-\delta N \rfloor}{N+1}\right) \approx C^{-1}\left(C(\eta)-\delta\right).
	\end{align}
	The variance satisfies:
	\begin{align}
		&\text{Var}(R_i)  \approx \frac{(\rho N + 1 - i) i}{(\rho N)^3} \left[\frac{1}{r\left(R^{-1}(\frac{\rho N + 1 - i}{\rho N+1})\right)}\right]^2.\\
		&\text{Var}(C_i) \approx \frac{i (N +1 - i) }{N^3}\left[\frac{1}{c\left(C^{-1}(\frac{i}{N+1})\right)}\right]^2.
	\end{align}

    Then $R_i$ and $C_i$ can be approximated by normal distribution when $N\rightarrow \infty$.\footnote{Based on asymptotic distribution of a central order statistics theorem. \url{http://www.math.ntu.edu.tw/~hchen/teaching/LargeSample/notes/noteorder.pdf}\url{https://ieeexplore.ieee.org/stamp/stamp.jsp?tp=&arnumber=9936687}} Then to estimate the probability of $R_i \geq C_i$, we can use the standard method to estimate the probability of one normal-distributed random variable larger than another normal-distributed random variable as follows:
    \begin{align}
    	&\mu_i \triangleq \mathbb{E}[R_i] - \mathbb{E}[C_i] =  R^{-1}\left(1-\frac{i}{\rho N+1}\right) - C^{-1}\left(\frac{i}{N+1}\right) .\\
    	&\sigma_i^2 \triangleq\text{Var}(R_i) + \text{Var}(C_i).
    \end{align}
    Given $\delta > 0 $, we have:
    \begin{equation}
    	\begin{aligned}
    		\mu_i = R^{-1}\left(1-\frac{i}{\rho N+1}\right) - C^{-1}\left(\frac{i}{N+1}\right) &\geq R^{-1}\left(1 - \frac{C(\eta)-\delta}{\rho}\right)- C^{-1}\left(C(\eta)-\delta\right)\\
    		& > R^{-1}\left(1 - \frac{C(\eta)}{\rho}\right)- C^{-1}\left(C(\eta)\right)\\
    		& = 0.
    	\end{aligned}
    \end{equation}
    Then for any $1 \leq i \leq \lfloor NC(\eta)-\delta N \rfloor$, we have 
    \begin{equation}\label{Eq:prob21}
    	\mathbb{P}(R_i \geq C_i)   = \mathbb{P}\Big(\frac{R_i - C_i - \mu_i}{\sigma_i} \geq -\frac{ \mu_i}{\sigma_i}\Big) = \mathbb{P}\Big(Z \geq -\frac{\mu_i}{\sigma_i}\Big),
    \end{equation}
     	where $Z$ is a standard normal variable. Moreover, we note that:
     	\begin{equation}
     		\begin{aligned}
     			& \hspace{5mm}\frac{\mu_i}{\sqrt{N}\sigma_i}\\
     			& \geq\frac{\mu_{\lfloor NC(\eta)-\delta N \rfloor}}{\sqrt{N}\sigma_{\lfloor NC(\eta)-\delta N \rfloor}}\\
     			& = \frac{R^{-1}\left(1 - \frac{C(\eta)-\delta}{\rho}\right) -  C^{-1}\left(C(\eta)-\delta\right)}{\sqrt{\frac{[C(\eta)-\delta][\rho-C(\eta)+\delta]}{\rho^3 }\left[\frac{1}{r\left(R^{-1}(1 - \frac{C(\eta)-\delta}{\rho})\right)}\right]^2 + [C(\eta)-\delta][1-C(\eta)+\delta]\left[\frac{1}{c\left(C^{-1}(C(\eta)-\delta)\right)}\right]^2}}.
     		\end{aligned}
     	\end{equation}
     and we define 
     \begin{equation}
     	\Delta \triangleq \frac{R^{-1}\left(1 - \frac{C(\eta)-\delta}{\rho}\right) -  C^{-1}\left(C(\eta)-\delta\right)}{\sqrt{\frac{[C(\eta)-\delta][\rho-C(\eta)+\delta]}{\rho^3 }\left[\frac{1}{r\left(R^{-1}(1 - \frac{C(\eta)-\delta}{\rho})\right)}\right]^2 + [C(\eta)-\delta][1-C(\eta)+\delta]\left[\frac{1}{c\left(C^{-1}(C(\eta)-\delta)\right)}\right]^2}},
     \end{equation}
     	where $\Delta >0$ is a constant given $\delta>0$. We use $\Phi$ to denote the CDF of standard normal distribution. For large $N$ and for any $1 \leq i \leq \lfloor NC(\eta)-\delta N \rfloor$, equation (\ref{Eq:prob21}) becomes:
     \begin{equation}\label{Eq:prob32}
     	\begin{aligned}
     		\mathbb{P}(R_i \geq C_i) &= \Phi\Big(\frac{\mu_i}{\sigma_i}\Big) \\
     		&\geq \Phi\Big(\frac{\mu_{\lfloor NC(\eta)-\delta N \rfloor}}{\sigma_{\lfloor NC(\eta)-\delta N \rfloor}}\Big) \\
     		& = \Phi\Big(\sqrt{N}\Delta\Big)\\
     		&\approx 1 - \frac{1}{\sqrt{2\pi N}\Delta} e^{-\frac{N\Delta^2}{2}}\\
     		&\approx 1.
     	\end{aligned}
     \end{equation}
 Hence, we have:
 \begin{equation}\label{Eq:swopt58}
 	\begin{aligned}
 		&\sum\limits_{i = 1}^{\lfloor NC(\eta)-\delta N \rfloor} \mathbb{P}(R_i \geq C_i) \mathbb{E}[R_i-C_i|R_i\geq C_i] \\
 		\approx &  \sum\limits_{i = 1}^{\lfloor NC(\eta)-\delta N \rfloor} \mathbb{E}[R_i-C_i] \\
 		\approx & \sum\limits_{i = 1}^{\lfloor NC(\eta)-\delta N \rfloor} R^{-1}\left(1-\frac{i}{\rho N+1}\right) - C^{-1}\left(\frac{i}{N+1}\right)\\
 		= & \left[(\rho N +1) \sum\limits_{i = 1}^{\lfloor NC(\eta)-\delta N \rfloor} \frac{R^{-1}\left(1-\frac{i}{\rho N+1}\right)}{\rho N +1}\right] - \left[(N +1) \sum\limits_{i = 1}^{\lfloor NC(\eta)-\delta N \rfloor} \frac{C^{-1}\left(\frac{i}{N+1}\right)}{N +1}\right]\\
 		\approx & \left[(\rho N +1) \int_{1-\frac{C(\eta)-\delta}{\rho}}^{1}R^{-1}(x) dx\right] - \left[(N +1) \int_{0}^{C(\eta)-\delta}C^{-1}(x) dx\right].
 	\end{aligned}
 \end{equation}
  
	\item Estimation of $\sum\limits_{i = \lfloor NC(\eta)-\delta N \rfloor+1}^{\lfloor NC(\eta)+\delta N \rfloor}  \mathbb{P}(R_i \geq C_i) \mathbb{E}[R_i-C_i|R_i\geq C_i]$. For this part, we have $\mathbb{P}(R_i \geq C_i) \leq 1$ and $\mathbb{E}[R_i-C_i|R_i\geq C_i] \leq 1$. Hence, we have
	 \begin{equation}\label{Eq:swopt59}
		\begin{aligned}
			&\sum\limits_{i = \lfloor NC(\eta)-\delta N \rfloor+1}^{\lfloor NC(\eta)+\delta N \rfloor}  \mathbb{P}(R_i \geq C_i) \mathbb{E}[R_i-C_i|R_i\geq C_i] 
			\leq  \sum\limits_{i = \lfloor NC(\eta)-\delta N \rfloor+1}^{\lfloor NC(\eta)+\delta N \rfloor} 1 \leq 2\delta N.
		\end{aligned}
	\end{equation}
	\item Estimation of $\sum\limits_{i = \lfloor NC(\eta)+\delta N \rfloor +1}^{\min\{\rho N,N\}} \mathbb{P}(R_i \geq C_i) \mathbb{E}[R_i-C_i|R_i\geq C_i]$. For $\lfloor NC(\eta)+\delta N \rfloor +1 \leq i \leq \min\{\rho N,N\}$, the expectation satisfies:
	\begin{align}
		&\mathbb{E}[R_i] \approx R^{-1}\left(1-\frac{i}{\rho N+1}\right) \leq  R^{-1}\left(1-\frac{\lfloor NC(\eta)+\delta N \rfloor+1}{\rho N+1}\right) \approx R^{-1}\left(1 - \frac{C(\eta)+\delta}{\rho}\right).\\
		&\mathbb{E}[C_i] \approx C^{-1}\left(\frac{i}{N+1}\right) \geq C^{-1}\left(\frac{\lfloor NC(\eta)+\delta N \rfloor +1}{N+1}\right) \approx C^{-1}\left(C(\eta)+\delta\right).
	\end{align}
	The variance satisfies:
	\begin{align}
		&\text{Var}(R_i)  \approx \frac{(\rho N + 1 - i) i}{(\rho N)^3} \left[\frac{1}{r\left(R^{-1}(\frac{\rho N + 1 - i}{\rho N+1})\right)}\right]^2.\\
		&\text{Var}(C_i) \approx \frac{i (N +1 - i) }{N^3}\left[\frac{1}{c\left(C^{-1}(\frac{i}{N+1})\right)}\right]^2.
	\end{align}
	Then $R_i$ and $C_i$ can be approximated by normal distribution when $N\rightarrow \infty$.\footnote{Based on asymptotic distribution of a central order statistics theorem. \url{http://www.math.ntu.edu.tw/~hchen/teaching/LargeSample/notes/noteorder.pdf}\url{https://ieeexplore.ieee.org/stamp/stamp.jsp?tp=&arnumber=9936687}} Then to estimate the probability of $R_i \geq C_i$, we can use the standard method to estimate the probability of one normal-distributed random variable larger than another normal-distributed random variable as follows:
	\begin{align}
		&\mu_i \triangleq \mathbb{E}[R_i] - \mathbb{E}[C_i] =  R^{-1}\left(1-\frac{i}{\rho N+1}\right) - C^{-1}\left(\frac{i}{N+1}\right) .\\
		&\sigma_i^2 \triangleq\text{Var}(R_i) + \text{Var}(C_i).
	\end{align}
	Given $\delta > 0 $, we have:
	\begin{equation}
		\begin{aligned}
			\mu_i = R^{-1}\left(1-\frac{i}{\rho N+1}\right) - C^{-1}\left(\frac{i}{N+1}\right) &\leq R^{-1}\left(1 - \frac{C(\eta)+\delta}{\rho}\right)- C^{-1}\left(C(\eta)+\delta\right)\\
			& < R^{-1}\left(1 - \frac{C(\eta)}{\rho}\right)- C^{-1}\left(C(\eta)\right)\\
			& = 0.
		\end{aligned}
	\end{equation}
    Then for any $\lfloor NC(\eta)+\delta N \rfloor +1 \leq i \leq \min\{\rho N,N\}$, we have 
    \begin{equation}\label{Eq:prob23}
    	\mathbb{P}(R_i \geq C_i)   = \mathbb{P}\Big(\frac{R_i - C_i - \mu_i}{\sigma_i} \geq -\frac{ \mu_i}{\sigma_i}\Big) = \mathbb{P}\Big(Z \geq -\frac{\mu_i}{\sigma_i}\Big),
    \end{equation}
    where $Z$ is a standard normal variable. Moreover, we note that:
    \begin{equation}
    	\begin{aligned}
    		& \hspace{5mm}-\frac{\mu_i}{\sqrt{N}\sigma_i}\\
    		& \geq-\frac{\mu_{\lfloor NC(\eta)+\delta N \rfloor}}{\sqrt{N}\sigma_{\lfloor NC(\eta)+\delta N \rfloor}}\\
    		& = \frac{-R^{-1}\left(1 - \frac{C(\eta)+\delta}{\rho}\right) +  C^{-1}\left(C(\eta)+\delta\right)}{\sqrt{\frac{[C(\eta)+\delta][\rho-C(\eta)-\delta]}{\rho^3 }\left[\frac{1}{r\left(R^{-1}(1 - \frac{C(\eta)+\delta}{\rho})\right)}\right]^2 + [C(\eta)+\delta][1-C(\eta)-\delta]\left[\frac{1}{c\left(C^{-1}(C(\eta)+\delta)\right)}\right]^2}}.
    	\end{aligned}
    \end{equation}
    and we define 
    \begin{equation}
    	\Delta_2 \triangleq \frac{-R^{-1}\left(1 - \frac{C(\eta)+\delta}{\rho}\right) +  C^{-1}\left(C(\eta)+\delta\right)}{\sqrt{\frac{[C(\eta)+\delta][\rho-C(\eta)-\delta]}{\rho^3 }\left[\frac{1}{r\left(R^{-1}(1 - \frac{C(\eta)+\delta}{\rho})\right)}\right]^2 + [C(\eta)+\delta][1-C(\eta)-\delta]\left[\frac{1}{c\left(C^{-1}(C(\eta)+\delta)\right)}\right]^2}},
    \end{equation}
	where $\Delta_2 >0$ is a constant given $\delta>0$. We use $\Phi$ to denote the CDF of standard normal distribution. For large $N$ and for any $\lfloor NC(\eta)+\delta N \rfloor +1 \leq i \leq \min\{\rho N,N\}$, equation (\ref{Eq:prob23}) becomes:
	\begin{equation}\label{Eq:prob43}
		\begin{aligned}
			\mathbb{P}(R_i \geq C_i) &= \Phi\Big(\frac{\mu_i}{\sigma_i}\Big) \\
			&\leq \Phi\Big(\frac{\mu_{\lfloor NC(\eta)+\delta N \rfloor}}{\sigma_{\lfloor NC(\eta)+\delta N \rfloor}}\Big) \\
			& = \Phi\Big(-\sqrt{N}\Delta_2\Big)\\
			& = 1 - \Phi\Big(\sqrt{N}\Delta_2\Big)\\
			&\approx \frac{1}{\sqrt{2\pi N}\Delta_2} e^{-\frac{N\Delta_2^2}{2}}\\
			&\approx 0.
		\end{aligned}
	\end{equation}
	Hence, we have:
\begin{equation}
	\sum\limits_{i = \lfloor NC(\eta)+\delta N \rfloor +1}^{\min\{\rho N,N\}} \mathbb{P}(R_i \geq C_i) \mathbb{E}[R_i-C_i|R_i\geq C_i] \approx 0.
\end{equation}
\end{enumerate}

To sum up, we estimate $sw^{\rm opt}$ as
\begin{equation}\label{Eq:swopt60}
	\begin{aligned}
		sw^{\rm opt} = &\sum\limits_{i = 1}^{\lfloor NC(\eta)-\delta N \rfloor} \mathbb{P}(R_i \geq C_i) \mathbb{E}[R_i-C_i|R_i\geq C_i] \\
		&+ \sum\limits_{i = \lfloor NC(\eta)-\delta N \rfloor+1}^{\lfloor NC(\eta)+\delta N \rfloor} \mathbb{P}(R_i \geq C_i) \mathbb{E}[R_i-C_i|R_i\geq C_i]\\
		&+ \sum\limits_{i = \lfloor NC(\eta)+\delta N \rfloor +1}^{\min\{\rho N,N\}} \mathbb{P}(R_i \geq C_i) \mathbb{E}[R_i-C_i|R_i\geq C_i]\\
		< & \left[(\rho N +1) \int_{1-\frac{C(\eta)-\delta}{\rho}}^{1}R^{-1}(x) dx\right] - \left[(N +1) \int_{0}^{C(\eta)-\delta}C^{-1}(x) dx\right] + 2\delta N.
	\end{aligned}
\end{equation}

\textbf{Next we estimate the social welfare when setting $A = \lfloor NC(\eta)+\delta N \rfloor$.}
The social welfare when setting $A = \lfloor NC(\eta)+\delta N \rfloor$ comprising of the following:
\begin{equation}\label{Eq:sw2}
	\begin{aligned}
		sw = &\sum\limits_{i = 1}^{\lfloor NC(\eta)-\delta N \rfloor} \mathbb{P}(R_i \geq C_i) \mathbb{E}[R_i-C_i|R_i\geq C_i] + \sum\limits_{i = \lfloor NC(\eta)-\delta N \rfloor+1}^{\lfloor NC(\eta)+\delta N \rfloor} \mathbb{P}(R_i \geq C_i) \mathbb{E}[R_i-C_i|R_i\geq C_i]- sw^{\rm loss}_1 - sw^{\rm loss}_2.
	\end{aligned}
\end{equation}

\textbf{The social welfare loss comes from two blocks $sw^{\rm loss}_1$ and $sw^{\rm loss}_2$. We will compute them one by one.}

\begin{itemize}
	\item The social welfare loss in block 1: The social welfare loss in block 1 comes from the situation that miners include some transactions to get fees, but these transactions reduces the social welfare.
	\begin{figure}[h]
		\centering
		{\includegraphics[width=10cm]{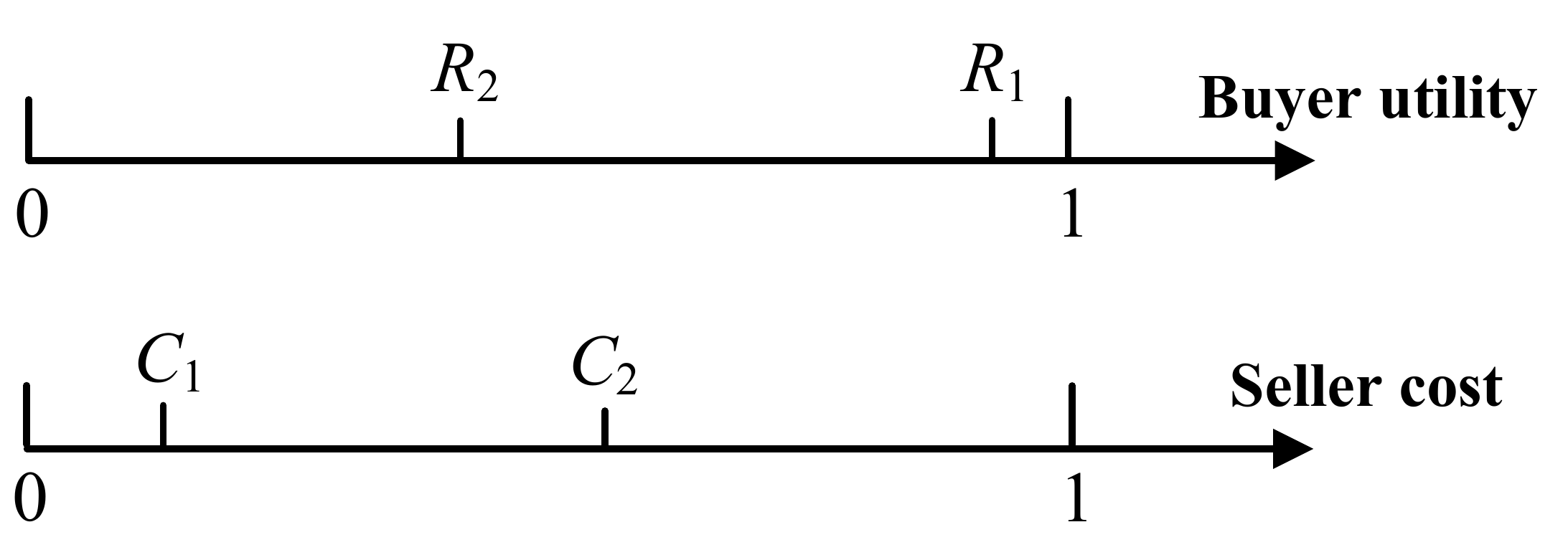}}
		\caption{Social welfare loss due to $R_2$ and $C_2$.}\label{Fig:loss7}
	\end{figure}
	
	Fig. \ref{Fig:loss7} illustrates the social welfare loss due to $R_2$ and $C_2$. The social optimum under such case is $R_1-C_1$. However, at Nash equilibrium, miners will match $R_1$ with $C_2$ and $R_2$ with $C_1$, such that they can collect all transactions' fees. This yields the social welfare of $R_1 + R_2-C_1 - C_2$. Hence, the social welfare loss compared with social optimum for such a situation is $C_2 - R_2$.
	
	Under $A =  \lfloor NC(\eta)+\delta N \rfloor$, the expected social welfare loss due to $R_2$ and $C_2$ is 
	\begin{equation}
		\begin{aligned}
			sw^{\rm loss}_{1,2} =& \mathbb{P}(C_1 \leq R_2 \leq C_2 \leq R_1) \mathbb{E}[C_2-R_2|C_1 \leq R_2 \leq C_2 \leq R_1]\\
			< & \mathbb{P}(R_2 \leq C_2) \mathbb{E}[C_2-R_2|R_2 \leq C_2].
		\end{aligned}
	\end{equation}
	
	\begin{figure}[h]
		\centering
		{\includegraphics[width=10cm]{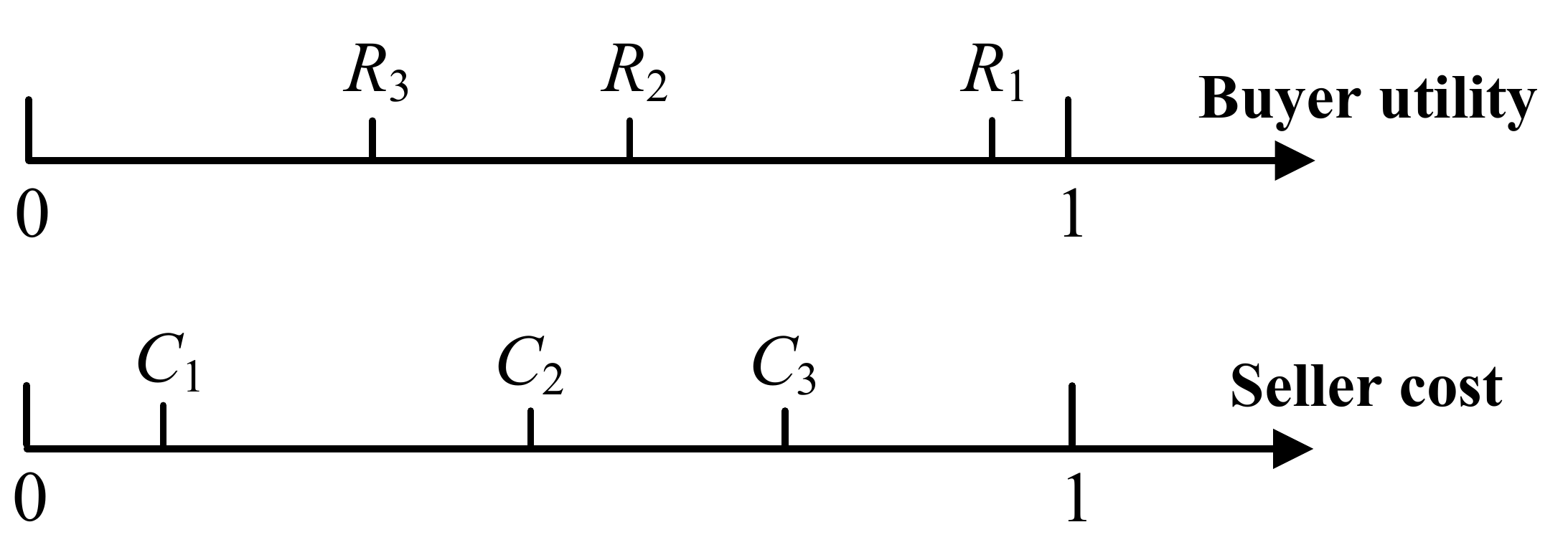}}
		\caption{Social welfare loss due to $R_3$ and $C_3$.}\label{Fig:loss8}
	\end{figure}
	Fig. \ref{Fig:loss8} illustrates the social welfare loss due to $R_3$ and $C_3$. The social optimum under such case is $R_1 + R_2-C_1 - C_2$. However, miners will match $R_1$ with $C_3$, $R_2$ with $C_2$, and $R_3$ with $C_1$. Such that they can collect all transactions' fees. This yields the social welfare of $R_1+R_2+R_3-C_1-C_2-C_3$. Hence, the social welfare loss compared with the social optimum for such a situation is $C_3 - R_3$.
	
	Under $A =  \lfloor NC(\eta)+\delta N \rfloor$, the expected social welfare loss due to $R_3$ and $C_3$ is 
	\begin{equation}
		\begin{aligned}
			sw^{\rm loss}_{1,3} =& \mathbb{P}(C_1 \leq R_3 \leq C_2 \leq R_2 \leq C_3 \leq R_1) \mathbb{E}[C_3-R_3|C_1 \leq R_3 \leq C_2 \leq R_2 \leq C_3 \leq R_1]\\
			< & \mathbb{P}(R_3 \leq C_3) \mathbb{E}[C_3-R_3 |R_3 \leq C_3].
		\end{aligned}
	\end{equation}

	We can use such method to calculate all the social welfare loss due to $R_i$ and $C_i$, where $i=2$ to $ \lfloor NC(\eta)+\delta N \rfloor$. Hence the total expected social welfare loss from block 1 is 
	\begin{equation}\label{eq:sw_loss3}
		\begin{aligned}
			sw^{\rm loss}_1 = & \sum_{i=2}^{\lfloor NC(\eta)+\delta N \rfloor}sw^{\rm loss}_{1,i}\\
			< & \sum_{i=2}^{\lfloor NC(\eta)+\delta N \rfloor}\mathbb{P}(R_i \leq C_i) \mathbb{E}[C_i-R_i |R_i \leq C_i]\\
			=&\sum_{i=2}^{\lfloor NC(\eta)-\delta N \rfloor}\mathbb{P}(R_i \leq C_i) \mathbb{E}[C_i-R_i |R_i \leq C_i] + \sum\limits_{i = \lfloor NC(\eta)-\delta N \rfloor+1}^{ \lfloor NC(\eta)+\delta N \rfloor} \mathbb{P}(R_i \leq C_i) \mathbb{E}[C_i-R_i |R_i \leq C_i] .
		\end{aligned}
	\end{equation}
	We estimate two parts one by one:
	\begin{enumerate}
		\item For $\sum_{i=2}^{\lfloor NC(\eta)-\delta N \rfloor}\mathbb{P}(R_i \leq C_i) \mathbb{E}[C_i-R_i |R_i \leq C_i]$, based on (\ref{Eq:prob32}), we have
		\begin{equation}
			\mathbb{P}(R_i \leq C_i) \approx 0, \quad\forall 2 \leq i\leq \lfloor NC(\eta)-\delta N \rfloor.
		\end{equation}
		Hence, we have
		\begin{equation}
			\sum_{i=2}^{\lfloor NC(\eta)-\delta N \rfloor}\mathbb{P}(R_i \leq C_i) \mathbb{E}[C_i-R_i |R_i \leq C_i] \approx 0.
		\end{equation}
		\item For $\sum\limits_{i = \lfloor NC(\eta)-\delta N \rfloor+1}^{\lfloor NC(\eta)+\delta N \rfloor} \mathbb{P}(R_i \leq C_i) \mathbb{E}[C_i-R_i |R_i \leq C_i] $, we have
		\begin{equation}
			\begin{aligned}
				&\sum\limits_{i = \lfloor NC(\eta)-\delta N \rfloor+1}^{\lfloor NC(\eta)+\delta N \rfloor} \mathbb{P}(R_i \leq C_i) \mathbb{E}[C_i-R_i |R_i \leq C_i]\\
				\leq & \sum\limits_{i = \lfloor NC(\eta)-\delta N \rfloor+1}^{\lfloor NC(\eta)+\delta N \rfloor} \mathbb{E}[C_i-R_i |R_i \leq C_i]\\
				\leq &  \sum\limits_{i = \lfloor NC(\eta)-\delta N \rfloor+1}^{\lfloor NC(\eta)+\delta N \rfloor} 1 \\
				= & 2\delta N.
			\end{aligned}
		\end{equation}
	\end{enumerate}
	\textbf{This completes the computation of $sw^{\rm loss}_1$.}
	
	\item The social welfare loss in block 2: Note that based on equation (\ref{Eq:prob43}), for any $\lfloor NC(\eta) + \delta N\rfloor+1 \leq i \leq \min\{\rho N,N\}$, we have 
	\begin{equation}
		\mathbb{P}(R_i \geq C_i) \approx 0.
	\end{equation} 
	In other words, when $N \rightarrow \infty$, the probability of there are transaction left for block 2 is zero. Hence, the social welfare loss due to block 2 is zero: 
	\begin{equation}
		sw^{\rm loss}_2 = 0.
	\end{equation}
\end{itemize}

Based on all these discussions, we can estimate the ratio between social optimum and social welfare, under block size $A =\lfloor NC(\eta) + \delta N\rfloor$ when $N \rightarrow \infty$:
\begin{equation}\label{ratio}
	\begin{aligned}
		&\hspace{4mm}\frac{sw^{\rm opt}}{sw} \\
		& = \frac{sw^{\rm opt}}{\sum\limits_{i = 1}^{\lfloor NC(\eta)-\delta N \rfloor} \mathbb{P}(R_i \geq C_i) \mathbb{E}[R_i-C_i|R_i\geq C_i] + \sum\limits_{i = \lfloor NC(\eta)-\delta N \rfloor+1}^{\lfloor NC(\eta)+\delta N \rfloor} \mathbb{P}(R_i \geq C_i) \mathbb{E}[R_i-C_i|R_i\geq C_i]- sw^{\rm loss}_1 - sw^{\rm loss}_2} \\
		& < \frac{sw^{\rm opt}}{\sum\limits_{i = 1}^{\lfloor NC(\eta)-\delta N \rfloor} \mathbb{P}(R_i \geq C_i) \mathbb{E}[R_i-C_i|R_i\geq C_i] - sw^{\rm loss}_1 - sw^{\rm loss}_2} \\
		& < \frac{ \left[(\rho N +1) \int_{1-\frac{C(\eta)-\delta}{\rho}}^{1}R^{-1}(x) dx\right] - \left[(N +1) \int_{0}^{C(\eta)-\delta}C^{-1}(x) dx\right] + 2\delta N}{\left[(\rho N +1) \int_{1-\frac{C(\eta)-\delta}{\rho}}^{1}R^{-1}(x) dx\right] - \left[(N +1) \int_{0}^{C(\eta)-\delta}C^{-1}(x) dx\right] - 2\delta N}\\
		& \approx \frac{ \rho\int_{1-\frac{C(\eta)-\delta}{\rho}}^{1}R^{-1}(x) dx - \int_{0}^{C(\eta)-\delta}C^{-1}(x) dx + 2\delta}{\rho \int_{1-\frac{C(\eta)-\delta}{\rho}}^{1}R^{-1}(x) dx - \int_{0}^{C(\eta)-\delta}C^{-1}(x) dx - 2\delta }.
	\end{aligned}
\end{equation}
\textbf{Then, we will use the Squeeze Theorem to prove that PoA is 1 when setting $\delta = N^{-\psi}$.} Based on (\ref{ratio}), the ratio between social optimum and social welfare satisfies:
\begin{equation}
	1 \leq \frac{sw^{\rm opt}}{sw}  \leq \frac{\rho \int_{1-\frac{C(\eta)-\delta}{\rho}}^{1}R^{-1}(x) dx - \int_{0}^{C(\eta)-\delta}C^{-1}(x) dx + 2\delta }{ \rho\int_{1-\frac{C(\eta)-\delta}{\rho}}^{1}R^{-1}(x) dx - \int_{0}^{C(\eta)-\delta}C^{-1}(x) dx - 2\delta} .
\end{equation}

First, for $1 \leq \frac{sw^{\rm opt}}{sw}$, we have
\begin{equation}\label{eq:squ1}
	\lim\limits_{\substack{N\rightarrow \infty \\ \delta = N^{-\psi}}} \frac{sw^{\rm opt}}{sw} \geq 1.
\end{equation}

Then, we analyze 
\begin{equation}\label{eq:alpha}
	\frac{sw^{\rm opt}}{sw}  \leq \frac{\rho \int_{1-\frac{C(\eta)-\delta}{\rho}}^{1}R^{-1}(x) dx - \int_{0}^{C(\eta)-\delta}C^{-1}(x) dx + 2\delta }{ \rho\int_{1-\frac{C(\eta)-\delta}{\rho}}^{1}R^{-1}(x) dx - \int_{0}^{C(\eta)-\delta}C^{-1}(x) dx - 2\delta}.
\end{equation}
Note that the term $\rho \int_{1-\frac{C(\eta)-\delta}{\rho}}^{1}R^{-1}(x) dx - \int_{0}^{C(\eta)-\delta}C^{-1}(x) dx$ in above equation must be higher than a strictly positive constant. The reason is as follows: based on equation (\ref{Eq:swopt58}), we have:
\begin{equation}\label{eq:positve}
	\begin{aligned}
		& \rho \int_{1-\frac{C(\eta)-\delta}{\rho}}^{1}R^{-1}(x) dx - \int_{0}^{C(\eta)-\delta}C^{-1}(x) dx\\
		=& \lim\limits_{N\rightarrow\infty}\frac{\left[(\rho N +1) \int_{1-\frac{C(\eta)-\delta}{\rho}}^{1}R^{-1}(x) dx\right] - \left[(N +1) \int_{0}^{C(\eta)-\delta}C^{-1}(x) dx\right]}{N} \\
		=& \lim\limits_{N\rightarrow\infty} \frac{\sum\limits_{i = 1}^{\lfloor NC(\eta)-\delta N \rfloor} R^{-1}\left(1-\frac{i}{\rho N+1}\right) - C^{-1}\left(\frac{i}{N+1}\right)}{N}\\
		= & \lim\limits_{N\rightarrow\infty} \frac{\left[\sum\limits_{i = 1}^{\lfloor \frac{NC(\eta)-\delta N}{2} \rfloor} R^{-1}\left(1-\frac{i}{\rho N+1}\right) - C^{-1}\left(\frac{i}{N+1}\right)\right] + \left[\sum\limits_{i = \lfloor \frac{NC(\eta)-\delta N}{2} \rfloor+1}^{\lfloor NC(\eta)-\delta N \rfloor} R^{-1}\left(1-\frac{i}{\rho N+1}\right) - C^{-1}\left(\frac{i}{N+1}\right)\right]}{N}\\
		> & \lim\limits_{N\rightarrow\infty} \frac{\lfloor \frac{NC(\eta)-\delta N}{2} \rfloor \left[R^{-1}\left(1-\frac{NC(\eta)-\delta N}{2(\rho N+1)}\right) - C^{-1}\left(\frac{NC(\eta)-\delta N}{2(N+1)}\right)\right]}{N}\\
		& +  \lim\limits_{N\rightarrow\infty}\frac{\lfloor \frac{NC(\eta)-\delta N}{2} \rfloor \left[R^{-1}\left(1-\frac{NC(\eta)-\delta N}{\rho N+1}\right) - C^{-1}\left(\frac{NC(\eta)-\delta N}{N+1}\right)\right]}{N}\\
		= & \frac{C(\eta)-\delta }{2} \left[R^{-1}\left(1-\frac{C(\eta)-\delta }{2\rho }\right) - C^{-1}\left(\frac{C(\eta)-\delta }{2}\right)\right] \\
		&+ \frac{C(\eta)-\delta}{2} \left[R^{-1}\left(1-\frac{C(\eta)-\delta }{\rho}\right) - C^{-1}\left(C(\eta)-\delta\right)\right] \\
		> & \frac{C(\eta)-\delta }{2} \left[R^{-1}\left(1-\frac{C(\eta) }{2\rho }\right) - C^{-1}\left(\frac{C(\eta) }{2}\right)\right] + \frac{C(\eta)-\delta}{2} \left[R^{-1}\left(1-\frac{C(\eta) }{\rho}\right) - C^{-1}\left(C(\eta)\right)\right] \\
		= & \frac{C(\eta)-\delta }{2} \left[R^{-1}\left(1-\frac{C(\eta) }{2\rho }\right) - C^{-1}\left(\frac{C(\eta) }{2}\right)\right].
	\end{aligned}
\end{equation}
Note that $\left[R^{-1}\left(1-\frac{C(\eta) }{2\rho }\right) - C^{-1}\left(\frac{C(\eta) }{2}\right)\right] > 0$. Moreover, when $\delta = N^{-\psi}$ and $N\rightarrow\infty$, $\delta$ must be smaller than $\frac{C(\eta)}{2}$. Hence, equation (\ref{eq:positve}) becomes:
\begin{equation}\label{eq:positve2}
	\begin{aligned}
		& \rho \int_{1-\frac{C(\eta)-\delta}{\rho}}^{1}R^{-1}(x) dx - \int_{0}^{C(\eta)-\delta}C^{-1}(x) dx\\
		>& \frac{C(\eta)-\delta }{2} \left[R^{-1}\left(1-\frac{C(\eta) }{2\rho }\right) - C^{-1}\left(\frac{C(\eta) }{2}\right)\right]\\
		>& \frac{C(\eta) }{4} \left[R^{-1}\left(1-\frac{C(\eta) }{2\rho }\right) - C^{-1}\left(\frac{C(\eta) }{2}\right)\right]\\
		> & 0.
	\end{aligned}
\end{equation}

Based on equations (\ref{eq:positve}) and (\ref{eq:positve2}), we have shown the term $\rho \int_{1-\frac{C(\eta)-\delta}{\rho}}^{1}R^{-1}(x) dx - \int_{0}^{C(\eta)-\delta}C^{-1}(x) dx$ must be strictly positive. Hence, for equation (\ref{eq:alpha}), we have
\begin{equation}\label{eq:squ2}
		\lim\limits_{\substack{N\rightarrow \infty \\ \delta = N^{-\psi}}}  \frac{sw^{\rm opt}}{sw} \leq \lim\limits_{\substack{N\rightarrow \infty \\ \delta = N^{-\psi}}} \frac{\rho \int_{1-\frac{C(\eta)-\delta}{\rho}}^{1}R^{-1}(x) dx - \int_{0}^{C(\eta)-\delta}C^{-1}(x) dx + 2\delta }{ \rho\int_{1-\frac{C(\eta)-\delta}{\rho}}^{1}R^{-1}(x) dx - \int_{0}^{C(\eta)-\delta}C^{-1}(x) dx - 2\delta} = 1.
\end{equation}

Based on equations (\ref{eq:squ1}) and (\ref{eq:squ2}). we have 
\begin{equation}
	\lim\limits_{\substack{N\rightarrow \infty \\ \delta = N^{-\psi}}} \frac{sw^{\rm opt}}{sw} = 1.
\end{equation}

This means that setting the block size as $\lfloor N\big(C(\eta) + N^{-\psi}\big) \rfloor$ always achieves the social optimum. Hence, the PoA of homogeneous trading quantity $\underline{b} = \overline{b}$ is PoA = 1.

This completes the proof of for homogeneous trading quantity $\underline{b} = \overline{b}$.

\vspace{30mm}

\vspace{10mm}
\textbf{Proof for heterogeneous trading quantities}: \textbf{We will first estimate the social optimum. Next we estimate the social welfare loss when setting $A = \lfloor NC(\eta)+\delta N \rfloor$.}

\textbf{Social optimum estimation}: We denote the buying utilities in a decreasing order as $R_1 \geq R_2\geq \cdots \geq R_{\rho N}$ and denote selling costs in an increasing order as $C_1 \leq C_2 \leq \cdots \leq C_N$. We also range to buying quantities and asking quantities in a decreasing order as $b_{1} \geq b_{2}\geq \cdots \geq b_{\rho N}$ and  $q_{1} \geq q_{2}\geq \cdots \geq q_{ N}$.

Note that for $R_i$, its buying quantities are not necessarily the $i$-th highest.
Hence, based on Rearrangement Inequality, the social optimum satisfy: 
\begin{equation}\label{Eq:swopt66}
	sw^{\rm opt} \leq  \sum\limits_{i = 1}^{\min\{\rho N,N\}} \mathbb{E}[\min\{b_{ i},q_{ i}\}]\mathbb{P}(R_i \geq C_i) \mathbb{E}[R_i-C_i|R_i\geq C_i].
\end{equation}
Note that $R_i$ is the $\rho N+1-i$-th order statistics of $\rho N$ iid random variables (i.e., buying utilities), $b_{i}$ is the $\rho N+1-i$-th order statistics of $\rho N$ iid random variables (i.e., buying quantities), $C_i$ is $i$-th order statistics of $N$ iid random variables (i.e., selling costs), and $q_{i}$ is the $N+1-i$-th order statistics of $N$ iid random variables (i.e., selling quantities).

\begin{figure}[h]
	\centering
	\includegraphics[width=10cm]{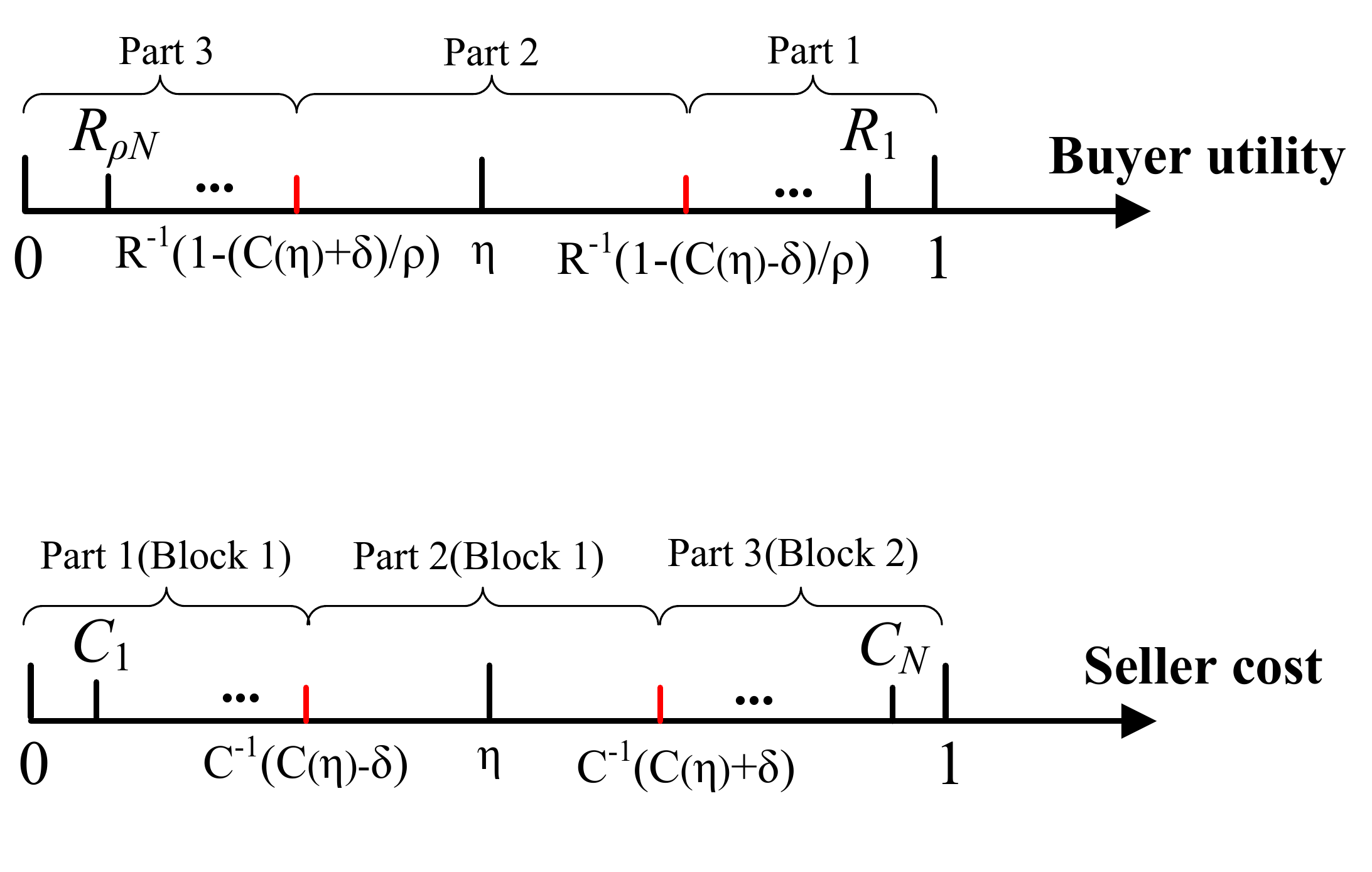}
	\caption{Three parts to estimate $sw^{\rm opt}$.}
	\label{Fig:large_N3}
\end{figure}

As illustrated in Fig. \ref{Fig:large_N3}, we estimate $sw^{\rm opt}$ in equation (\ref{Eq:swopt66}) by three parts. Our estimation is to prove that when $N$ is large, part 1 are matched with probability 1, part 2's match can be approximated, and part 3 are matched with probability of 0.

The detailed estimation is as follows:
\begin{equation}\label{Eq:swopt67}
	\begin{aligned}
		sw^{\rm opt} \leq &\sum\limits_{i = 1}^{\lfloor NC(\eta)-\delta N \rfloor} \mathbb{E}[\min\{b_{ i},q_{ i}\}]\mathbb{P}(R_i \geq C_i) \mathbb{E}[R_i-C_i|R_i\geq C_i] \\
		&+ \sum\limits_{i = \lfloor NC(\eta)-\delta N \rfloor+1}^{\lfloor NC(\eta)+\delta N \rfloor} \mathbb{E}[\min\{b_{ i},q_{ i}\}]\mathbb{P}(R_i \geq C_i) \mathbb{E}[R_i-C_i|R_i\geq C_i]\\
		&+ \sum\limits_{i = \lfloor NC(\eta)+\delta N \rfloor +1}^{\min\{\rho N,N\}} \mathbb{E}[\min\{b_{ i},q_{ i}\}]\mathbb{P}(R_i \geq C_i) \mathbb{E}[R_i-C_i|R_i\geq C_i].
	\end{aligned}
\end{equation}
Here we estimate three parts of equation (\ref{Eq:swopt67}) as follows:
\begin{enumerate}
	\item Estimation of $\sum\limits_{i = 1}^{\lfloor NC(\eta)-\delta N \rfloor} \mathbb{E}[\min\{b_{ i},q_{ i}\}]\mathbb{P}(R_i \geq C_i) \mathbb{E}[R_i-C_i|R_i\geq C_i]$. We first estimate $\mathbb{E}[\min\{b_{ i},q_{ i}\}]$ then we estimate $\mathbb{P}(R_i \geq C_i) \mathbb{E}[R_i-C_i|R_i\geq C_i]$.
	
	For $1 \leq i \leq \lfloor NC(\eta)-\delta N \rfloor$, the expectations of $b_{ i}$ and $q_{ i}$ satisfy:
	\begin{align}
		& \mathbb{E}[b_{ i}] \approx B^{-1}\left(1-\frac{i}{\rho N+1}\right).\\
		& \mathbb{E}[q_{ i}] \approx Q^{-1}\left(1-\frac{i}{N+1}\right).
	\end{align}
	Moreover, since $\min\{X,Y\}$ is a concave function, Jensen's inequality tells us:
	\begin{equation}
		\mathbb{E}[\min\{b_{ i},q_{ i}\}] \leq \min\{\mathbb{E}[b_{ i}],\mathbb{E}[q_{ i}]\} = \min\left\{B^{-1}\left(1-\frac{i}{\rho N+1}\right),Q^{-1}\left(1-\frac{i}{N+1}\right)\right\}.
	\end{equation}
	For $1 \leq i \leq \lfloor NC(\eta)-\delta N \rfloor$, the expectations of $R_i$ and $C_i$ satisfy:
	\begin{align}
		&\mathbb{E}[R_i] \approx R^{-1}\left(1-\frac{i}{\rho N+1}\right) \geq  R^{-1}\left(1-\frac{\lfloor NC(\eta)-\delta N \rfloor}{\rho N+1}\right) \approx R^{-1}\left(1 - \frac{C(\eta)-\delta}{\rho}\right).\\
		&\mathbb{E}[C_i] \approx C^{-1}\left(\frac{i}{N+1}\right) \leq C^{-1}\left(\frac{\lfloor NC(\eta)-\delta N \rfloor}{N+1}\right) \approx C^{-1}\left(C(\eta)-\delta\right).
	\end{align}
	The variance satisfies:
	\begin{align}
		&\text{Var}(R_i)  \approx \frac{(\rho N + 1 - i) i}{(\rho N)^3} \left[\frac{1}{r\left(R^{-1}(\frac{\rho N + 1 - i}{\rho N+1})\right)}\right]^2.\\
		&\text{Var}(C_i) \approx \frac{i (N +1 - i) }{N^3}\left[\frac{1}{c\left(C^{-1}(\frac{i}{N+1})\right)}\right]^2.
	\end{align}

	Then $R_i$ and $C_i$ can be approximated by normal distribution when $N\rightarrow \infty$.\footnote{Based on asymptotic distribution of a central order statistics theorem. \url{http://www.math.ntu.edu.tw/~hchen/teaching/LargeSample/notes/noteorder.pdf}\url{https://ieeexplore.ieee.org/stamp/stamp.jsp?tp=&arnumber=9936687}} Then to estimate the probability of $R_i \geq C_i$, we can use the standard method to estimate the probability of one normal-distributed random variable larger than another normal-distributed random variable as follows:
	\begin{align}
		&\mu_i \triangleq \mathbb{E}[R_i] - \mathbb{E}[C_i] =  R^{-1}\left(1-\frac{i}{\rho N+1}\right) - C^{-1}\left(\frac{i}{N+1}\right) .\\
		&\sigma_i^2 \triangleq\text{Var}(R_i) + \text{Var}(C_i).
	\end{align}
	Given $\delta > 0 $, we have:
	\begin{equation}
		\begin{aligned}
			\mu_i = R^{-1}\left(1-\frac{i}{\rho N+1}\right) - C^{-1}\left(\frac{i}{N+1}\right) &\geq R^{-1}\left(1 - \frac{C(\eta)-\delta}{\rho}\right)- C^{-1}\left(C(\eta)-\delta\right)\\
			& > R^{-1}\left(1 - \frac{C(\eta)}{\rho}\right)- C^{-1}\left(C(\eta)\right)\\
			& = 0.
		\end{aligned}
	\end{equation}
	Then for any $1 \leq i \leq \lfloor NC(\eta)-\delta N \rfloor$, we have 
	\begin{equation}\label{Eq:prob41}
		\mathbb{P}(R_i \geq C_i)   = \mathbb{P}\Big(\frac{R_i - C_i - \mu_i}{\sigma_i} \geq -\frac{ \mu_i}{\sigma_i}\Big) = \mathbb{P}\Big(Z \geq -\frac{\mu_i}{\sigma_i}\Big),
	\end{equation}
	where $Z$ is a standard normal variable. Moreover, we note that:
	\begin{equation}
		\begin{aligned}
			& \hspace{5mm}\frac{\mu_i}{\sqrt{N}\sigma_i}\\
			& \geq\frac{\mu_{\lfloor NC(\eta)-\delta N \rfloor}}{\sqrt{N}\sigma_{\lfloor NC(\eta)-\delta N \rfloor}}\\
			& = \frac{R^{-1}\left(1 - \frac{C(\eta)-\delta}{\rho}\right) -  C^{-1}\left(C(\eta)-\delta\right)}{\sqrt{\frac{[C(\eta)-\delta][\rho-C(\eta)+\delta]}{\rho^3 }\left[\frac{1}{r\left(R^{-1}(1 - \frac{C(\eta)-\delta}{\rho})\right)}\right]^2 + [C(\eta)-\delta][1-C(\eta)+\delta]\left[\frac{1}{c\left(C^{-1}(C(\eta)-\delta)\right)}\right]^2}}.
		\end{aligned}
	\end{equation}
	and we define 
	\begin{equation}
		\Delta \triangleq \frac{R^{-1}\left(1 - \frac{C(\eta)-\delta}{\rho}\right) -  C^{-1}\left(C(\eta)-\delta\right)}{\sqrt{\frac{[C(\eta)-\delta][\rho-C(\eta)+\delta]}{\rho^3 }\left[\frac{1}{r\left(R^{-1}(1 - \frac{C(\eta)-\delta}{\rho})\right)}\right]^2 + [C(\eta)-\delta][1-C(\eta)+\delta]\left[\frac{1}{c\left(C^{-1}(C(\eta)-\delta)\right)}\right]^2}},
	\end{equation}
	where $\Delta >0$ is a constant given $\delta>0$. We use $\Phi$ to denote the CDF of standard normal distribution. For large $N$ and for any $1 \leq i \leq \lfloor NC(\eta)-\delta N \rfloor$, equation (\ref{Eq:prob41}) becomes:
	\begin{equation}\label{Eq:prob42}
		\begin{aligned}
			\mathbb{P}(R_i \geq C_i) &= \Phi\Big(\frac{\mu_i}{\sigma_i}\Big) \\
			&\geq \Phi\Big(\frac{\mu_{\lfloor NC(\eta)-\delta N \rfloor}}{\sigma_{\lfloor NC(\eta)-\delta N \rfloor}}\Big) \\
			& = \Phi\Big(\sqrt{N}\Delta\Big)\\
			&\approx 1 - \frac{1}{\sqrt{2\pi N}\Delta} e^{-\frac{N\Delta^2}{2}}\\
			&\approx 1.
		\end{aligned}
	\end{equation}
	To sum up, we have:
	\begin{equation}\label{Eq:swopt68}
		\begin{aligned}
			&\sum\limits_{i = 1}^{\lfloor NC(\eta)-\delta N \rfloor} \mathbb{E}[\min\{b_{ i},q_{ i}\}]\mathbb{P}(R_i \geq C_i) \mathbb{E}[R_i-C_i|R_i\geq C_i] \\
			\leq &  \sum\limits_{i = 1}^{\lfloor NC(\eta)-\delta N \rfloor} \min\{\mathbb{E}[b_{ i}],\mathbb{E}[q_{ i}]\}\mathbb{E}[R_i-C_i] \\
			\approx & \sum\limits_{i = 1}^{\lfloor NC(\eta)-\delta N \rfloor} \min\left\{B^{-1}\left(1-\frac{i}{\rho N+1}\right),Q^{-1}\left(1-\frac{i}{N+1}\right)\right\} \left(R^{-1}\left(1-\frac{i}{\rho N+1}\right) - C^{-1}\left(\frac{i}{N+1}\right)\right)\\
			\approx  & (N +1) \int_{0}^{C(\eta)-\delta} \min\Big\{B^{-1}(1-\frac{x}{\rho}),Q^{-1}(1-x)\Big\}\left(R^{-1}(1-\frac{x}{\rho}) - C^{-1}(x)\right) dx.
		\end{aligned}
	\end{equation}
	
	\item Estimation of $\sum\limits_{i = \lfloor NC(\eta)-\delta N \rfloor+1}^{\lfloor NC(\eta)+\delta N \rfloor}  \mathbb{E}[\min\{b_{ i},q_{ i}\}]\mathbb{P}(R_i \geq C_i) \mathbb{E}[R_i-C_i|R_i\geq C_i]$. For this part, we have $\mathbb{E}[\min\{b_{ i},q_{ i}\}] \leq 1$, $\mathbb{P}(R_i \geq C_i) \leq 1$ and $\mathbb{E}[R_i-C_i|R_i\geq C_i] \leq 1$. Hence, we have
	\begin{equation}\label{Eq:swopt69}
		\begin{aligned}
			&\sum\limits_{i = \lfloor NC(\eta)-\delta N \rfloor+1}^{\lfloor NC(\eta)+\delta N \rfloor}  \mathbb{E}[\min\{b_{ i},q_{ i}\}]\mathbb{P}(R_i \geq C_i) \mathbb{E}[R_i-C_i|R_i\geq C_i] 
			\leq  \sum\limits_{i = \lfloor NC(\eta)-\delta N \rfloor+1}^{\lfloor NC(\eta)+\delta N \rfloor} 1 \leq 2\delta N.
		\end{aligned}
	\end{equation}
	\item Estimation of $\sum\limits_{i = \lfloor NC(\eta)+\delta N \rfloor +1}^{\min\{\rho N,N\}} \mathbb{P}(R_i \geq C_i) \mathbb{E}[R_i-C_i|R_i\geq C_i]$. For $\lfloor NC(\eta)+\delta N \rfloor +1 \leq i \leq \min\{\rho N,N\}$, the expectations of $R_i$ and $C_i$ satisfy:
	\begin{align}
		&\mathbb{E}[R_i] \approx R^{-1}\left(1-\frac{i}{\rho N+1}\right) \leq  R^{-1}\left(1-\frac{\lfloor NC(\eta)+\delta N \rfloor+1}{\rho N+1}\right) \approx R^{-1}\left(1 - \frac{C(\eta)+\delta}{\rho}\right).\\
		&\mathbb{E}[C_i] \approx C^{-1}\left(\frac{i}{N+1}\right) \geq C^{-1}\left(\frac{\lfloor NC(\eta)+\delta N \rfloor +1}{N+1}\right) \approx C^{-1}\left(C(\eta)+\delta\right).
	\end{align}
	The variances satisfy:
	\begin{align}
		&\text{Var}(R_i)  \approx \frac{(\rho N + 1 - i) i}{(\rho N)^3} \left[\frac{1}{r\left(R^{-1}(\frac{\rho N + 1 - i}{\rho N+1})\right)}\right]^2.\\
		&\text{Var}(C_i) \approx \frac{i (N +1 - i) }{N^3}\left[\frac{1}{c\left(C^{-1}(\frac{i}{N+1})\right)}\right]^2.
	\end{align}
	Then $R_i$ and $C_i$ can be approximated by normal distribution when $N\rightarrow \infty$.\footnote{Based on asymptotic distribution of a central order statistics theorem. \url{http://www.math.ntu.edu.tw/~hchen/teaching/LargeSample/notes/noteorder.pdf}\url{https://ieeexplore.ieee.org/stamp/stamp.jsp?tp=&arnumber=9936687}} Then to estimate the probability of $R_i \geq C_i$, we can use the standard method to estimate the probability of one normal-distributed random variable larger than another normal-distributed random variable as follows:
	\begin{align}
		&\mu_i \triangleq \mathbb{E}[R_i] - \mathbb{E}[C_i] =  R^{-1}\left(1-\frac{i}{\rho N+1}\right) - C^{-1}\left(\frac{i}{N+1}\right) .\\
		&\sigma_i^2 \triangleq\text{Var}(R_i) + \text{Var}(C_i).
	\end{align}
	Given $\delta > 0 $, we have:
	\begin{equation}
		\begin{aligned}
			\mu_i = R^{-1}\left(1-\frac{i}{\rho N+1}\right) - C^{-1}\left(\frac{i}{N+1}\right) &\leq R^{-1}\left(1 - \frac{C(\eta)+\delta}{\rho}\right)- C^{-1}\left(C(\eta)+\delta\right)\\
			& < R^{-1}\left(1 - \frac{C(\eta)}{\rho}\right)- C^{-1}\left(C(\eta)\right)\\
			& = 0.
		\end{aligned}
	\end{equation}
	Then for any $\lfloor NC(\eta)+\delta N \rfloor +1 \leq i \leq \min\{\rho N,N\}$, we have 
	\begin{equation}\label{Eq:prob44}
		\mathbb{P}(R_i \geq C_i)   = \mathbb{P}\Big(\frac{R_i - C_i - \mu_i}{\sigma_i} \geq -\frac{ \mu_i}{\sigma_i}\Big) = \mathbb{P}\Big(Z \geq -\frac{\mu_i}{\sigma_i}\Big),
	\end{equation}
	where $Z$ is a standard normal variable. Moreover, we note that:
	\begin{equation}
		\begin{aligned}
			& \hspace{5mm}-\frac{\mu_i}{\sqrt{N}\sigma_i}\\
			& \geq-\frac{\mu_{\lfloor NC(\eta)+\delta N \rfloor}}{\sqrt{N}\sigma_{\lfloor NC(\eta)+\delta N \rfloor}}\\
			& = \frac{-R^{-1}\left(1 - \frac{C(\eta)+\delta}{\rho}\right) +  C^{-1}\left(C(\eta)+\delta\right)}{\sqrt{\frac{[C(\eta)+\delta][\rho-C(\eta)-\delta]}{\rho^3 }\left[\frac{1}{r\left(R^{-1}(1 - \frac{C(\eta)+\delta}{\rho})\right)}\right]^2 + [C(\eta)+\delta][1-C(\eta)-\delta]\left[\frac{1}{c\left(C^{-1}(C(\eta)+\delta)\right)}\right]^2}}.
		\end{aligned}
	\end{equation}
	and we define 
	\begin{equation}
		\Delta_2 \triangleq \frac{-R^{-1}\left(1 - \frac{C(\eta)+\delta}{\rho}\right) +  C^{-1}\left(C(\eta)+\delta\right)}{\sqrt{\frac{[C(\eta)+\delta][\rho-C(\eta)-\delta]}{\rho^3 }\left[\frac{1}{r\left(R^{-1}(1 - \frac{C(\eta)+\delta}{\rho})\right)}\right]^2 + [C(\eta)+\delta][1-C(\eta)-\delta]\left[\frac{1}{c\left(C^{-1}(C(\eta)+\delta)\right)}\right]^2}},
	\end{equation}
	where $\Delta_2 >0$ is a constant given $\delta>0$. We use $\Phi$ to denote the CDF of standard normal distribution. For large $N$ and for any $\lfloor NC(\eta)+\delta N \rfloor +1 \leq i \leq \min\{\rho N,N\}$, equation (\ref{Eq:prob44}) becomes:
	\begin{equation}\label{Eq:prob53}
		\begin{aligned}
			\mathbb{P}(R_i \geq C_i) &= \Phi\Big(\frac{\mu_i}{\sigma_i}\Big) \\
			&\leq \Phi\Big(\frac{\mu_{\lfloor NC(\eta)+\delta N \rfloor}}{\sigma_{\lfloor NC(\eta)+\delta N \rfloor}}\Big) \\
			& = \Phi\Big(-\sqrt{N}\Delta_2\Big)\\
			& = 1 - \Phi\Big(\sqrt{N}\Delta_2\Big)\\
			&\approx \frac{1}{\sqrt{2\pi N}\Delta_2} e^{-\frac{N\Delta_2^2}{2}}\\
			&\approx 0.
		\end{aligned}
	\end{equation}
	Hence, we have:
	\begin{equation}
		\sum\limits_{i = \lfloor NC(\eta)+\delta N \rfloor +1}^{\min\{\rho N,N\}} \mathbb{P}(R_i \geq C_i) \mathbb{E}[R_i-C_i|R_i\geq C_i] \approx 0.
	\end{equation}
\end{enumerate}

To sum up, we estimate $sw^{\rm opt}$ as
\begin{equation}\label{Eq:swopt70}
	\begin{aligned}
		sw^{\rm opt} = &\sum\limits_{i = 1}^{\lfloor NC(\eta)-\delta N \rfloor} \mathbb{P}(R_i \geq C_i) \mathbb{E}[R_i-C_i|R_i\geq C_i] \\
		&+ \sum\limits_{i = \lfloor NC(\eta)-\delta N \rfloor+1}^{\lfloor NC(\eta)+\delta N \rfloor} \mathbb{P}(R_i \geq C_i) \mathbb{E}[R_i-C_i|R_i\geq C_i]\\
		&+ \sum\limits_{i = \lfloor NC(\eta)+\delta N \rfloor +1}^{\min\{\rho N,N\}} \mathbb{P}(R_i \geq C_i) \mathbb{E}[R_i-C_i|R_i\geq C_i]\\
		< & (N +1) \int_{0}^{C(\eta)-\delta} \min\Big\{B^{-1}(1-\frac{x}{\rho}),Q^{-1}(1-x)\Big\}\left(R^{-1}(1-\frac{x}{\rho}) - C^{-1}(x)\right) dx + 2\delta N.
	\end{aligned}
\end{equation}

\textbf{Next we estimate the social welfare when setting $A = \lfloor NC(\eta)+\delta N \rfloor$.} Since miners will randomly match transactions, the expected quantity of trading is the minimal of buying quantity and selling quantity. The CDF of the minimal is
\begin{equation}
	H_{\rm trade}(\kappa) = B(\kappa) + Q(\kappa) - B(\kappa)Q(\kappa).
\end{equation}
The expectation of trading quantity is denoted as 
\begin{equation}
	\begin{aligned}
		\mathbb{E}[\kappa] & = \int_0^1 \kappa[b(\kappa) + q(\kappa) - b(\kappa)Q(\kappa) - B(\kappa)q(\kappa)] d\kappa \\
		& = \mathbb{E}[b] + \mathbb{E}[q]  - \int_0^1 \kappa[b(\kappa)Q(\kappa) + B(\kappa)q(\kappa)] d\kappa \\
	\end{aligned}
\end{equation}

We denote 
The social welfare when setting $A = \lfloor NC(\eta)+\delta N \rfloor$ comprising of the following:
\begin{equation}\label{Eq:sw3}
	\begin{aligned}
		sw = &\sum\limits_{i = 1}^{\lfloor NC(\eta)-\delta N \rfloor} \mathbb{E}[\kappa]\mathbb{P}(R_i \geq C_i) \mathbb{E}[R_i-C_i|R_i\geq C_i] + \sum\limits_{i = \lfloor NC(\eta)-\delta N \rfloor+1}^{\lfloor NC(\eta)+\delta N \rfloor} \mathbb{E}[\kappa]\mathbb{P}(R_i \geq C_i) \mathbb{E}[R_i-C_i|R_i\geq C_i]\\
		&- sw^{\rm loss}_1 - sw^{\rm loss}_2.
	\end{aligned}
\end{equation}

\textbf{We will compute the terms in (\ref{Eq:sw3}) one by one.}

\begin{itemize}
	\item $\sum\limits_{i = 1}^{\lfloor NC(\eta)-\delta N \rfloor} \mathbb{E}[\kappa]\mathbb{P}(R_i \geq C_i) \mathbb{E}[R_i-C_i|R_i\geq C_i]$: based on the proof of Theorem 3, we have 
	\begin{equation}
		\begin{aligned}
			&\sum\limits_{i = 1}^{\lfloor NC(\eta)-\delta N \rfloor} \mathbb{E}[\kappa]\mathbb{P}(R_i \geq C_i) \mathbb{E}[R_i-C_i|R_i\geq C_i]\\
		   =& \mathbb{E}[\kappa] \left[(\rho N +1) \int_{1-\frac{C(\eta)-\delta}{\rho}}^{1}R^{-1}(x) dx - (N +1) \int_{0}^{C(\eta)-\delta}C^{-1}(x) dx\right].
		\end{aligned}
	\end{equation}

    \item $\sum\limits_{i = \lfloor NC(\eta)-\delta N \rfloor+1}^{\lfloor NC(\eta)+\delta N \rfloor} \mathbb{E}[\kappa]\mathbb{P}(R_i \geq C_i) \mathbb{E}[R_i-C_i|R_i\geq C_i]$: This term must be strictly positive, we neglect this term and treat it as zero, since we derive a lower bound of the approximation ratio.
	\item $sw^{\rm loss}_1$: $sw^{\rm loss}_1$ denotes the social welfare loss in block 1, which comes from the situation that miners include some transactions to get fees, but these transactions reduces the social welfare.
	\begin{figure}[h]
		\centering
		{\includegraphics[width=10cm]{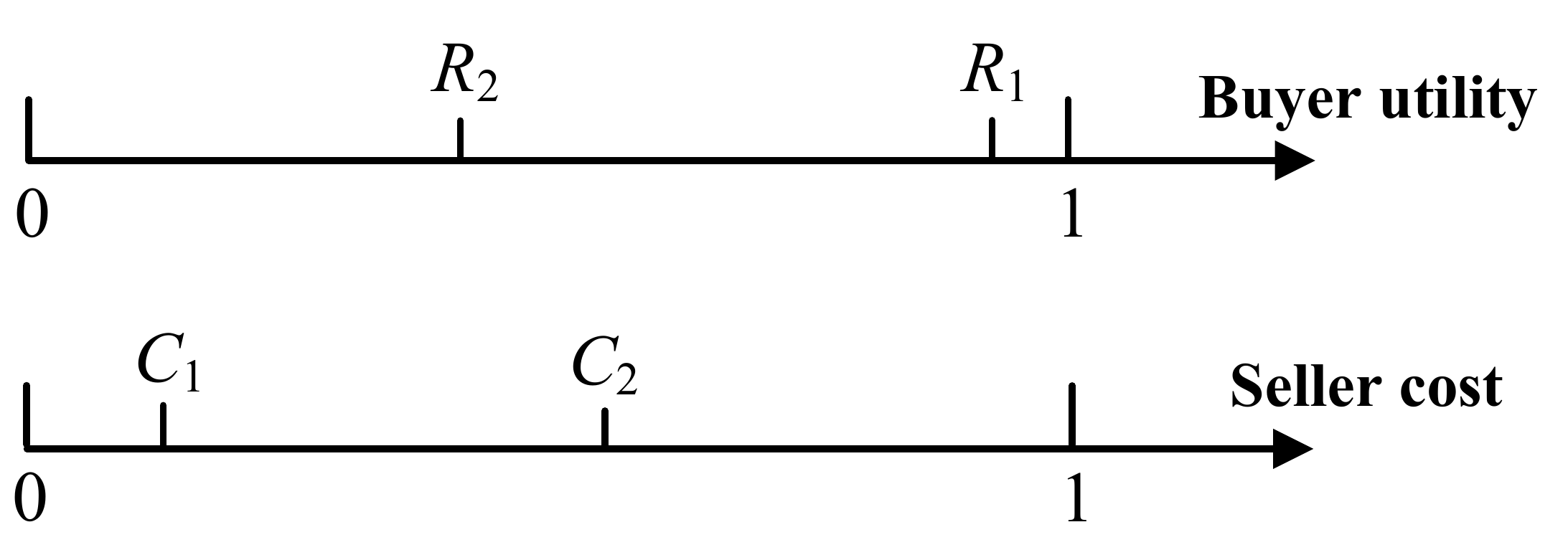}}
		\caption{Social welfare loss due to $R_2$ and $C_2$.}\label{Fig:loss9}
	\end{figure}
	
	Fig. \ref{Fig:loss9} illustrates the social welfare loss due to $R_2$ and $C_2$. The social optimum under such case is $R_1-C_1$. However, at Nash equilibrium, miners will match $R_1$ with $C_2$ and $R_2$ with $C_1$, such that they can collect all transactions' fees. This yields the social welfare of $R_1 + R_2-C_1 - C_2$. Hence, the social welfare loss compared with social optimum for such a situation is $C_2 - R_2$.
	
	Under $A =  \lfloor NC(\eta)+\delta N \rfloor$, the expected social welfare loss due to $R_2$ and $C_2$ is 
	\begin{equation}
		\begin{aligned}
			sw^{\rm loss}_{1,2} =& \mathbb{E}[\kappa]\mathbb{P}(C_1 \leq R_2 \leq C_2 \leq R_1) \mathbb{E}[C_2-R_2|C_1 \leq R_2 \leq C_2 \leq R_1]\\
			< & \mathbb{E}[\kappa]\mathbb{P}(R_2 \leq C_2) \mathbb{E}[C_2-R_2|R_2 \leq C_2].
		\end{aligned}
	\end{equation}
	
	\begin{figure}[h]
		\centering
		{\includegraphics[width=10cm]{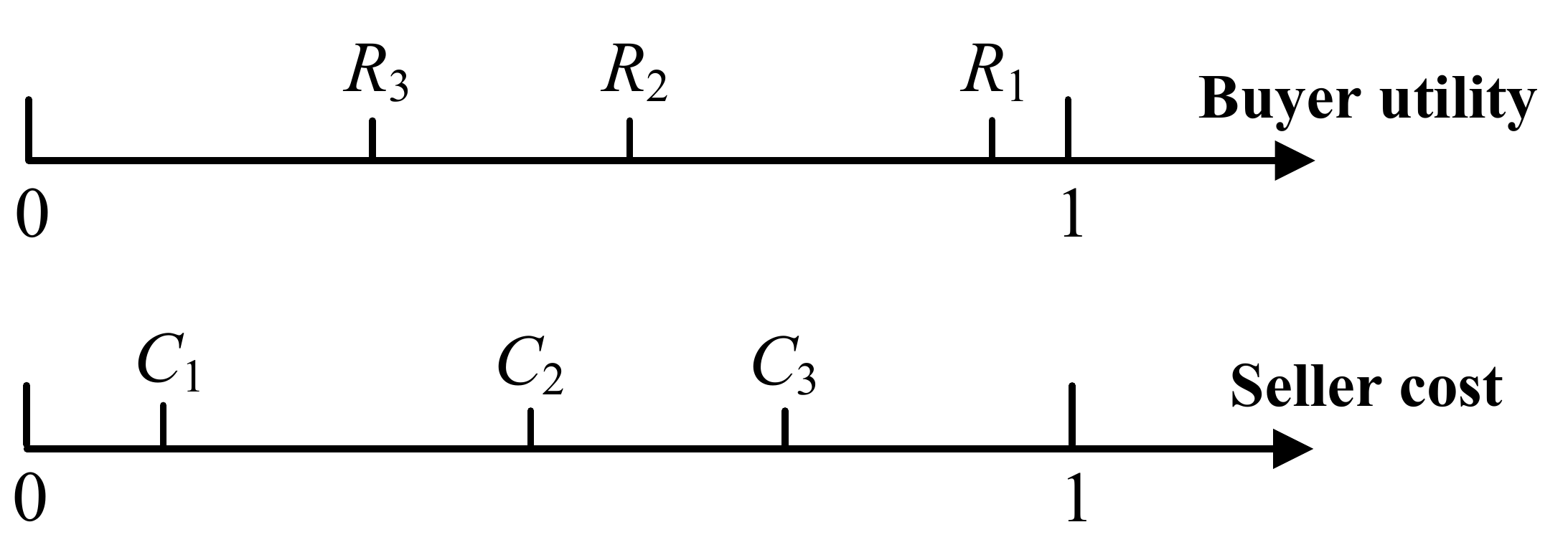}}
		\caption{Social welfare loss due to $R_3$ and $C_3$.}\label{Fig:loss10}
	\end{figure}
	Fig. \ref{Fig:loss10} illustrates the social welfare loss due to $R_3$ and $C_3$. The social optimum under such case is $R_1 + R_2-C_1 - C_2$. However, miners will match $R_1$ with $C_3$, $R_2$ with $C_2$, and $R_3$ with $C_1$. Such that they can collect all transactions' fees. This yields the social welfare of $R_1+R_2+R_3-C_1-C_2-C_3$. Hence, the social welfare loss compared with the social optimum for such a situation is $C_3 - R_3$.
	
	Under $A =  \lfloor NC(\eta)+\delta N \rfloor$, the expected social welfare loss due to $R_3$ and $C_3$ is 
	\begin{equation}
		\begin{aligned}
			sw^{\rm loss}_{1,3} =& \mathbb{E}[\kappa]\mathbb{P}(C_1 \leq R_3 \leq C_2 \leq R_2 \leq C_3 \leq R_1) \mathbb{E}[C_3-R_3|C_1 \leq R_3 \leq C_2 \leq R_2 \leq C_3 \leq R_1]\\
			< & \mathbb{E}[\kappa]\mathbb{P}(R_3 \leq C_3) \mathbb{E}[C_3-R_3 |R_3 \leq C_3].
		\end{aligned}
	\end{equation}

	We can use such method to calculate all the social welfare loss due to $R_i$ and $C_i$, where $i=2$ to $ \lfloor NC(\eta)+\delta N \rfloor$. Hence the total expected social welfare loss from block 1 is 
	\begin{equation}\label{eq:sw_loss4}
		\begin{aligned}
			sw^{\rm loss}_1 = & \sum_{i=2}^{\lfloor NC(\eta)+\delta N \rfloor}sw^{\rm loss}_{1,i}\\
			< & \sum_{i=2}^{\lfloor NC(\eta)+\delta N \rfloor}\mathbb{E}[\kappa]\mathbb{P}(R_i \leq C_i) \mathbb{E}[C_i-R_i |R_i \leq C_i]\\
			=&\sum_{i=2}^{\lfloor NC(\eta)-\delta N \rfloor}\mathbb{E}[\kappa]\mathbb{P}(R_i \leq C_i) \mathbb{E}[C_i-R_i |R_i \leq C_i] + \sum\limits_{i = \lfloor NC(\eta)-\delta N \rfloor+1}^{ \lfloor NC(\eta)+\delta N \rfloor} \mathbb{E}[\kappa]\mathbb{P}(R_i \leq C_i) \mathbb{E}[C_i-R_i |R_i \leq C_i] .
		\end{aligned}
	\end{equation}
	We estimate two parts one by one:
	\begin{enumerate}
		\item For $\sum_{i=2}^{\lfloor NC(\eta)-\delta N \rfloor}\mathbb{E}[\kappa]\mathbb{P}(R_i \leq C_i) \mathbb{E}[C_i-R_i |R_i \leq C_i]$, based on (\ref{Eq:prob42}), we have
		\begin{equation}
			\mathbb{P}(R_i \leq C_i) \approx 0, \quad\forall 2 \leq i\leq \lfloor NC(\eta)-\delta N \rfloor.
		\end{equation}
		Hence, we have
		\begin{equation}
			\sum_{i=2}^{\lfloor NC(\eta)-\delta N \rfloor}\mathbb{E}[\kappa]\mathbb{P}(R_i \leq C_i) \mathbb{E}[C_i-R_i |R_i \leq C_i] \approx 0.
		\end{equation}
		\item For $\sum\limits_{i = \lfloor NC(\eta)-\delta N \rfloor+1}^{\lfloor NC(\eta)+\delta N \rfloor} \mathbb{P}(R_i \leq C_i) \mathbb{E}[C_i-R_i |R_i \leq C_i] $, we have
		\begin{equation}
			\begin{aligned}
				&\sum\limits_{i = \lfloor NC(\eta)-\delta N \rfloor+1}^{\lfloor NC(\eta)+\delta N \rfloor} \mathbb{E}[\kappa]\mathbb{P}(R_i \leq C_i) \mathbb{E}[C_i-R_i |R_i \leq C_i]\\
				\leq & \sum\limits_{i = \lfloor NC(\eta)-\delta N \rfloor+1}^{\lfloor NC(\eta)+\delta N \rfloor} \mathbb{E}[\kappa]\mathbb{E}[C_i-R_i |R_i \leq C_i]\\
				\leq &  \sum\limits_{i = \lfloor NC(\eta)-\delta N \rfloor+1}^{\lfloor NC(\eta)+\delta N \rfloor} \mathbb{E}[\kappa] \\
				= & 2\delta N\mathbb{E}[\kappa].
			\end{aligned}
		\end{equation}
	\end{enumerate}
	
	\item $sw^{\rm loss}_2$: $sw^{\rm loss}_2$ denotes the social welfare loss in block 2: Note that based on equation (\ref{Eq:prob53}), for any $\lfloor NC(\eta) + \delta N\rfloor+1 \leq i \leq \min\{\rho N,N\}$, we have 
	\begin{equation}
		\mathbb{P}(R_i \geq C_i) \approx 0.
	\end{equation} 
	In other words, when $N \rightarrow \infty$, the probability of there are transaction left for block 2 is zero. Hence, the social welfare loss due to block 2 is zero: 
	\begin{equation}
		sw^{\rm loss}_2 = 0.
	\end{equation}
\end{itemize}

Based on all these discussions, we can estimate the ratio between the social optimum and social welfare under block size $A =\lfloor NC(\eta) + \delta N\rfloor$ when $N \rightarrow \infty$:
\begin{equation}\label{eq:alpha2}
	\begin{aligned}
		&\frac{sw^{\rm opt}}{sw}\\
		=& \frac{sw^{\rm opt}}{\sum\limits_{i = 1}^{\lfloor NC(\eta)-\delta N \rfloor} \mathbb{E}[\kappa]\mathbb{P}(R_i \geq C_i) \mathbb{E}[R_i-C_i|R_i\geq C_i] + \sum\limits_{i = \lfloor NC(\eta)-\delta N \rfloor+1}^{\lfloor NC(\eta)+\delta N \rfloor} \mathbb{E}[\kappa]\mathbb{P}(R_i \geq C_i) \mathbb{E}[R_i-C_i|R_i\geq C_i] sw^{\rm loss}_1 - sw^{\rm loss}_2}\\
		< & \frac{sw^{\rm opt}}{\sum\limits_{i = 1}^{\lfloor NC(\eta)-\delta N \rfloor} \mathbb{E}[\kappa]\mathbb{P}(R_i \geq C_i) \mathbb{E}[R_i-C_i|R_i\geq C_i] - sw^{\rm loss}_1 - sw^{\rm loss}_2} \\
		< & \frac{(N +1) \int_{0}^{C(\eta)-\delta} \min\Big\{B^{-1}(1-\frac{x}{\rho}),Q^{-1}(1-x)\Big\}\left(R^{-1}(1-\frac{x}{\rho}) - C^{-1}(x)\right) dx + 2\delta N}{\mathbb{E}[\kappa]\left[(\rho N +1) \int_{1-\frac{C(\eta)-\delta}{\rho}}^{1}R^{-1}(x) dx - (N +1) \int_{0}^{C(\eta)-\delta}C^{-1}(x) dx- 2\delta N\right] }\\
		\approx & \frac{\int_{0}^{C(\eta)-\delta} \min\Big\{B^{-1}(1-\frac{x}{\rho}),Q^{-1}(1-x)\Big\}\left(R^{-1}(1-\frac{x}{\rho}) - C^{-1}(x)\right) dx + 2\delta}{\mathbb{E}[\kappa]\left[\rho \int_{1-\frac{C(\eta)-\delta}{\rho}}^{1}R^{-1}(x) dx - \int_{0}^{C(\eta)-\delta}C^{-1}(x) dx- 2\delta \right] }\\
		= & \frac{\int_{0}^{C(\eta)-\delta} \min\Big\{B^{-1}(1-\frac{x}{\rho}),Q^{-1}(1-x)\Big\}\left(R^{-1}(1-\frac{x}{\rho}) - C^{-1}(x)\right) dx + 2\delta}{\mathbb{E}[\kappa]\left[\int_{0}^{C(\eta)-\delta}R^{-1}(1-\frac{x}{\rho}) - C^{-1}(x) dx- 2\delta \right] },
	\end{aligned}
\end{equation}
where 
\begin{equation}
		\mathbb{E}[\kappa] = \mathbb{E}[b] + \mathbb{E}[q]  - \int_0^1 \kappa[b(\kappa)Q(\kappa) + B(\kappa)q(\kappa)] d\kappa. \\
\end{equation}
As the buying and selling quantities are within $[\underline{b},\overline{b}]$, we have $\min\Big\{B^{-1}(1-\frac{x}{\rho}),Q^{-1}(1-x)\Big\} \leq \overline{b}$ and $	\mathbb{E}[\kappa] \geq \underline{b}$. Hence, (\ref{eq:alpha2}) becomes
\begin{equation}\label{eq:alpha3}
	\begin{aligned}
		&\frac{sw^{\rm opt}}{sw}\\
		< & \frac{\int_{0}^{C(\eta)-\delta} \min\Big\{B^{-1}(1-\frac{x}{\rho}),Q^{-1}(1-x)\Big\}\left(R^{-1}(1-\frac{x}{\rho}) - C^{-1}(x)\right) dx + 2\delta}{\mathbb{E}[\kappa]\left[\int_{0}^{C(\eta)-\delta}R^{-1}(1-\frac{x}{\rho}) - C^{-1}(x) dx- 2\delta \right] }\\
		= & \frac{\overline{b}\left[\int_{0}^{C(\eta)-\delta} \left(R^{-1}(1-\frac{x}{\rho}) - C^{-1}(x)\right) dx + 2\delta\right]}{\underline{b}\left[\int_{0}^{C(\eta)-\delta}R^{-1}(1-\frac{x}{\rho}) - C^{-1}(x) dx- 2\delta \right] }\\
		= &  \frac{\overline{b}}{\underline{b}} \cdot\frac{\int_{0}^{C(\eta)-\delta} \left(R^{-1}(1-\frac{x}{\rho}) - C^{-1}(x)\right) dx + 2\delta}{\int_{0}^{C(\eta)-\delta}R^{-1}(1-\frac{x}{\rho}) - C^{-1}(x) dx- 2\delta  }\\
		= &  \frac{\overline{b}}{\underline{b}} \cdot \frac{\rho \int_{1-\frac{C(\eta)-\delta}{\rho}}^{1}R^{-1}(x) dx - \int_{0}^{C(\eta)-\delta}C^{-1}(x) dx + 2\delta }{ \rho\int_{1-\frac{C(\eta)-\delta}{\rho}}^{1}R^{-1}(x) dx - \int_{0}^{C(\eta)-\delta}C^{-1}(x) dx - 2\delta}.
	\end{aligned}
\end{equation}

We take the limit on both side of (\ref{eq:alpha3}), then we have:
\begin{equation}\label{eq:alpha4}
		\lim\limits_{\substack{N\rightarrow \infty \\ \delta = N^{-\psi}}} \frac{sw^{\rm opt}}{sw}\leq    \lim\limits_{\substack{N\rightarrow \infty \\ \delta = N^{-\psi}}}\frac{\overline{b}}{\underline{b}} \cdot \frac{\rho \int_{1-\frac{C(\eta)-\delta}{\rho}}^{1}R^{-1}(x) dx - \int_{0}^{C(\eta)-\delta}C^{-1}(x) dx + 2\delta }{ \rho\int_{1-\frac{C(\eta)-\delta}{\rho}}^{1}R^{-1}(x) dx - \int_{0}^{C(\eta)-\delta}C^{-1}(x) dx - 2\delta}.
\end{equation}
Based on (\ref{eq:squ2}), we have: 
\begin{equation}\label{eq:alpha5}
	\lim\limits_{\substack{N\rightarrow \infty \\ \delta = N^{-\psi}}} \frac{sw^{\rm opt}}{sw}\leq    \lim\limits_{\substack{N\rightarrow \infty \\ \delta = N^{-\psi}}}\frac{\overline{b}}{\underline{b}} \cdot \frac{\rho \int_{1-\frac{C(\eta)-\delta}{\rho}}^{1}R^{-1}(x) dx - \int_{0}^{C(\eta)-\delta}C^{-1}(x) dx + 2\delta }{ \rho\int_{1-\frac{C(\eta)-\delta}{\rho}}^{1}R^{-1}(x) dx - \int_{0}^{C(\eta)-\delta}C^{-1}(x) dx - 2\delta}= \frac{\overline{b}}{\underline{b}}.
\end{equation}

Hence, the ratio between the social optimum and social welfare for heterogeneous trading quantity is upper bounded by $\frac{\overline{b}}{\underline{b}}$. The PoA, as the maximum ratio, is upper bounded by $\frac{\overline{b}}{\underline{b}}$.

\textbf{This completes the proof of heterogeneous-quantity trading.}

\textbf{This completes the proof of Theorem 4.}


\newpage
{\color{blue}
\section{Experiment}

This section provides a comprehensive introduction to the implementation of our Adjustable Block Size (ABS) mechanism in the Ethereum blockchain and the experiment. The ABS mechanism aims to dynamically adjust the block size, which in Ethereum is represented by the gas limit of each block.

\subsection{Experiment Overview} 
 In Ethereum, the block size is determined by the gas limit. To implement our ABS mechanism, we adjust the gas limit of each block to control the block size. We utilize Geth version 1.11.2 \cite{Eth_client} as the Ethereum client and modify its source code to implement the gas limit adjustment. After the implementation, we run the blockchain protocol and generate the buying and selling transactions to match.

\subsection{Experiment Details}
 The experiment includes four steps:

\subsubsection{Modifying Geth Source Code}
To adjust the gas limit of each block, we modify the ``block\_validator.go'' file in the Geth source code. Specifically, we update the function responsible for calculating the gas limit ``CalcGasLimit''. In order to set the block size to accommodate approximately 100 transactions, we configure the gas limit to 2,100,000 since each typical transaction consumes approximately 21,000 gas. The code modification is shown in Figure \ref{Fig:gas}, where we add comments to the original code of ``CalcGasLimit'' and include our modifications. Additionally, we update the gas limit in the ``genesis.json'' file to the following value: 
\begin{verbatim}
	"gasLimit": "0x200b20", 
\end{verbatim}
This change is made because the gas limit is specified in hexadecimal format.

\begin{figure}[h]
	\centering
	{\includegraphics[width=14cm]{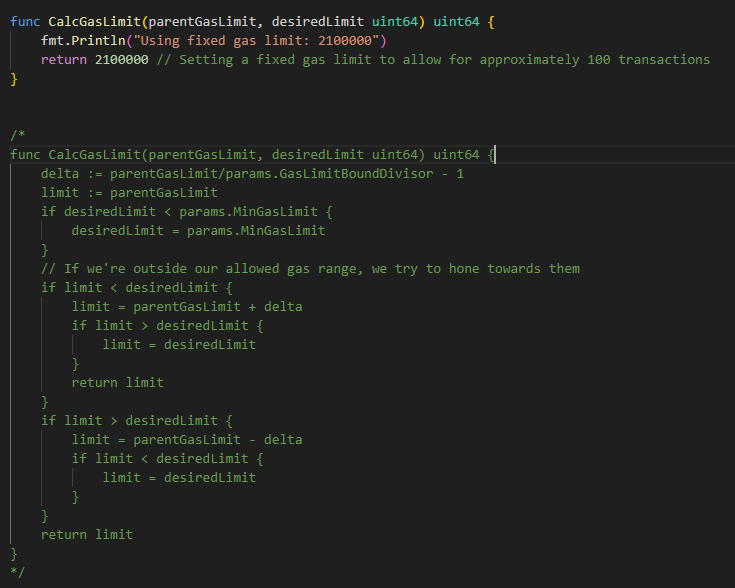}}
	\vspace{-3mm}
	\caption{Gas limit adjustment.}\label{Fig:gas}
	\vspace{-6mm}
\end{figure}

\newpage
\vspace{-10mm}
\subsubsection{ Running the Modified Ethereum Client}
After modifying the gas limit, we run a local instance of the Ethereum blockchain using the modified Geth client in a Cygwin64 terminal. The steps to achieve this are outlined below:

\begin{enumerate}
	\item Navigate to the Geth client directory and compile the file. The commands are as follows:
	\begin{verbatim}
		cd /cygdrive/d/Program\ Files/go-ethereum-1.11.2.2
		
		make geth
	\end{verbatim}
	\item Remove blockchain data from any previous experiments (if it exists). The command is:
	\begin{verbatim}
		geth removedb
	\end{verbatim}
	\item Initialize the genesis file and run the modified client. The commands are:
	\begin{verbatim}
		./build/bin/geth init "D:\Program Files\go-ethereum-1.11.2.2
		\cmd\devp2p\internal\ethtest\testdata\genesis.json"
		
		./build/bin/geth --networkid 12345 --nodiscover --mine --
		miner.etherbase 0xF90FDcB13069361f27e3aC5a2A1aca66E37437f2 
		--miner.threads 1 --http --http.port 8545 --http.api web3,
		eth,net,personal
	\end{verbatim}
\end{enumerate}
 Figure \ref{Fig:blockchain_history} shows the operation result of blockchain execution after running our modified client. It shows the mining history of blocks, and the red rectangle in the figure shows that we have modified the fixed gas limit to 2,100,000.

\begin{figure}[h]
	\centering
	{\includegraphics[width=18cm]{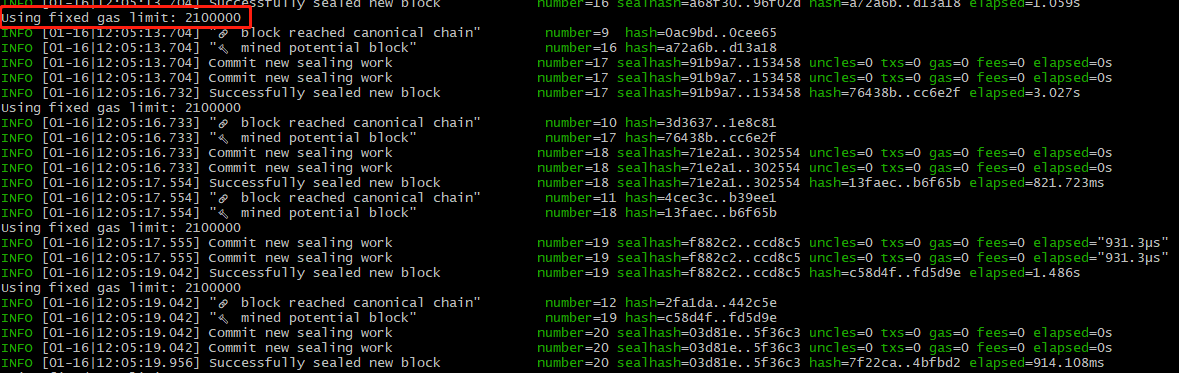}}
	\caption{Ethereum blockchain mining.}\label{Fig:blockchain_history}
\end{figure}

\subsubsection{ Verification of Gas Limit:}
We open a new Cygwin64 terminal to verify the implementation by inspecting the gas limit of the mined blocks. After navigating to the appropriate directory, we retrieved the block information. The code executed is as follows:
\begin{verbatim}
	cd /cygdrive/d/Program\ Files/go-ethereum-1.11.2.2
	
	curl -X POST --data "{\"jsonrpc\":\"2.0\",\"method\":\"eth_
		getBlockByNumber\",\"params\":[\"latest\",true],\"id\":1}" 
	-H "Content-Type: application/json" http://localhost:8545
\end{verbatim}

 Figure \ref{Fig:block_info} displays the results of executing the code. The red rectangle in the figure highlights that the block's gas limit is indeed set to 0x200b20 in hexadecimal, which is equivalent to 2,100,000 and matches the value we configured. This confirms that the ABS mechanism has been successfully implemented.\\

\begin{figure}[h]
	\centering
	{\includegraphics[width=18cm]{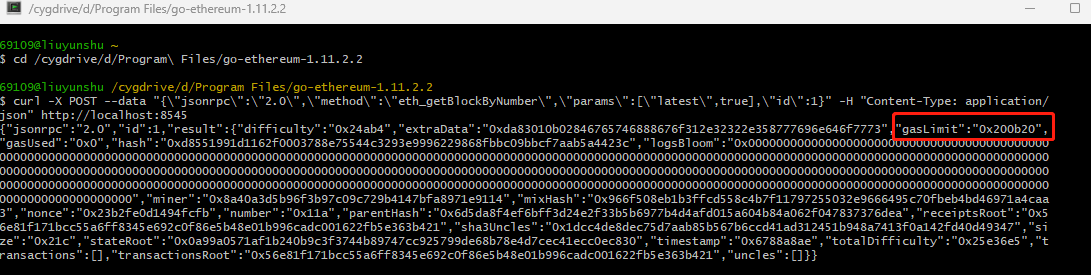}}
	\caption{A block's gas limit.}\label{Fig:block_info}
\end{figure}

\subsubsection{Transaction Generation:} We open a new cmd to generate the buying and selling transactions. First, we navigate to the folder and install Web3.js with specific version:

\begin{verbatim}
	cd /d D:\Program Files\go-ethereum-1.11.2.2\
	ethereum-transaction
	
	npm install web3@1.6.1
\end{verbatim}
Next is to run the code
\begin{verbatim}
	node sendTransaction3.js
\end{verbatim}
The code of ``sendTransaction3.js'' is as follows:
\begin{verbatim}
	const Web3 = require('web3');
	const web3 = new Web3('http://localhost:8545');  // Initialize Web3 with HTTP provider
	
	// Your account and private key
	const account = '0xF90FDcB13069361f27e3aC5a2A1aca66E37437f2';
	const privateKey = '0xec65482032b5cf235cffc47aacd4d9f2e1fc71d670e82c9110d17a4776ea6dde';
	
	const recipientAccounts = [
	'0xeA8EC02a56D1a334C61036032Bb9ad349FB6C160',
	'0xc4cc8Fe4B613C7774157FEd7d7F898C52b415eF2',
	'0x303bDCeD279e89bb3B0f10A9c2f4738Bdc8496e3',
	'0xcb1E67065D274DD8A2bb03fa77C798fC902ec670',
	'0x18F3F58D9F408723f2E2EFAaAC756dDf0Fc05869',
	'0x65B2E2E97B2076B27aE15713D2a0916694E9220F',
	'0xAd3F680F5d594A49c4b39926A40c3Af832Ecfacb',
	'0xEF9835504AD5A42CdAABeE58c7f8299534Ff14DA',
	'0xaBeCed76e700790cC03EFD2691fAc9D82706e97e',
	'0x49591Eaaa296180ba24ad7568Ccb30946Fd6d7c0',
	'0xf36137dEA116d4296485534CE90c3D88A56d13E2',
	'0xF241E66491f58F0C0FAf82a7F7eB0EF857742550',
	'0x61f244C8D7B354B09fa657b2DA6C36571fcac821',
	'0x7b369bFC4b3BFAc833c84521424C4f41c037B3a8',
	'0xdc332E90Ed666eed9330Cf72740f14B17A9ae36c',
	'0x84796451696BdBA236E9b82B19B06Be831F41Ca8',
	'0x259648887B2A2322007C06B8C33382B460d1017F',
	'0xa9F6ac59Eb8ECF113D2e05FBDa8a567fe66774c9',
	'0x43f01580C9E2f38548b9687930db349cfD8c6B89',
	'0x006498deE48fd6dd4b300fb25438b0C741C61613'
	];
	
	// Ensure the address is checksummed
	const checksumAddress = web3.utils.toChecksumAddress(account);
	
	// Add the private key to Web3's wallet
	web3.eth.accounts.wallet.add(privateKey);  // Add private key to Web3 wallet
	
	// Function to generate a random number within a range
	const getRandomInRange = (min, max) => {
		return Math.floor(Math.random() * (max - min + 1)) + min;
	};
	
	// Function to send a single transaction
	const sendSingleTransaction = async (index, nonce) => {
		// Generate a random gas price between 10 gwei and 100 gwei (for example)
		const gasPrice = web3.utils.toWei(getRandomInRange(10, 100).toString(), 'gwei');
		
		// Select a random recipient from the list
		const recipient = recipientAccounts[index % recipientAccounts.length];  // Ensure we don't go out of bounds
		
		// Prepare the transaction details
		const tx = {
			from: checksumAddress,
			to: recipient,  // Randomized recipient address
			value: web3.utils.toWei('0.1', 'ether'),  // Transaction value in Ether (0.1 ETH for each transaction)
			gas: 21000,  // Gas limit
			gasPrice: gasPrice,  // Randomized gas price
			nonce: nonce  // Unique nonce for each transaction
		};
		
		// Sign and send the transaction
		try {
			const signedTx = await web3.eth.accounts.signTransaction(tx, privateKey);
			const receipt = await web3.eth.sendSignedTransaction(signedTx.rawTransaction);
			console.log(`Transaction ${index + 1} successful:`, receipt);
		} catch (err) {
			console.error(`Error sending transaction ${index + 1}:`, err);
		}
	};
	
	// Function to send 10 transactions concurrently with unique nonces
	const sendTransactionsConcurrently = async () => {
		// Get the current nonce for the sender's account
		const currentNonce = await web3.eth.getTransactionCount(checksumAddress);
		
		const transactionPromises = [];
		
		for (let i = 0; i < 10; i++) {
			// Add each transaction promise to the array with unique nonce
			transactionPromises.push(sendSingleTransaction(i, currentNonce + i));  // Increment nonce for each transaction
		}
		
		// Wait for all transactions to complete concurrently
		await Promise.all(transactionPromises);
	};
	
	// Execute the function to send 10 transactions concurrently
	sendTransactionsConcurrently().then(() => {
		console.log('All transactions have been sent.');
	}).catch(err => {
		console.error('Error in sending transactions:', err);
	});
\end{verbatim}

Figure \ref{Fig:tx_info} displays the results that miners successfully record the transactions in the blockchain. This completes the experiment.\\

\begin{figure}[h]
	\centering
	{\includegraphics[width=18cm]{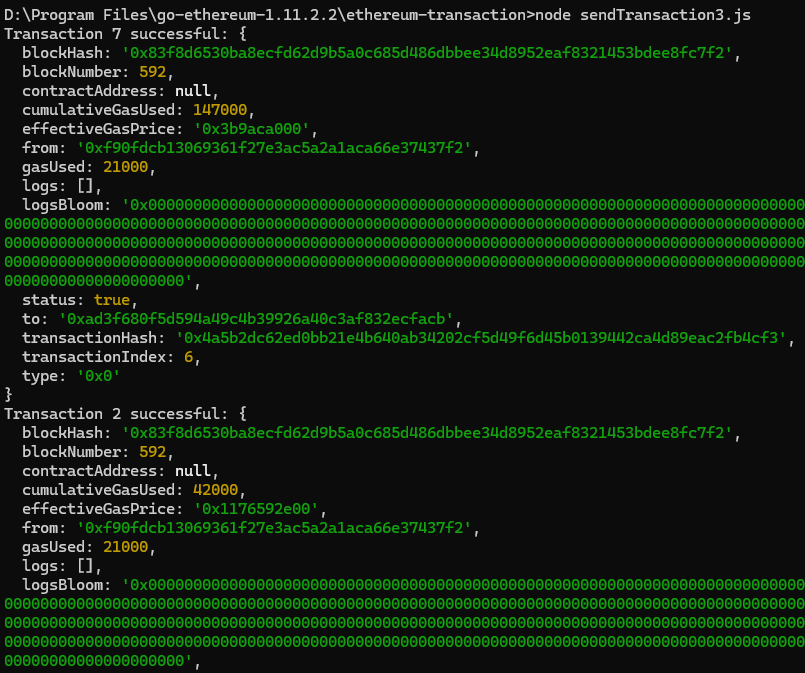}}
	\caption{Transaction recorded in blockchain.}\label{Fig:tx_info}
\end{figure}

Through the experiment, we successfully adjusted the block size in the Ethereum blockchain by modifying the gas limit of each block and complete matching. This implementation demonstrates the feasibility and practicality of ABS mechanism.}

\bibliography{Info_2023}
\end{document}